 \documentclass[]{aa} 

\usepackage{graphicx}
\usepackage{txfonts}
\usepackage{color,ulem}
\newcommand{\sub }[1]{_{\mathrm{#1}}}
\newcommand{\Mstar}{M_{\star}}
\newcommand{\Lstar}{L_{\star}}
\newcommand{\Rstar}{R_{\star}}
\newcommand{\Teff}{T\sub{eff}}
\newcommand{\Mdot}{\dot{M}\sub{acc}}
\newcommand{\sacc}{s\sub{acc}}
\newcommand{\tacc}{t\sub{acc}}

\newcommand{\Mrad}{M\sub{BCZ}}

\newcommand{\Msun}{  M_\sun }

\newcommand{\Rsun}{  R_\sun }
\newcommand{\Lsun}{  L_\sun}

\newcommand{\Mearth}{ M_\oplus}

 \newcommand{ \Ys }{ Y\sub{surf} }
 \newcommand{ \Zs }{ Z\sub{surf} }
 \newcommand{ \ZXs }{ (Z/X)\sub{surf} }
 \newcommand{ \Zc }{ Z\sub{center} }
 \newcommand{ \Xc }{ X\sub{center} }
 \newcommand{ \Yc }{ Y\sub{center} }
 \newcommand{ \Tc }{ T\sub{center} }
 \newcommand{ \Xacc }{ X\sub{acc} } 
 \newcommand{ \Yacc }{ Y\sub{acc} } 
 \newcommand{ \Zacc }{ Z\sub{acc} } 
 \newcommand{ \Xaccini }{ X\sub{proto} } 
 \newcommand{ \Yaccini }{ Y\sub{proto} } 
 \newcommand{ \Yaccmin }{ Y\sub{acc,min} } 
 \newcommand{ \Zaccini }{ Z\sub{proto} }
 \newcommand{ \Zaccmax }{ Z\sub{acc,max} }
 \newcommand{ \DZacc }{ \Delta \Zacc }
 \newcommand{ \fov }{ f\sub{overshoot} }
 \newcommand{ \amlt }{ \alpha\sub{MLT} }
 \newcommand{ \Hp }{ H_p }
 \newcommand{ \fopa }{ \delta_\kappa }
 \newcommand{ \csobs }{ c_{s,\rm obs} }
 \newcommand{ \delcs }{ \delta c_s }
 
 \newcommand{ \cs }{ c_s }
 \newcommand{ \RCZ }{ R\sub{CZ} }
 \newcommand{\Mlost}{M_{XY,\mathrm{lost}}}
 \newcommand{\Mpl}{M_{Z,\mathrm{planet}}}
 \newcommand{\Xproto}{X\sub{proto}}
 \newcommand{\Yproto}{Y\sub{proto}}
 \newcommand{\Zproto}{Z\sub{proto}}
 \newcommand{\Zcloud}{\Zproto}
\newcommand{ \FULL }{MZvar}
\newcommand{ \FULLnoov }{MZvar-noov}
\newcommand{ \kapa }{K23}
\newcommand{ \kapb }{K2}
\newcommand{ \kapc }{K23$'$}
\newcommand{ \kappla }{K2-MZvar}
\newcommand{ \kapplb }{K23-MZvar}
\newcommand{ \unused }{ --- }
\newcommand{ \fixed }[1]{ \textit{#1} }

\defcitealias{GS98}{GS98}
\defcitealias{Asplund+09}{AGSS09}
\usepackage[colorlinks = true,
        linkcolor = blue,
        urlcolor  = blue,
        citecolor = blue,
        anchorcolor = blue
]{hyperref}
\makeatletter
\renewcommand*\aa@pageof{, page \thepage{} of \pageref*{LastPage}}
\makeatother

\bibpunct{(}{)}{;}{a}{}{,} 

\begin{document}

    \title{Imprint of planet formation in the deep interior of the Sun\thanks{
   Supplemental materials are available at the CDS via anonymous ftp to \texttt{cdsarc.u-strasbg.fr (130.79.128.5)} or via \url{http://cdsarc.u-strasbg.fr/viz-bin/qcat?J/A+A/***/***} or at \url{https://doi.org/10.5281/zenodo.5506424}
   }}

   \subtitle{}

   \author{Masanobu Kunitomo
          \inst{1}
          \and
          Tristan Guillot\inst{2}
          }

   \institute{Department of Physics, School of Medicine, Kurume University, 67 Asahimachi, Kurume, Fukuoka 830-0011, Japan\label{inst1}\\
		\email{kunitomo.masanobu@gmail.com}
   		\and
		Laboratoire Lagrange, UMR 7293,
        Universit\'e C\^ote d'Azur,
        CNRS,
        Observatoire de la C\^ote d'Azur,
        06304 Nice CEDEX 04, France\label{inst2}
}
   \date{Received 6 May 2021 / Accepted 4 August 2021}
   \authorrunning{M. Kunitomo \& T. Guillot}

 
  \abstract
{
In protoplanetary disks, the growth and inward drift of dust lead to the generation of a temporal ``pebble wave'' of increased metallicity. This phase must be followed by a phase in which the exhaustion of the pebbles in the disk and the formation of planets lead to the accretion of metal-poor gas. At the same time, disk winds may lead to the selective removal of hydrogen and helium from the disk. Hence, stars grow by accreting gas that has an evolving composition. In this work, we investigated how the formation of the Solar System may have affected the composition and structure of the Sun, and whether it plays any role in solving the so-called solar abundance problem, that is, the fact that standard models with up-to-date lower-metallicity abundances reproduce helioseismic constraints significantly more poorly than those with old higher-metallicity abundances.
We simulated the evolution of the Sun from the protostellar phase to the present age and attempted to reproduce spectroscopic and helioseismic constraints. We performed chi-squared tests to optimize our input parameters, which we extended by adding secondary parameters. These additional parameters accounted for the variations in the composition of the accreted material and an increase in the opacities. 
We confirmed that, for realistic models, planet formation occurs when the solar convective zone is still massive; thus, the overall changes due to planet formation are too small to significantly improve the chi-square fits. 
We found that solar models with up-to-date abundances require an opacity increase of 12\% to 18\% centered at $T=10^{6.4}$\,K to reproduce the available observational constraints. This is slightly higher than, but is qualitatively in good agreement with, recent measurements of higher iron opacities. 
These models result in better fits to the observations than those using old abundances; therefore, they are a promising solution to the solar abundance problem. 
Using these improved models, we found that planet formation processes leave a small imprint in the solar core, whose metallicity is enhanced by up to 5\%. This result can be tested by accurately measuring the solar neutrino flux. 
In the improved models, the protosolar molecular cloud core is characterized by a primordial metallicity in the range $\Zproto=0.0127$--0.0157 and a helium mass fraction in the range $\Yproto=0.268$--0.274.
}


   \maketitle
%

\section{Introduction}
\label{sec:intro}

Young stellar objects grow mainly by accreting mass from a circumstellar or protoplanetary disk. Planet formation theories predict that the composition of the disk should vary with time owing to the growth of dust grains into centimeter-sized grains, or ``pebbles'', that drift toward the central star, the formation of planetesimals and planets, and the presence of disk winds. In our previous studies, we provided constraints on the evolutionary models of protostellar and pre-main-sequence (pre-MS) stars, including accretion \citep{Kunitomo+17}, and also investigated the impact of the evolving composition of the accreting material on the composition of the stellar surface \citep{Kunitomo+18}. We showed that some of the anomalies observed in the chemical composition of stellar clusters, $\lambda$ Boo stars, and binary stars can be explained by planet formation processes. Moreover, our studies revealed that the large deficit in the Sun's refractory elements compared to that in solar twins, as found by \citet[][]{Melendez+09}, may be explained by the formation of giant planets in the Solar System with a very high rock-to-ice ratio \citep{Kunitomo+18} \citep[see][for an alternative explanation]{Booth+Owen20}.
Given the detailed information and constraints available regarding the composition and structure of the Sun, it would be interesting to investigate the imprint of planet formation processes in the Sun that may be detected by observations.

The main characteristics of the Sun, namely, its radius, luminosity, and age, are precisely known. However, the theoretical modeling of the Sun's atmospheric composition, which is determined using high-resolution spectroscopy, has significantly evolved in the past decade. This is owing to the replacement of old one-dimensional atmospheric models by more accurate three-dimensional models with new atomic data that account for convection and nonlocal thermodynamic equilibrium effects, which substantially improved the fit to the observed spectral lines \citep{Asplund+05,Asplund+09}.
As a consequence, the inferred present-day solar surface metallicity decreases significantly from $\Zs=0.018$ \citep[][hereafter \citetalias{GS98}]{GS98} to $\Zs=0.013$ \citep[][hereafter \citetalias{Asplund+09}]{Asplund+09}.

The best constraints on the internal structure of the Sun are obtained from helioseismology, which helps determine not only the location of the base of the convective zone (CZ) but also the sound speed from the radius $r \approx 0.1\,\Rsun$ to the solar surface, where $\Rsun$ is the solar radius \citep[e.g.,][]{Basu16}. However, it has been shown that the old high-$Z$ solar models result in significantly better fits to the observed sound speed compared to the low-$Z$ solar models. This is the so-called solar abundance problem \citep[see reviews by][]{Asplund+09,Serenelli16}, which has been the focus of many studies. Potential solutions to the solar abundance problem include the consideration of increased opacities \citep{Bahcall+05,Christensen-Dalsgaard+09,Villante10,Ayukov+Baturin13, Bailey+15,Vinyoles+17}, increased efficiencies of diffusion \citep[so-called extra mixing;][]{Christensen-Dalsgaard+18,Buldgen+19}, diffusion due to the solar rotation \citep[so-called rotational mixing;][]{Yang19}, helium-poor accretion \citep{Zhang+19}, an updated solar composition \citep{Young18}, and revised nuclear reaction rates \citep{Ayukov+Baturin17}. Accretion models with a time-dependent composition, in particular, low-$Z$ accretion in the late pre-MS phase, have also been investigated but were found to be unsuccessful in solving the solar abundance problem \citep{Castro+07,Guzik+Mussack10,Serenelli+11}.

To investigate the effect of planet formation on the structure and composition of the Sun requires a quantitative analysis of: (1) various planet formation processes and (2) solar evolution models that consider the accretion of matter with a time-dependent composition to reproduce all observational constraints. In the present work, which is an extension of our previous works \citep{Kunitomo+17, Kunitomo+18}, we conducted a thorough search for solutions to the solar abundance problem by applying our accretion models with a time-dependent composition to optimized solar evolution models. 

The remainder of this paper is organized as follows. 
In Sect.\,\ref{sec:context}, we discuss how the growth of dust grains, the radial drift of pebbles, and planet formation affect the composition of the gas accreted by the proto-Sun during its protostellar and pre-MS evolution. In Sect.\,\ref{sec:method}, we describe the method used to compute our stellar evolution models and the procedure followed to optimize these models to reproduce the observational constraints.
In Sect.\,\ref{sec:results}, we present the results of different solar evolution models and compare them with observations; in addition, we show that an increased opacity appears to be the most likely solution for the solar abundance problem. Using these models, in Sect.\,\ref{sec:discussion-planets}, we examine the effect of planet formation on the composition and structure of the present-day Sun. 
Our results are summarized in Sect.\,\ref{sec:conclusion}.

\section{Context}
\label{sec:context}

\subsection{Evolution of the composition of protoplanetary disks}
\label{sec:planets}

During the collapse of a molecular cloud core, a large fraction of the material first forms a circumstellar disk around the central star (or stars) \citep{Inutsuka12}. Next, the transport of angular momentum from the inner region of the disk to the outer region causes the matter to be accreted by the central star(s) \citep{Lynden-Bell+Pringle74} on a timescale of order 1--10\,Myr \citep[e.g.,][]{Hartmann+16}. During this time, initially micron-sized dust grains grow larger in size, as indicated by the dust emission at millimeter wavelengths \citep[e.g.,][]{Beckwith+90}. A comparison of the mass of heavy elements present in exoplanetary systems and that inferred to be present in protoplanetary disks indicates that these grains must grow further in size to form planetesimals and even planets during the early stages of disk evolution \citep{Manara+18,Tychoniec+20}.
Dust growth in the Class 0/I phases of young stellar objects has also been investigated in several theoretical studies \citep[e.g.,][]{Tsukamoto+17, Tanaka+Tsukamoto19}. 

The formation of planetesimals and planets is likely to lead to a decrease in the metallicity of the gas accreted by the central star(s). \citet{Kunitomo+18} estimated that approximately 97 to $168\,\Mearth$ of heavy elements in the Solar System were either used to form planets or were ejected from the system. 
This is a small value compared to the total mass of heavy elements present in the Sun, which is estimated to be approximately $5000\,\Mearth$; however, it can still cause the metallicity of the accreted gas to change, particularly during the later stages of stellar accretion. Moreover, it has been proposed that the peculiar composition of $\lambda$ Boo stars can be explained by the presence of giant planets in the surrounding protoplanetary disks \citep{Kama+15}. 

It is also worth noting that as the grains grow in size, they drift rapidly inward in the protoplanetary disk \citep{Adachi+76, Weidenschilling77a}. This can create a wave of solids that may be lost to the central star in the absence of any other processes \citep{Stepinski+Valageas96, Garaud07}. Circumstellar disks have been proposed to be large (100\,au or more in size) and expanding \citep[e.g.,][]{Hueso+Guillot05}, which implies that the outer part of the disk acts as a reservoir in which the dust grains grow and from which they are gradually released. By assuming the standard theory of grain growth and adopting the $\alpha$-disk model \citep[][]{Shakura+Sunyaev73}, \citet{Garaud07} demonstrated that the timescale in which the solids are released from the outer disk is determined by a balance between the outward propagation of the disk and the growth of the dust grains. 

The metallicity $\Zacc$ of the gas accreted by the proto-Sun can be calculated from the ratio between the mass fluxes of the dust $\dot{M}_{\rm d}$ and gas $\Mdot$, such that
\begin{equation}
    \Zacc=\frac{\dot{M}_{\rm d}}{\Mdot}\,.
\end{equation}
The above equation assumes that the metallicity is equal to the amount of condensing material. We note that because most elements heavier than hydrogen and helium condense in the protoplanetary disk, the above assumption is justified in our case.

\begin{figure}[tbp]
  \begin{center}
        \includegraphics[width=\hsize,keepaspectratio]{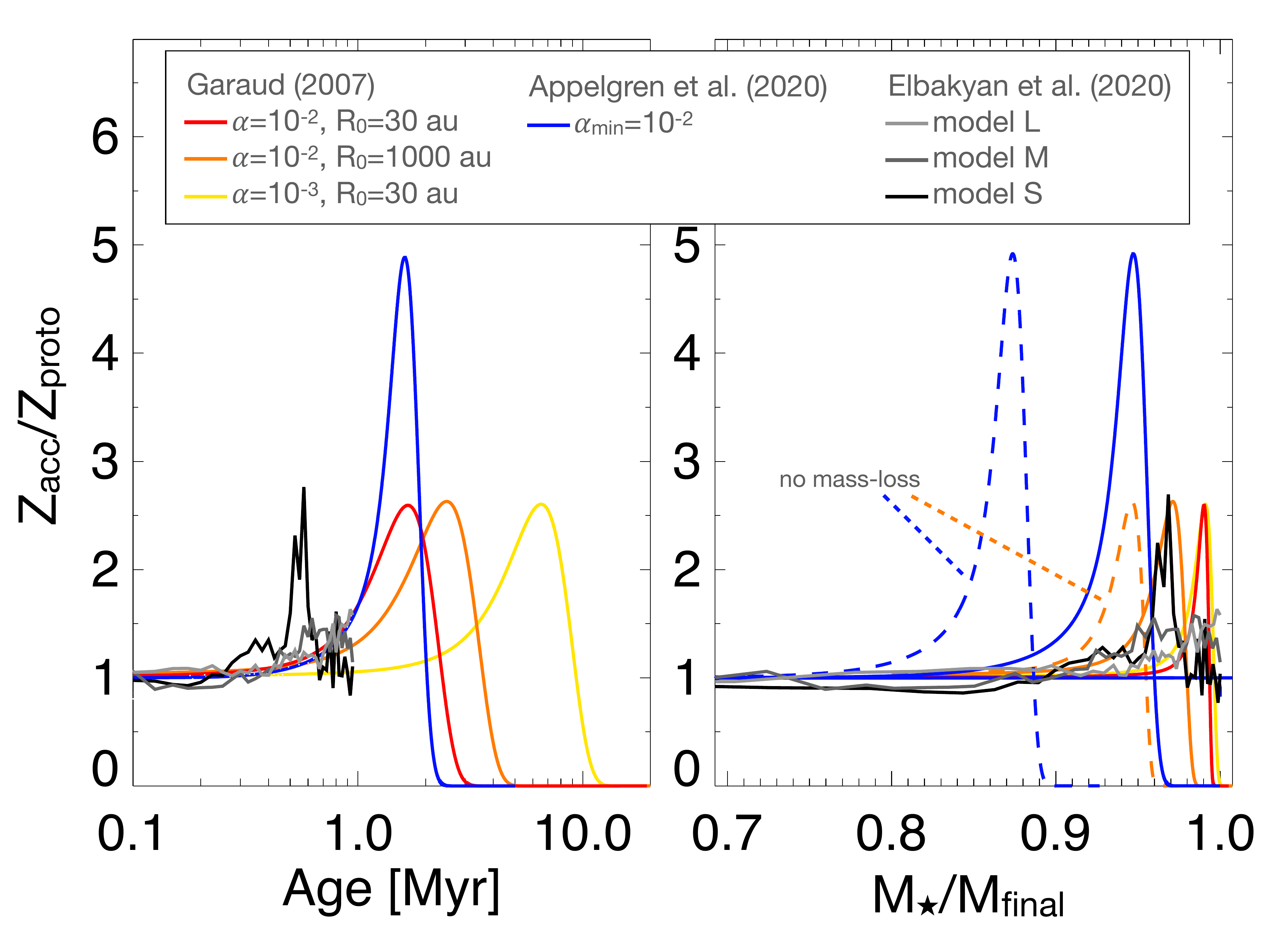}
        \caption{\small{
        Evolution of the metallicity $\Zacc$ of the accreted gas in units of the initial metallicity $\Zaccini$ as a function of time (left panel) and mass of the proto-Sun (right panel). Several evolutionary models of circumstellar disks that include both gas and dust are shown (see text for details). The protostar mass was calculated assuming some mass loss (e.g., due to partial photoevaporation of the circumstellar disk) except in two cases, which are represented by dashed lines.}}
        \label{fig:Eacc}
    \end{center}
\end{figure}

Figure~\ref{fig:Eacc} shows the evolution of $\Zacc$ with time and the mass of the proto-Sun, as obtained from various studies on the evolution of dust in protoplanetary disks. In the beginning, $\Zacc=\Zcloud$, where $\Zcloud$ is the primordial metallicity of the molecular cloud core. Eventually, $\Zacc$ increases with time due to the incoming pebble wave before decreasing sharply owing to the removal of all or most of the solids from the disk.

The first set of disk models shown in Fig.~\ref{fig:Eacc} corresponds to the analytical solutions obtained by \citet{Garaud07} for $\alpha=10^{-2}$ and $10^{-3}$ and $R_0=30$\,au and 1000\,au, where $\alpha$ is the dimensionless turbulent viscosity parameter and $R_0$ is the initial disk radius. The second model corresponds to the numerical solution obtained by \citet{Appelgren+20} for a standard one-dimensional disk model. In this case, the authors assumed a variable $\alpha$ such that its value increased in the presence of gravitational instabilities but had a fixed minimum value of $\alpha_{\rm min}=10^{-2}$. Finally, the third set of models corresponds to the solutions obtained by \citet{Elbakyan+20} for a two-dimensional disk model. The accretion is highly variable in this case, and hence, we plot only the mean value over a timescale of $2.5\times 10^4$ years. 

The left panel of Fig.~\ref{fig:Eacc} shows that the pebble wave appears at less than 1 Myr and lasts until approximately 10 Myr. This timescale is extremely difficult to estimate because it depends on several factors, including the turbulent viscosity in the disk, initial angular momentum of the disk material, overall structure of the disk, and the model used to calculate the grain growth. The peak value of $\Zacc/\Zcloud$ varies from 1.5, for models L and M of \citet{Elbakyan+20}, to approximately 5, for the model proposed by \citet{Appelgren+20}. It should be noted that \cite{Mousis+19} found an even higher metallicity peak in the range 10--20. The peak in metallicity is followed by a sudden drop to a value close to the initial metallicity in the models proposed by \citet{Elbakyan+20} and to zero in the remaining models; however, in reality, the final metallicity value should lie somewhere in between. This is because we know from observations that protoplanetary disks are never completely cleared of dust \citep[e.g.,][]{Dullemond+Dominik05}, indicating that the perfect depletion obtained by \cite{Garaud07} and \cite{Appelgren+20} is due to the simplifications adopted in their models to account for the grain size distribution. Conversely, these models do not consider planetesimal and planet formation processes, which can remove solids from the system and thus lead to an additional decrease in the metallicity of the accreted gas. In addition, we note that fully formed planets can efficiently filter out pebbles and prevent them from reaching the protostar \citep{Guillot+14}. 

For the purpose of this study, it is also important to estimate how the metallicity of the accreted gas varies as a function of the mass of the protostar. The disk models above neglect the change in the mass of the protostar owing to disk accretion. Therefore, we estimated the mass of the protostar as a function of time, as follows:
\begin{equation}
\Mstar(t)=M_{\rm final}+\Mlost-{M}_{\rm g}(t),    
\end{equation}
where $t$ is time, $M_{\rm g}(t)$ is the gas mass in the disk, $M_{\rm final}$ is the final mass of the star (note that $ M_{\rm final}=1\,\Msun$ for the Sun), and $\Mlost$ is the mass ejected from the disk \citep[e.g., via photoevaporation or magnetohydrodynamic (MHD) winds; see][]{Hollenbach+00, Suzuki+Inutsuka09,Alexander+14}.
We stop the calculation of $\Mstar$ at a time $t_{\rm final}$ when $M_{\rm g}(t=t_{\rm final})=\Mlost$.
Following \citet{Guillot+Hueso06}, we assumed that predominantly hydrogen and helium were lost from the disk \citep[also, see][]{Gorti+15,Miyake+16}. We note that photoevaporation is more effective than MHD disk winds in the selective removal of hydrogen and helium during the later stages of accretion \citep{Kunitomo+20}. Although the total mass of the disk lost via winds is not well determined, we expect $\Mlost\la0.05\,\Msun$.

The right panel of Fig.~\ref{fig:Eacc} shows that the peak in the metallicity of the accreted gas occurs after approximately 90\% of the mass has been accreted onto the protostar. For most of the models, the peak occurs at a high value of the protostellar mass, namely, between 95\% and 98\%. For the disk model proposed by \cite{Appelgren+20} with no mass loss, a pebble wave occurs when the mass of the protostar is approximately 85\% of the final mass of the star. This is because in their model the disk extends beyond 1000\,au. However, such a large disk is likely to get photoevaporated rapidly due to external radiation \citep[e.g.,][]{Guillot+Hueso06,Anderson+13}. Thus, for our simulations of the Sun, we considered a situation in which the metallicity starts to increase after $0.85\,\Msun$ of the gas has been accreted, peaks between $0.9\,\Msun$ and $0.98\,\Msun$, and finally decreases to a very low value.

\subsection{Internal structure of the pre-main-sequence Sun}
\label{sec:interior}

\begin{figure*}[!tb]
    \sidecaption
        \includegraphics[width=12cm,keepaspectratio]{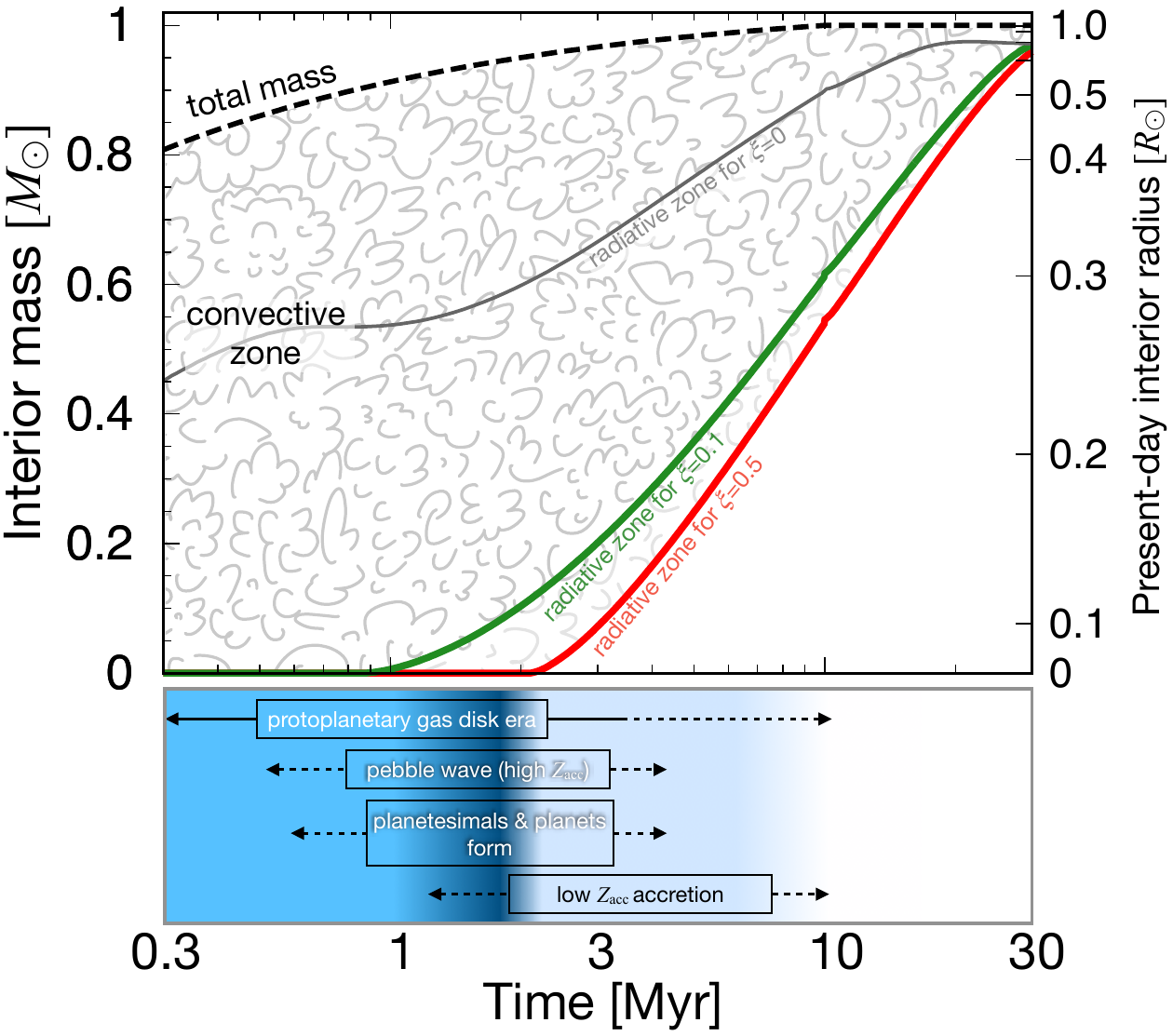}
        \caption{Early evolution of the solar interior. The total mass of the accreting proto-Sun is indicated by the dashed line. The surface CZ is depicted as a cloudy region \citep[see][]{Kippenhahn+Weigert90} delimited by a radiative zone that grows outward from the center of the Sun. Three evolution models are shown corresponding to different values of the accretion efficiency parameter $\xi$: a classical evolution model with $\xi=0.5$ (red line), a cold accretion model with $\xi=0.1$ (green line) that is a lower limit of the $\xi$ value in the observations of young clusters \citep{Kunitomo+17}, and a model with $\xi=0$ (gray line) corresponding to a theoretical proto-Sun formed in the absence of any accretion heat. The right-hand $y$-axis provides the radius corresponding to the mass on the left-hand $y$-axis of the present-day Sun (which is 4.567\,Gyr old). The bottom panel highlights the key physical processes that occur during the growth of the proto-Sun: the presence of a circumstellar gas disk (during the first million years); an increase in the metallicity $\Zacc$ of the accreted gas due to a pebble wave; the concomitant formation of planetesimals and planets; and a sudden decrease in $\Zacc$.
        }
        \label{fig:t-MCZ}
\end{figure*}

One may expect the metallicity of the accreted gas to have a significant impact on the composition of the Sun's atmosphere \citep[see, e.g.,][]{Chambers10}. However, this is not the case because of the internal structure of the proto-Sun. Studies on the evolution of the Sun, which is in agreement with the evolution of stars in young clusters of similar mass as the Sun \citep{Kunitomo+17}, indicate that the Sun must have been almost fully convective for the first 1 to 2\,Myr. Then, the CZ starts to recede slowly as the Sun evolves toward the main sequence (MS; which takes approximately 40\,Myr), finally reaching a stage where it is only 2.5\% of the total mass of the present Sun. 

Figure~\ref{fig:t-MCZ} shows the first 30\,Myr of the evolution of the solar CZ for three models adopted from \cite{Kunitomo+18}: a standard accretion model with $\xi=0.5$, which corresponds to the maximum accretion efficiency; a limiting model with $\xi=0.1$, which is in agreement with the observational data from young clusters but corresponds to an accretion efficiency of only 10\%; and a cold accretion model with $\xi=0$ for reference \citep[see][for a complete discussion]{Kunitomo+17}.
The $\xi$ value corresponds to the ratio of the accretion heat injected into the protostar to the liberated gravitational energy of accreting materials \citep{Kunitomo+17}.
In the present work, we adopted models equivalent to the standard model with $\xi=0.5$ (see Sect.\,\ref{sec:Mdot} for details).

The lifetime of protoplanetary disks is still uncertain but is thought to be less than 10\,Myr \citep[see, e.g.,][]{Haisch+01, Kennedy+Kenyon09, Fedele+10, Hartmann+16}. In fact, the magnetism observed in meteorites indicates that the lifetime of the protoplanetary disk in the Solar System was $\simeq4$\,Myr \citep[see, e.g.,][]{Wang+17,Weiss+21}. Thus, during the protoplanetary disk phase, the interior of the Sun was largely convective, with the CZ encompassing at least 50\% of the Sun's mass, whereas the typical disk mass is $\sim 0.01\,\Msun$ \citep[see, e.g.,][]{Williams+Cieza11}. This implies that any anomalies in the solar composition due to the accretion of the protoplanetary disk gas with the pebble wave or with the depletion of heavy elements by the formation of planetesimals and planets were likely to be suppressed by the large CZ. 

Thus, we can expect two consequences from the above discussion: (1) the signature left by both the high-$Z$ and low-$Z$ accretion is suppressed by roughly one to two orders of magnitude and (2) convective mixing leads to a uniform composition in a large fraction of the solar interior. As seen in Fig.~\ref{fig:t-MCZ}, a mass coordinate of $0.5\,\Msun$ corresponds to approximately 27\% of the present-day solar radius, which indicates that the signatures of the planet formation processes are buried deep in the solar interior (mostly in its nuclear burning core).

\section{Computation method}
\label{sec:method}

In this section, we describe how we (i) simulate stellar evolution, (ii) compare our results with observations, and (iii) minimize the total $\chi^2$ value by changing the input parameters.

\subsection{Stellar evolution with accretion}
\label{sec:stellarevol}

We used the one-dimensional stellar evolution code \texttt{MESA} version 12115 \citep{Paxton+11,Paxton+13,Paxton+15,Paxton+18,Paxton+19}.
For details of the computational method used in this work, we refer the readers to \citet{Kunitomo+17}, \citet{Kunitomo+18}, and the series of papers by Paxton et al.
Below we briefly summarize the method and the various parameters used.

\subsubsection{Initial conditions}
\label{sec:init}

Stars are formed via the gravitational collapse of a molecular cloud core. A protostar (or second hydrostatic core) forms after the formation of a transient hydrostatic object \citep[the so-called first core; see][]{Larson69,Inutsuka12}. Radiation hydrodynamic simulations have suggested that the initial mass of a protostar is typically $\sim0.003\,\Msun$ \citep{Masunaga+Inutsuka00,Vaytet+Haugbolle17}.
In this study, we used a stellar seed of mass $0.1\,\Msun$, radius $4\,\Rsun$, and metallicity $Z=0.02$, to avoid numerical convergence issues at very low protostellar masses in the new version of the \texttt{MESA} code.
Recent studies have shown that the thermal evolution of the protostar depends on the entropy $\sacc$ of the accreted material \citep[e.g.,][]{Hartmann+97,BCG09,Hosokawa+11,Tognelli+15,Kunitomo+17,Kuffmeier+18}.
The initial seed corresponds to a protostar of mass $0.1\,\Msun$ that has evolved via hot accretion from its birth mass of $\sim0.003\,\Msun$\footnote{
We note that we chose a hot stellar seed because the accretion rate and $\sacc$ are expected to be high during the early stages of protostellar evolution \citep{Machida+10,Hosokawa+11,BVC12,Tomida+13}.
According to \citet[][see their Fig.\,D.1]{Kunitomo+17}, in hot accretion models, the stellar radius converges before the protostar reaches a mass of $0.1\,\Msun$.
}.

For the non-accreting cases (see Sect.\,\ref{sec:noacc}), we used an initial stellar seed of mass $1\,\Msun$ and a central temperature of $3\times10^5$\,K (i.e., corresponding to the top of the Hayashi track).
The seed has a uniform composition. The mass fractions of hydrogen, helium, and metals of the seed are $\Xproto$, $\Yproto$, and $\Zproto$, respectively.
We note that the Kelvin-Helmholtz timescale at the top of the Hayashi track is short \citep[see, e.g.,][]{Stahler+Palla04}, and hence, the pre-MS evolution without accretion is not sensitive to the choice of the initial central temperature.

\subsubsection{Mass accretion}
\label{sec:Mdot}

\begin{figure}[!t]
  \begin{center}
        \includegraphics[width=\hsize,keepaspectratio]{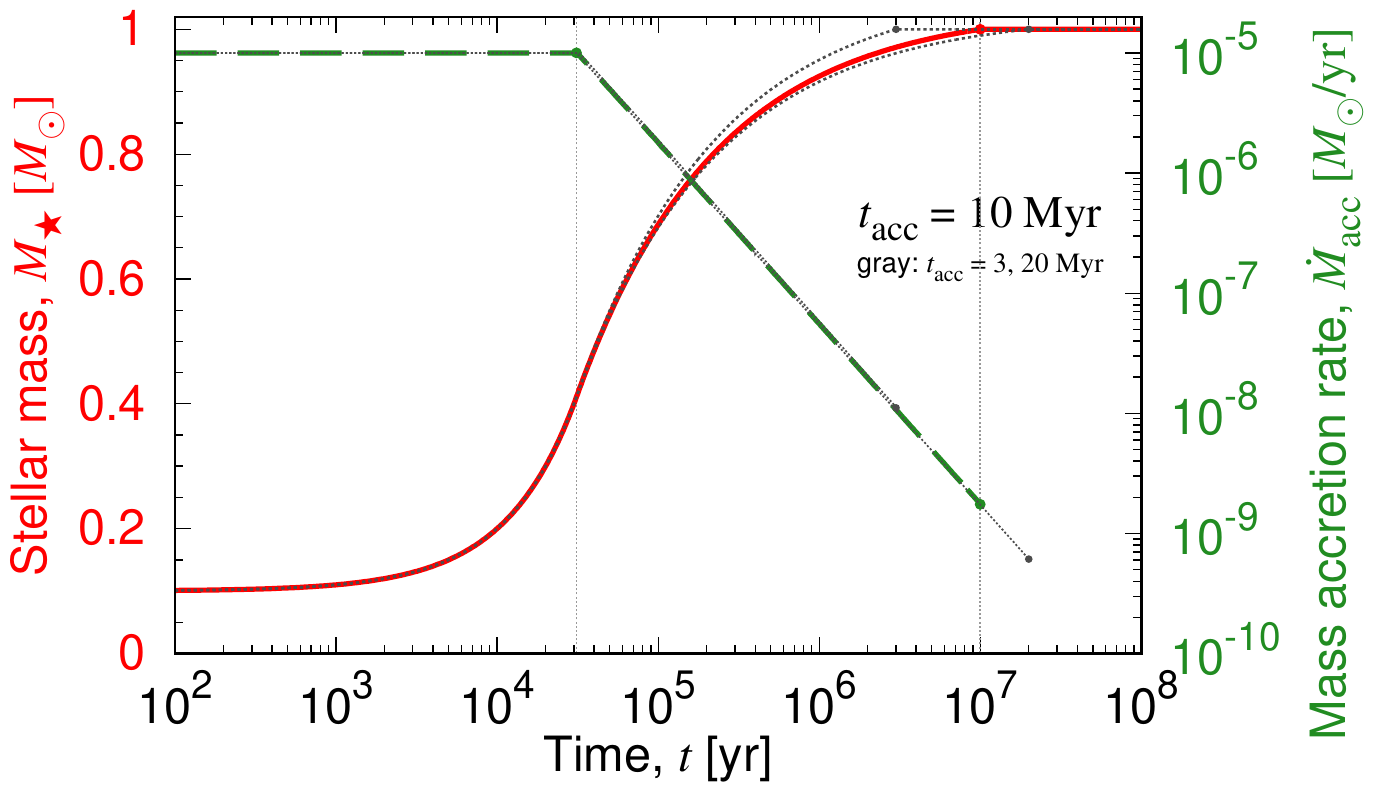}
        \caption{\small{
        Evolution of stellar mass $\Mstar$ (red solid line) and accretion rate $\Mdot$ (green dashed line) with time in the fiducial case.
        The gray vertical lines indicate $t_1$ and $\tacc$ in the fiducial case.
        The gray dotted lines indicate $\Mstar$ and $\Mdot$ for $\tacc=3$ and 20\,Myr.
        }}
        \label{fig:Mdot}
    \end{center}
\end{figure}

We adopted the following mass accretion rate $\Mdot$ in the protostellar and pre-MS phases\footnote{In this study, we refer to the main accretion phase (i.e., the class 0/I phase) as the protostellar phase, which is just before the protostellar mass becomes equal to the final mass of the star (i.e., $1\,\Msun$). A pre-MS star, in contrast, has $\Mstar\simeq 1\,\Msun$ but has not yet reached the zero-age MS.
}:
\begin{align}
    \Mdot = 
    \begin{cases}
    10^{-5}\,\Msun/{\rm{yr}}    &  {\rm{for}}\,\,  t \leq t_1\,, \\
    10^{-5}\,\Msun/{\rm{yr}}\,\times (t/t_1)^{-1.5} &  {\rm{for}}\,\,  t>t_1\,
    \end{cases}
    \label{eq:Mdot}
\end{align}
\citep[see][for details about the exponent $-1.5$]{Hartmann+98}.
In our fiducial model, we set $t_1=31,160\,$yr so that $\Mstar$ could reach $1\,\Msun$ at the end of the accretion phase, that is, at $\tacc=10$\,Myr. The evolution of $\Mstar$ and $\Mdot$ in our fiducial model is shown in Fig.\,\ref{fig:Mdot}.
In some of our simulations, we changed the accretion timescale $\tacc$ and $t_1$, but we always used the same exponent (i.e., $-1.5$) for $t>t_1$. As discussed in Sect.\,\ref{sec:interior}, accretion generally ends several million years; therefore, $\tacc=10\,$Myr corresponds to the upper limit of the observational constraint.

We note that previous studies, such as \citet[][]{Serenelli+11} and \citet{Zhang+19}, considered accretion only in the pre-MS phase,
whereas, in this study, we considered accretion in the protostellar phase, which led to a larger mass of the accreted material.
Moreover, \citet[][see their section\,3.2.3]{Serenelli+11} and \citet[][see their section 2.4]{Zhang+19} introduced accretion in their models after a certain time\footnote{
\citet[][]{Serenelli+11} varied this time $\tau\sub{ac,i}$ between 5 and 30\,Myr, while \citet{Zhang+19} fixed it to 2\,Myr.} to allow the pre-MS Sun to develop a radiative core.
These studies needed a non-accreting phase because they adopted arbitrary initial conditions, whereas our model is based on the current understanding of star formation.
We also note that the non-accreting timescale in some of the models of \citet{Serenelli+11} exceeded the typical disk lifetime (i.e., several million years).

In the present work, we did not consider mass loss.
Vigorous stellar winds have been known to flow out from young stars \citep{Wood+05,Suzuki+13}; moreover, \citet{Zhang+19} suggested that mass loss has the potential to alter the surface composition of the star. Although in this study we focused only on the effects of planet formation and opacity enhancement on the stellar composition, the effect of mass loss should be considered in future studies.

We used the same accretion model characterized by the parameter $\sacc$ as in \citet{Kunitomo+17,Kunitomo+18}, who assumed that the accretion heat injected into the protostar $L\sub{add}$ is a fraction of the gravitational energy liberated by accretion. 
Consequently, we assumed that $L\sub{add}=\xi G\Mstar\Mdot /\Rstar$, where $\xi$ is a dimensionless parameter, $G$ is the gravitational constant, and $\Rstar$ is the stellar radius. The accretion heat is assumed to be uniformly distributed throughout the star \citep[see related discussion in][]{Kunitomo+17}.
Although we set $\xi=0.1$ \citep[see][]{Kunitomo+17} in our accretion models, we chose a high-entropy initial seed (see Sect.\,\ref{sec:init}). This implies that our thermal evolution calculations are equivalent to those of standard accretion models (\citet{SST80I}; $\xi=0.5$ models in \citet{Kunitomo+17}).
Thus, the evolution of the CZ in our models corresponds to the $\xi=0.5$ line shown in Fig.\,\ref{fig:t-MCZ} (also, see Appendix\,\ref{app:non-accreting}).
In this context, we can expect the calculations with $\xi=0.1$ and a low-entropy initial seed to capture a slightly larger effect due to planet formation processes. We leave this for future work.

\subsubsection{Convection}
\label{sec:conv}

We adopted the mixing-length theory for convection proposed by \citet{Cox+Giuli68}. 
We assumed the ratio of the mixing length to the local pressure scale height ($\Hp$) to be $\amlt$.
We also considered the effect of the composition gradient on the convective stability (i.e., the Ledoux criterion).

Previous studies have suggested that additional mixing (e.g., convective overshooting and tachocline circulation), especially at the base of the CZ, plays an important role in solar sound-speed anomalies \citep[see, e.g.,][]{Christensen-Dalsgaard+18, Buldgen+19, Zhang+19}. 
However, these processes are poorly understood and are not well constrained. 
\citet{Christensen-Dalsgaard+18} showed that extending the Sun's CZ or adding extra mixing in the tachocline region leads to a better agreement between theoretical models and helioseismic constraints. 
\citet{Zhang+19} included convective overshoot in their models by adopting an exponentially decreasing diffusion coefficient and considering a change in the luminosity due to kinetic energy transfer in the overshooting region. They showed that the radial location $\RCZ$ of the base of the CZ is sensitive to the underlying overshooting model.

In this study, we considered the conventional convective overshooting model proposed by \citet{Herwig00}, in which the diffusion coefficient decreases exponentially from the convective--radiative boundary.
We assumed the $e$-folding length of the diffusion coefficient to be $\fov\Hp$.
In addition, we adopted the same overshooting parameters below and above the CZ, irrespective of the presence of nuclear burning.
Although we also considered semiconvection in our models, we confirmed that it has little effect on the results.
In this study, $\amlt$ and $\fov$ are the two free parameters. Based on the results of \citet{Zhang+19} and given the fact that a detailed exploration of convective overshooting models is beyond the scope of the present work, we slightly relaxed the constraint on $\RCZ$ when seeking the best solutions (see Sect.\,\ref{sec:obs}).

\subsubsection{Abundance tables}
\label{sec:abund}

We used the abundance table presented in \citetalias{Asplund+09} unless otherwise mentioned. 
We note that recent studies have suggested some modifications to the tables in \citetalias{Asplund+09}\footnote{
\citet[][]{Steffen+15}, \citet{Young18}, and \citet{Lodders19} suggested modifications to the O, Ne, Th, and U abundances.
In addition, \citet{Caffau+11} and recently \citet{Asplund+21} suggested different abundance tables.
Although some studies \citep[e.g.,][]{Serenelli+09,Vinyoles+17} adopted the abundances of refractory elements from CI chondrites, we used the original table of solar photospheric abundances presented in \citetalias{Asplund+09}.
}.
However, these modifications are relatively minor considering the goals of the present study and are therefore not considered here \cite[see][for the effect of modified solar abundances]{Buldgen+19}.

We also performed simulations using the abundance table presented in \citetalias{GS98} for comparison with previous studies. We adopted the table in which the abundances of refractory elements were modified using meteorites.

\subsubsection{Opacity}
\label{sec:kap}

In stellar evolution calculations, the Rosseland-mean opacity $\kappa$ is determined by interpolating the opacities listed in standard opacity tables using local thermodynamic and compositional quantities, such that $\kappa=\kappa(\rho, T, X, Z)$, where $\rho$ is the density, $T$ is the temperature, $X$ is the hydrogen mass fraction, and $Z$ is the metallicity.
We adopted the OPAL opacity table \citep{Iglesias+Rogers96} for the cases using the \citetalias{Asplund+09} composition and the OP opacity table \citep{Seaton05} for the cases using the \citetalias{GS98} composition. The opacity table presented in \citet{Ferguson+05} was used to determine the opacities in the low-temperature regions for both cases.
We refer the readers to \citet{Paxton+11} for more details.
We confirmed that the simulations with the OP table yielded very similar results to those with the OPAL table for the \citetalias{Asplund+09} composition \citep[][also, see Appendix\,\ref{app:non-accreting}]{Buldgen+19}.

As mentioned in Sect.\,\ref{sec:intro}, we aim to investigate the effect of opacity enhancement.
We modeled the opacity enhancement using a dimensionless factor $\fopa$, such that 
\begin{align}\label{eq:kap}
    \kappa' &= \kappa (1 + \fopa)\,.
\end{align}

Below, we describe our $\fopa$ model.
Laboratory experiments have not been able to reproduce the real conditions at the base of the solar CZ, which has led to uncertainties in the opacity.
\citet{Bailey+15} performed experiments with conditions close to those in the real Sun and showed that the wavelength-dependent opacity of iron was 30\%--400\% higher than previously thought, and also, the Rosseland-mean opacity was increased by $7\pm3\%$ \citep[see also][]{Nagayama+19}.
\citet[][see their Figure 2]{LePennec+15} obtained the contribution of each element to the opacity. Iron was shown to have three peaks at $\log (T/{\rm K})=5.66, 6.45$, and 7.18\footnote{In this study, $\log\equiv \log_{10}$.
}. We fitted the contribution of iron to the opacity by the sum of three Gaussian functions of $T$ and we used the same function to model $\fopa$, considering the possibility that further uncertainties may remain in the iron opacity and assuming that the iron abundance is inhomogeneous in the solar interior.
Thus, we modeled the opacity increase as
\begin{align}\label{eq:delkap}
    \fopa = \sum_{i=1}^3 A_i\,\exp \left[ -\frac{ \left(  \log (T/{\rm K}) -b_i\right)^2 }{2c_i^2} \right]\,,
\end{align}
where $A_i$ is a free parameter, 
{and $b_i$ and $c_i$ were derived by fitting Figure\,2 of \citet{LePennec+15}, leading to} $(b_1, b_2, b_3)=(5.66, 6.45, 7.18)$ and
$(c_1, c_2, c_3)=(0.22, 0.18, 0.25)$.

We note that \citet{Villante10} suggested another $\fopa$ model that increases linearly with $T$.
We note that our opacity increase model is based on actual data of the contribution to the opacity, whereas the model in \citet{Villante10} is more ad hoc in nature.
Although in this study we only present the results using Eq.\,\eqref{eq:delkap}, we confirmed that both opacity-increase formalisms improve the solar evolution model in a similar manner.

\subsubsection{Composition of accreted material}
\label{sec:Zacc}

For all the models with protostellar accretion, we fixed the mass fraction of deuterium to $X\sub{D}=28\,$ppm \citep[see][and references therein]{Kunitomo+17} and that of $\element[][3]{He}$ also to 28\,ppm \citep{Mahaffy+98}.
In the non-accreting cases, we set $X\sub{D}=0$, assuming that the deuterium was completely depleted in the protostellar phase.

For the models with accretion, we considered three models to account for the composition of the accreted material (i.e., $\element[][1]{H}$, $\element[][4]{He}$, and metals): homogeneous accretion, metal-poor accretion, and helium-poor accretion (see Table\,\ref{tab:chi2}).
In the first model, the helium mass fraction $\Yacc$ and metallicity of the accreted materials $\Zacc$ are constant with time; therefore, $\Yacc=\Yaccini$ and $\Zacc=\Zaccini$, where the subscript ``proto'' indicates the initial value.

\begin{table*}[!ht]
	\begin{center}
	\caption{Parameter settings of the chi-squared simulations.}
	\label{tab:chi2}
           \begin{tabular}{llllllll}
            \hline
            \hline
            \noalign{\smallskip}
            \# & Model name & Opacity$^a$ & $A_2$ & Abundance & $\tacc$$^b$ & Composition & $\fov$  \\
             & & & & table & [Myr] & of accreted & \\
             & & & &       & & material$^c$ & \\
            \noalign{\smallskip}
            \hline
            \noalign{\smallskip}
            \multicolumn{7}{l}{ [\textit{Optimized non-accreting models}] } &\\
            \noalign{\smallskip}
            1 & noacc & S & -- & \citetalias{Asplund+09} &  -- & -- & variable\\
            2 & noacc-GS98 & S & -- & \citetalias{GS98} &  -- & --& variable \\
            3 & noacc-noov & S & -- & \citetalias{Asplund+09} &  -- & -- & 0 \\
            4 & noacc-GS98-noov & S & -- & \citetalias{GS98} &  -- & -- & 0  \\
            \noalign{\smallskip}
            \hline
            \noalign{\smallskip}
            \multicolumn{7}{l}{ [\textit{Optimization with time-dependent $\Yacc$ (He-poor accretion)}] } &\\
            \noalign{\smallskip}
            5 & He12Myr & S & -- & \citetalias{Asplund+09} &  12 & Y & variable\\
            6 & He20Myr & S & -- & \citetalias{Asplund+09} &  20 & Y & variable \\
            \noalign{\smallskip}
            \hline
            \noalign{\smallskip}
            \multicolumn{7}{l}{ [\textit{Optimization with time-dependent $\Zacc$ (pebble wave and planet formation)}] } &\\
            \noalign{\smallskip}
            7 & \FULL & S & -- & \citetalias{Asplund+09} &  10 & Z & variable \\
            8 & \FULLnoov & S & -- & \citetalias{Asplund+09} &  10 & Z & 0 \\
            \noalign{\smallskip}
            \hline
            \noalign{\smallskip}
            \multicolumn{7}{l}{ [\textit{Optimization with opacity enhancement and homogeneous $\Zacc$ and $\Yacc$ }] } &\\
            \noalign{\smallskip}
            9 & \kapb & K$_2$ & [0, 0.22] & \citetalias{Asplund+09} &  10 & H & variable\\
            10 & \kapa & K$_{23}$ & [0, 0.22] & \citetalias{Asplund+09} &  10 & H & variable \\
            11 & \kapc & K$_{23}'$ & 0.10 & \citetalias{Asplund+09} &  10 & H & variable \\
            12 & K2$'$ & K$_2$ & 0.12/0.15/0.18 & \citetalias{Asplund+09} &  3/5/10 & H & 0/0.01/0.025 \\
            \noalign{\smallskip}
            \hline
            \noalign{\smallskip}
            \multicolumn{7}{l}{ [\textit{Optimization with opacity enhancement and time-dependent $\Zacc$}] } &\\
            \noalign{\smallskip}
            13 & \kappla & K$_{2}$ & 0.12/0.15/0.18 & \citetalias{Asplund+09} &  10 & Z & variable \\
            14 & \kapplb & K$_{23}$ & 0.12/0.15/0.18 & \citetalias{Asplund+09} &  10 & Z & variable \\
            15 & K2-MZ & K$_2$ & 0.12 & \citetalias{Asplund+09} &  5 & Z$'$ & 0.01 \\
            16 & K2-MZ1$'$/MZ8$'$ & K$_2$ & 0.12/0.15/0.18 & \citetalias{Asplund+09} &  3/5/10 & Z$'$ & 0/0.01/0.025 \\
            \noalign{\smallskip}
            \hline
            \noalign{\smallskip}
            \end{tabular} 
         \end{center}
         \tablefoot{
         ($^a$) S: Standard (i.e., no modification to the opacity).
         K$_2$: Enhanced opacity ($\kappa$) using Gaussian functions with $A_1=A_3=0$ and $A_2$ fixed to a value between 0 and 0.22 (see Sect.\,\ref{sec:kap}).
         K$_{23}$: Same as K$_{2}$ but $A_3$ varies in the simplex method.
         K$_{23}'$: $A_2=0.10$ (fixed) and $A_3$ is fixed to either $0.05$ or $-0.05$.
         ($^b$) Accretion timescale (see Sect.\,\ref{sec:Mdot}).
         ($^c$) H: homogeneous composition with time (i.e., $\Yacc=\Yaccini$ and $\Zacc=\Zaccini$). 
         Y: He-poor accretion.
         Z: Time-dependent $\Zacc$ (see Sect.\,\ref{sec:Zacc}).
         Z$'$: $M_1$, $M_2$, and $\Delta \Zacc$ are fixed (see Table\,\ref{tab:MZ}).
         }
\end{table*}

\begin{figure}[!t]
  \begin{center}
        \includegraphics[width=\hsize,keepaspectratio]{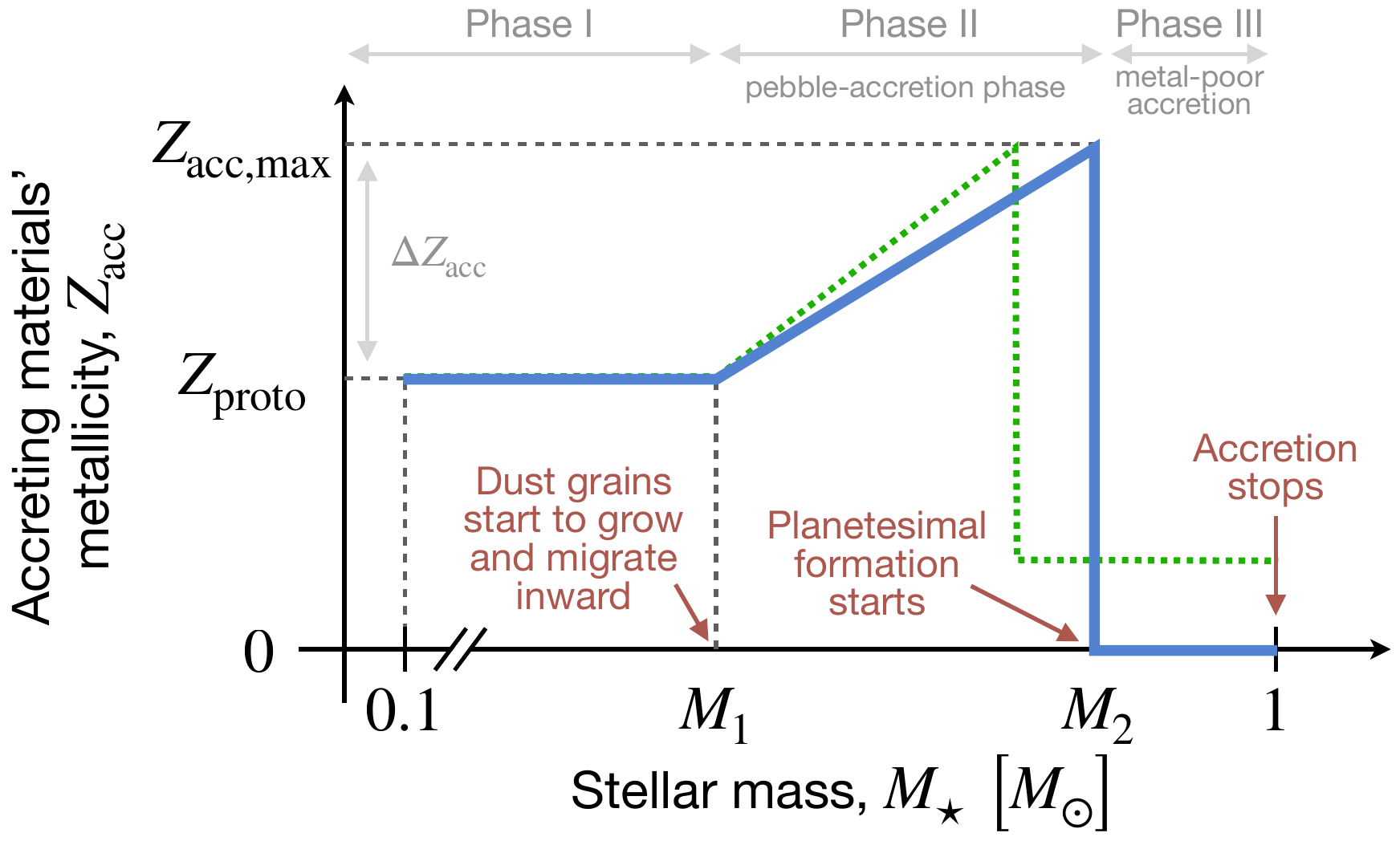}
        \caption{\small{
        Sketch of the evolution of the metallicity $\Zacc$ of the accreted material with stellar mass $\Mstar$.
        In this work, we used the model depicted by the blue solid line and assumed that the difference between this model and the one depicted by the green dotted line has little impact on the internal structure of the Sun (see text for details).
        }}
        \label{fig:Zacc}
    \end{center}
\end{figure}

In the metal-poor accretion model, $\Zacc$ is time-dependent because of the underlying planet formation processes.
We adopted a simple model for $\Zacc$ to capture the following two processes: pebble drift and planetesimal formation (see Sect.\,\ref{sec:planets}).
Figure\,\ref{fig:Zacc} shows the evolution model of $\Zacc$ used in this work.
In the early phase (Phase I; $\Mstar \leq M_1$), $\Zacc=\Zaccini$ is constant.
Once the Stokes number of dust grains increases and the grains start to migrate inward, $\Zacc$ increases monotonically with time. We refer to this phase as Phase II or the pebble-accretion phase.
Planetesimals begin to form when $\Mstar=M_2$. Subsequently, in Phase III (i.e., the metal-poor accretion phase), drifting dust grains are captured by the planetesimals and are not allowed to reach the proto-Sun, causing $\Zacc$ to abruptly decrease to 0.
When $\Mstar$ reaches $1\,\Msun$, the accretion stops abruptly (i.e., $\Mdot=0$). After the accretion phase, the proto-Sun evolves to become the Sun (Phase IV).

When $dZ\equiv\Zacc-\Zaccini\ne 0$ (i.e., $\Mstar>M_1$), 
we assumed that $\Xacc=\Xaccini -0.7dZ$ and $\Yacc=\Yaccini -0.3dZ$, where $\Xacc$ is the hydrogen mass fraction of accreted materials and $\Xaccini$ is the initial value of $\Xacc$.
We note that because $X\sub{D}$ and the $\element[][3]{He}$ abundance were kept constant, only the $\element[][1]{H}$ and $\element[][4]{He}$ abundances changed with time.

We point out that our assumption of $\Zacc=0$ in Phase III is a major simplification. In reality, grains are not completely filtered by planets and planetesimals. Some of the grains collide with the planetesimals leading to a small but nonzero $\Zacc$, which corresponds to the green dotted line in Fig.\,\ref{fig:Zacc} (see also Sect.\,\ref{sec:planets}). However, because the mass of the solar CZ is large during the accretion phase (see Fig.\,\ref{fig:t-MCZ}), the difference between these two models has little impact on the results.

In the He-poor accretion model, the helium abundance $\Yacc$ of the accreted material was assumed to vary with time, such that
\begin{align}
    \Yacc = 
    \begin{cases}
    \Yaccini    &  {\rm{for}}\,\,  \Mstar\leq M_1\,, \\
    \Yaccmin &  {\rm{for}}\,\,  \Mstar > M_1\,.
    \end{cases}
    \label{eq:Yacc}
\end{align}
The He-poor accretion was proposed by \citet{Zhang+19} as a solution to the solar sound-speed anomaly.
\citet{Zhang+19} considered a constant $\Yacc$ but had a larger initial stellar mass in their simulations than in the present study (see Sect.\,\ref{sec:Mdot}); thus, their simulations are equivalent to ours in the late phase (i.e., $\Mstar > M_1$).
In the He-poor accretion models, we assumed that $\Zacc=\Zaccini$ and $\Xacc=\Xaccini -dY$, where $dY\equiv \Yacc - \Yaccini$.

\subsubsection{Other input physics}
\label{sec:other_inputs}

We adopted the OPAL equation-of-state tables updated in 2005 \citep{Rogers+Nayfonov02}.
We used the nuclear reaction rates in \citet{Caughlan+Fowler88} and \citet{Angulo+99} with certain modifications \citep[see][for details]{Paxton+11}.
The prescription for element diffusion presented in \citet{Thoul+94} was also used.
We note that the JINA reaction table is also available in \texttt{MESA}; however, we confirmed that the results are insensitive to the choice of the reaction table (see Appendix\,\ref{app:non-accreting}).

The outer boundary condition used in this study was given by the atmospheric tables \citep[see Sect.2.7 of ][]{Kunitomo+18}.
Although previous studies adopted different boundary conditions\footnote{
For example, \citet{Vinyoles+17} and \citet{Zhang+19} adopted the atmospheric model proposed by \citet{Krishna_Swamy66}.
}, we confirmed that the outer boundary conditions did not have a significant impact on the results.

\subsubsection{Summary of the input parameters}
\label{sec:summary_inputs}

\begin{table*}
\caption{Input parameters.}             
\label{tab:inputs}      
\centering                          
\begin{tabular}{lll}        
\hline\hline                 
Parameter & Description & Reference \\    
\hline                        
  $\amlt$ & Mixing-length parameter & Sect.\,\ref{sec:conv} \\      
  $\fov$ & Overshooting parameter & Sect.\,\ref{sec:conv} \\
  $A_1$, $A_2$, $A_3$ & Amplitudes of opacity enhancement & Sect.\,\ref{sec:kap} \\
  $M_1$, $M_2$ & Stellar masses when $\Zacc$ starts to increase and when it becomes zero, respectively & Sect.\,\ref{sec:Zacc}, Fig.\,\ref{fig:Zacc} \\
  $\Zaccini$ & Metallicity of accreted material when $\Mstar\leq M_1$ & Sect.\,\ref{sec:Zacc}, Fig.\,\ref{fig:Zacc} \\
  $\Zaccmax$ & Maximum metallicity when $\Mstar=M_2$ & Sect.\,\ref{sec:Zacc}, Fig.\,\ref{fig:Zacc} \\
  $\Yaccini$ & He abundance of accreted materials when $\Mstar\leq M_1$ & Sect.\,\ref{sec:Zacc} \\
  $\Yaccmin$ & He abundance of accreted materials when $\Mstar>M_1$ in Runs 5 and 6 & Sect.\,\ref{sec:Zacc} \\
\hline                                   
\end{tabular}
\tablefoot{For most of the cases, we set $A_1=A_3=0$. 
}
\end{table*}

Table\,\ref{tab:inputs} summarizes the input parameters used in this work.
In the runs with opacity enhancement and $\Zacc$ evolution, the number of input parameters can be up to 10.

\subsection{Comparison with observations}
\label{sec:obs}

\begin{table*}
\caption{Target quantities.}             
\label{tab:targets}      
\centering                          
\begin{tabular}{lllll}        
\hline\hline                 
Observed parameter & Description & Value & Uncertainty & References \\    
\hline                        
  $\ZXs$ & Abundance ratio of metals to hydrogen \tablefootmark{a}                & 0.0181 & $10^{-3}$ & 1 \\
  & & 0.02292 & $10^{-3}$ & 2 \\
  $\Ys$ & Surface helium abundance                                  & 0.2485 & 0.0035 & 3 \\
  $\RCZ$ & Location of the convective--radiative boundary [$\Rsun$]  & 0.713 & 0.01 \tablefootmark{b} & \\
  rms($\delcs$) & Root-mean-square sound speed                        & 0 & $10^{-3}$ \tablefootmark{c} & \\
  $\log\Lstar$ & Bolometric luminosity $[\Lsun]$                    & 0 & 0.01\,dex \tablefootmark{c} &  \\
  $\Teff$ & Effective temperature [K]                               & 5777 & 10 \tablefootmark{c} & \\
\hline                                   
\end{tabular}
\tablefoot{
In this work, $\Lsun=3.8418\times10^{33}\,\rm erg/s$ and $\Rsun=6.9598\times10^{10}\,\rm cm$ \citep{Bahcall+05}.
\tablefoottext{a}{The $\ZXs$ value depends on the adopted abundance tables (see Sect.\,\ref{sec:abund}).}
\tablefoottext{b}{\citet{Bahcall+05} suggested $0.713\pm0.001\,\Rsun$; however, we used a larger value for convergence purposes (see text for details).}
\tablefoottext{c}{Arbitrarily small (nonzero) uncertainty for convergence purposes (see text for details).}
\tablebib{(1) \citetalias{Asplund+09}, (2) \citetalias{GS98}, (3) \citet{Basu+Antia04}.}
}
\end{table*}

When the elapsed time of the stellar evolution simulations described in Sect.\,\ref{sec:stellarevol} reached the solar age, we compared the simulation results with spectroscopic and helioseismic observations of: 
the ratio of the surface metallicity to the surface hydrogen abundance $\ZXs$; 
the surface helium abundance $\Ys$; 
the location of the convective--radiative boundary $\RCZ$; 
the root mean square (rms) of $\delcs$ (see Eq.\,\eqref{eq:delcs});
the bolometric luminosity $\Lstar$;
and the effective temperature $\Teff$.
We define the observed minus calculated sound speed as
\begin{eqnarray} \label{eq:delcs}
\delcs\equiv (\csobs-\cs)/\csobs\,.
\end{eqnarray}
To obtain the rms$(\delcs)$, we compared our simulated results with the observed data provided in \citet[][see their Table 3]{Basu+09}.
We interpolated the $\cs$ profile obtained at the solar age (typically $\sim3000$ grids) to the locations of 37 points given in \citet{Basu+09} and thus calculated the rms value of $\delcs$ (see discussions in Sect.\,\ref{sec:solarproblem}).

Table\,\ref{tab:targets} summarizes the observed parameters and their $1\sigma$ uncertainties with three exceptions.
First, although the uncertainty of $\RCZ$ in the helioseismic observations is $0.001\,\Rsun$ in \citet{Bahcall+05}, we relaxed this constraint by a factor of 10. This is because the $\RCZ$ value is likely to be affected by the uncertainties in the convective overshooting model, which are beyond the scope of this study \citep[see Section 3.1 of ][and Sect.\,\ref{sec:conv} in this work]{Zhang+19}.
Second, we chose the rms($\delcs$) uncertainty ($10^{-3}$) based on the results obtained with the \citetalias{GS98} composition presented in \citet{Serenelli+09} (see their Table 2), rather than from the helioseismic observations.
Finally, the uncertainties in $\Lstar$ and $\Teff$ are arbitrary; however, we confirmed that our results are not sensitive to these values because all the models reproduce these quantities well ($<0.001$\,dex and 3\,K, respectively).

We assumed the solar age to be 4.567\,Gyr \citep{Amelin+02} after Ca-Al-rich inclusions (CAIs) were condensed.
We note that there could be a time difference between the formation of CAIs and time taken for $\Mstar$ to reach $0.1\,\Msun$.
Assuming that CAIs were probably formed in the protosolar nebula, this time difference is at most several million years, and therefore, we neglect it in the present study.

\subsection{Chi-squared tests}
\label{sec:chi2}

By comparing our simulation results with observations, we derived the $\chi^2$ value, which is given by
\begin{equation}\label{eq:chi2}
    \chi^2 = \frac{\sum_{i=1}^N \left[  (q_i-q_{i,\rm target})/\sigma(q_i)\right]^2}{N}\,,
\end{equation}
where $q_i$ is the simulated value of a particular quantity, $q_{i,\rm target}$ is the corresponding observed (or target) value, $N=6$ is the total number of target quantities, and $\sigma$ is the uncertainty of each quantity listed in Table\,\ref{tab:targets}.
We aimed to search for a set of input parameters that minimized the $\chi^2$ value.
To do so, we used the downhill simplex method \citep{Nelder+Mead65}.
Typically, $\sim200$ simulations were needed for the minimization.
We performed such simulations under a variety of settings.

In the simplex algorithm, we imposed constraints on the parameter space composed of $\Zaccmax$, $M_1$, and $M_2$. We did not consider any cases where the total metal mass of the accreted material was outside the range [0.5, 2] $\times \Zaccini (1\,\Msun-0.1\,\Msun)$.

It should be noted that in general standard solar models are optimized in the non-accreting case with three input parameters, namely $\amlt$, $\Yproto$, and $\Zproto$, and three target quantities, namely $\Lstar$, either $\Teff$ or $\Rstar$, and $\ZXs$ \citep[see, e.g.,][]{Vinyoles+17}. \citet[][see their Tables 1 and 2]{Farag+20} showed that, if the above three input and target quantities are used, then the optimized solutions using the \texttt{MESA} code are similar to those in \citet{Vinyoles+17}. We confirmed that, using these three target quantities, we could reproduce exactly the results of \citet{Farag+20}. However, we found that in some cases this can result in values of $\Ys$ outside the range of observed values \citep[see, e.g., Table~4 of ][]{Vinyoles+17}. To search for the solutions that explain all the observable quantities, we used as constraints the six quantities in Table~\ref{tab:targets}.

We also performed simulations in which the 37 points used to calculate the $\delcs$ values (see Sect.\,\ref{sec:obs}) were considered as independent and were added to the five other target values in Table~\ref{tab:targets} \citep[thus $N=42$; see, e.g., ][]{Villante+14}. We found that the models that poorly fit the available constraints were not improved and that our best models were not changed by this new approach.

We note that the solutions with the simplex method sometimes fall into local minima. Although we carefully chose the initial values of the input parameters to avoid this problem, future simulations using the Markov chain Monte Carlo method are encouraged.

\section{Solar models fitting the observational constraints}
\label{sec:results}

In this section, we present the results of the simulations optimized using the simplex method for different conditions. 
First, in Sects.\,\ref{sec:noacc} and \ref{sec:Hepoor}, we show that our results are in agreement with those of previous studies.
Next, we present our results for the metal-poor accretion (Sect.\,\ref{sec:Zpoor}) and opacity increase (Sect.\,\ref{sec:kap-results}) models.
The parameter settings are summarized in Table\,\ref{tab:chi2}.
The detailed results are provided in Tables\,\ref{tab:chi2-results-input} and \ref{tab:chi2-results-output}.

\subsection{Non-accreting models}
\label{sec:noacc}

To date, most studies have performed solar evolution simulations without including accretion. In this work, we simulated such non-accreting models to ensure consistency with the previous studies.

We performed four sets of simulations (Runs 1--4 in Table\,\ref{tab:chi2}) adopting either the \citetalias{GS98} or \citetalias{Asplund+09} abundance tables, and with and without convective overshooting.
We started the simulations with a $1\,\Msun$ star having a fully convective structure (see Sect.\,\ref{sec:init}).
The initial stellar composition, $\amlt$, and $\fov$ were optimized using the simplex method.

\begin{figure*}[!ht]
  \begin{center}
        \includegraphics[width=\hsize,keepaspectratio]{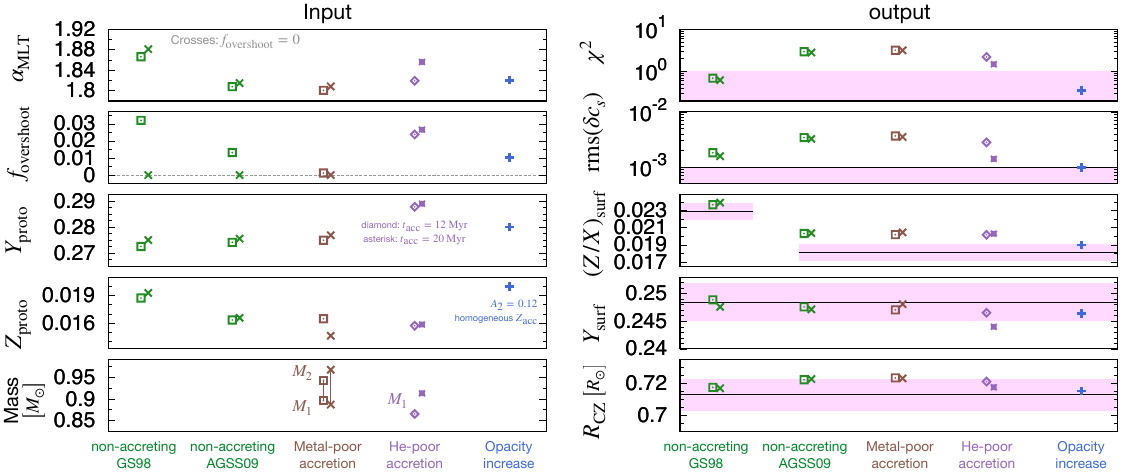}
        \caption{\small{
        Optimized input parameters (left) and results at the solar age (right).
        The left panels show $\amlt$, $\fov$, $\Yaccini$, $\Zaccini$, and stellar mass ($M_1$ and $M_2$) for nine models from top to bottom.
        The right panels show the $\chi^2$ value (see Eq.\,\eqref{eq:chi2}), rms$(\delcs)$, $\ZXs$, $\Ys$ and $\RCZ$ from top to bottom.
        The five models shown in each panel are: non-accreting models with \citetalias{GS98} composition (Runs 2 and 4); non-accreting models with \citetalias{Asplund+09} composition (Runs 1 and 3); accreting models with time-dependent $\Zacc$ (Runs 7 and 8); models with He-poor accretion (Runs 5 and 6); and \kapb\ model (Run 9) with $A_2=0.12$, from left to right (see Table\,\ref{tab:chi2} for details).
        Squares and crosses represent the cases with and without overshooting, respectively.
        Diamonds and asterisks represent the cases with He-poor accretion for $\tacc=12$ and 20\,Myr, respectively.
        The shaded regions in the right panels indicate $\chi^2< 1$ or the $1\sigma$ uncertainties (see Table\,\ref{tab:targets}).
        }}
        \label{fig:other}
    \end{center}
\end{figure*}

We found that the converged $\cs$ profiles are in good agreement with those of previous studies (see Fig.\,\ref{fig:non-accreting}), which validates our simulations with the \texttt{MESA} code.
The optimized input parameters and corresponding best results are shown in Fig.\,\ref{fig:other}.
The maximum values of $\delcs$ for the cases with the \citetalias{GS98} and \citetalias{Asplund+09} composition were approximately 0.004 and 0.009, respectively, irrespective of the inclusion of overshooting (see Fig.\,\ref{fig:non-accreting}).
The rms($\delcs$) value for the \citetalias{Asplund+09} composition was found to be in the range (3.2--3.4)$\times10^{-3}$, which is approximately twice as large as that for the \citetalias{GS98} composition, that is, (1.6--1.8)$\times10^{-3}$ (see Table\,\ref{tab:chi2-results-output}).
The other results obtained with the \citetalias{Asplund+09} composition (i.e., surface composition and $\RCZ$) were also found to be worse than those obtained with the \citetalias{GS98} composition, as indicated by the four times larger total $\chi^2$ value in the \citetalias{Asplund+09} case.

\subsection{Helium-poor accretion}
\label{sec:Hepoor}

Next, we performed simulations with the He-poor accretion following \citet{Zhang+19}.
In this case, we started simulations with a $0.1\,\Msun$ seed (see Sect.\,\ref{sec:init}) and simulated its evolution from the protostellar to the solar age.
The input parameters optimized using the simplex method were $\amlt, \fov$, initial composition of the accreted material, $M_1$, and $\Yaccmin$ (see Eq.\,\eqref{eq:Yacc}).
Considering that \citet{Zhang+19} required a longer accretion timescale of $\tacc\geq 12$\,Myr\footnote{
\citet[][see their Section 3.2]{Zhang+19} claimed the duration of their accretion phase was $\tacc\geq 10$\,Myr; however, they started accretion at 2\,Myr (see Sect.\,\ref{sec:Mdot} of this work). We note that $t=0$ in this study corresponds to the protostellar phase (i.e., $\Mstar=0.1\,\Msun$). }, we set $\tacc=12$ and 20\,Myr in this work. We note that our fiducial model corresponds to $\tacc=10$\,Myr (see Table\,\ref{tab:chi2}).

Figure\,\ref{fig:other} shows that He-poor accretion improves the rms($\delcs$) value, as claimed by \citet{Zhang+19}. The rms($\delcs$) value was found to be $2.8\times10^{-3}$ and $1.4\times10^{-3}$ for $\tacc=12$ and 20\,Myr, respectively. \citet{Zhang+19} obtained even better (but qualitatively similar) rms($\delcs$) values by considering mass loss by stellar winds and a more complex overshooting model. 

Despite the above results, we regard the possibility of He-poor accretion to be unlikely. First, the giant planets in our Solar System, which also captured gas in the protoplanetary disk, do not show a large depletion in their helium abundance \citep[see, e.g.,][]{Guillot+Gautier2015}. Jupiter and Saturn's atmospheres are characterized by slightly lower-than-protosolar helium abundances, which is consistent with the helium settling in their interiors from an initial protosolar value \citep{Mankovich+Fortney2020}. In contrast, Uranus and Neptune appear to have a helium abundance compatible with the protosolar value \citep[][]{Guillot+Gautier2015}. 

Second, \citet{Zhang+19} surmised that the high first ionization potential of helium might lead to the accumulation of helium at the inner edge of the disk followed by He-poor accretion onto the proto-Sun. However, this would lead to $\simeq0.015\,\Msun$ of the helium remaining in the disk (see their Table 4).
It is difficult to understand why such a large amount of helium would not have eventually accreted onto the proto-Sun.

\subsection{Metal-poor accretion}
\label{sec:Zpoor}

The possibility that metal-poor accretion may affect the structure of the Sun was first suggested by \citet{Guzik+05} and then tested by \citet{Castro+07}, \citet{Guzik+Mussack10}, \citet{Serenelli+11}, and \citet{Hoppe+20}.
Although these studies observed some improvement due to the inclusion of metal-poor accretion, they failed to find a solution as good as the models with high-$Z$ abundances \citep[i.e.,][\citetalias{GS98}]{Grevesse+Noels93} with respect to $\delcs, \RCZ$, and $\Ys$.

In this work, we revisited metal-poor accretion in a larger parameter space and in the framework of recent planet formation theories (see Sect.\,\ref{sec:planets}).
While the aforementioned studies considered metal-poor accretion onto a pre-MS or MS Sun, we considered metal-poor accretion during the protostellar phase.
We assumed the composition of the accreted material to vary with time, as shown in Fig.\,\ref{fig:Zacc}. We ran simulations with model \FULL\ that had seven input parameters: $\amlt, \fov, \Yaccini, \Zaccini, \Zaccmax, M_1$, and $M_2$.
We also performed a series of calculations without including overshooting (model \FULLnoov).

Figure\,\ref{fig:other} shows the results obtained with models \FULL\ and \FULLnoov.
We found that the results are almost identical to those of the non-accreting models, with no significant improvement in the $\chi^2$ value. This indicates that metal-poor accretion cannot be a solution to the solar abundance problem. This is consistent with the findings of previous studies \citep[e.g.,][]{Serenelli+11} and can be easily explained. As shown in Fig.~\ref{fig:t-MCZ}, the proto-Sun has a nearly fully convective interior, implying that the effect of metal-poor accretion is heavily suppressed therein. This, in turn, limits the signatures of metal-poor accretion on the internal structure of the present-day Sun. 

Although metal-poor accretion does not have an impact on the $\chi^2$ test, it affects the metallicity profile of the solar interior. We will revisit this issue in Sect.\,\ref{sec:discussion-planets}.

\subsection{Opacity increase}
\label{sec:kap-results}

\begin{figure*}[!ht]
  \begin{center}
        \includegraphics[width=\hsize,keepaspectratio]{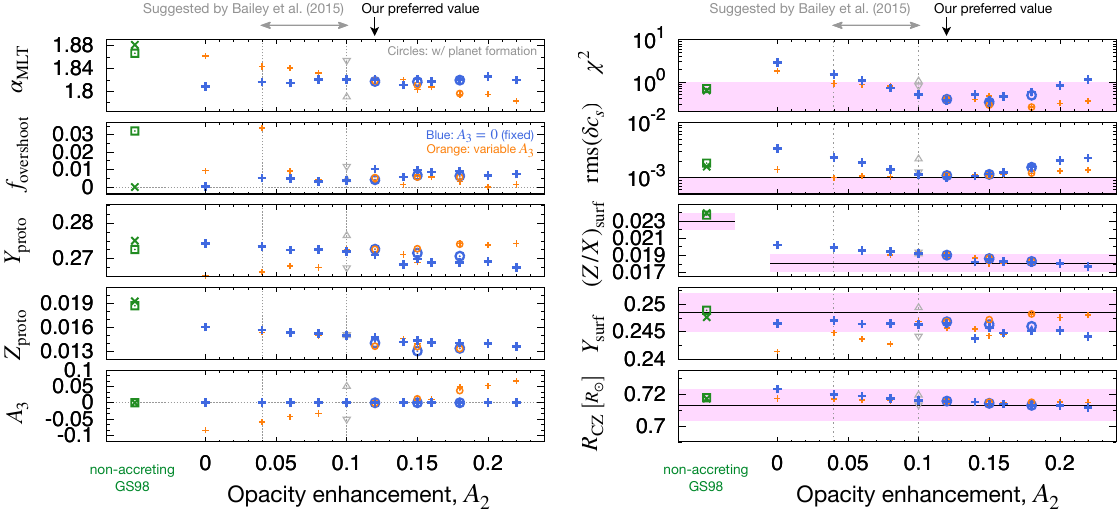}
        \caption{\small{
        Similar to Fig.\,\ref{fig:other} but showing the dependence on $A_2$ in Runs 9, 10, 11, 13, and 14; in addition, the left bottom panel shows $A_3$.
        The orange and blue colors indicate the cases with varying $A_3$ and $A_3=0$, respectively.
        The circles and plus signs denote the cases with and without planet formation, respectively (i.e., $\Zacc$ either evolves with time or is a constant).
        The gray triangles at $A_2=0.1$ denote the cases with $A_3=\pm0.05$.
        The two gray vertical lines demarcate the opacity enhancement (i.e., $7\pm3\%$) suggested by \citet{Bailey+15}.
        }}
        \label{fig:A2}
    \end{center}
\end{figure*}

Before the experiments conducted by \citet{Bailey+15} showed that iron opacities in stellar interiors were underestimated, solar models with a $\sim10$--30\% ad hoc increase in the opacities had already been examined and have even been shown to partly solve the solar abundance problem \citep[e.g.,][]{Christensen-Dalsgaard+09,Serenelli+09,Christensen-Dalsgaard+Houdek10,Villante10}.
Guided by the results of \citet{Bailey+15}, we adopted a physically motivated opacity-increase factor $\fopa$ that depends on three adjustable parameters $A_1$, $A_2$, and $A_3$, as defined by Eqs.\,\eqref{eq:kap} and \eqref{eq:delkap}. Plausible values of these parameters were obtained by minimizing the $\chi^2$ value using the simplex method.

First, we obtained results that were insensitive to the value of the low-temperature parameter $A_1$. This is because this parameter modifies the opacities in the CZ, where temperature changes are insensitive to the opacity increase. Therefore, we set $A_1=0$ in our models. 

Next, we constructed models with different values of $A_2$ in the range 0 to 0.22 and five variable input parameters for the simplex method: $\amlt, \fov, \Yaccini$, $\Zaccini$, and $A_3$ (corresponding to model \kapa\ in Table\,\ref{tab:chi2}). 
Another series of models was constructed by setting $A_3=0$ (model \kapb).

Figure\,\ref{fig:A2} shows the results as a function of $A_2$. When $A_2=A_3=0$, the results obtained with the \citetalias{Asplund+09} abundances were found to be clearly much worse than those obtained with the \citetalias{GS98} abundances. However, when $A_2$ was increased (in both models \kapa\ and \kapb), the quality of the fit to the observational constraints improved significantly, with the best results being obtained for $A_2\simeq0.12$--0.18. Modifying $A_3$ in addition to $A_2$ (as in model \kapa) was found to be helpful but not necessary. The $\chi^2$ and rms$(\delcs)$ values decreased in this case because of the additional degree of freedom. However, additional simulations with $A_3=\pm0.05$ and $A_2=0.10$ (model \kapc) confirmed that the effect of $A_3$ was marginal. Therefore, we did not explore the effect of $A_3$ further and simulated instead the \kapb\ models (with $A_3=0$).

The iron opacities measured in laboratory experiments at a temperature of $\sim 2\times 10^6$\,K should correspond to an increase of $7\pm3\%$ in the Rosseland-mean opacities over standard opacities \citep{Bailey+15}. This temperature is consistent with an increase in $A_2$ of similar magnitude \citep[see Sect.\,\ref{sec:kap} and][]{LePennec+15}. Based on this and the best-fit results from Fig.\,\ref{fig:A2}, we adopted $A_2=0.12$ for our fiducial model. We note that this value of $A_2$ is also consistent with that used in models with increased opacities in previous studies \citep[][]{Serenelli+09,Buldgen+19}.

We observed a clear correlation between $\Zaccini$ and $A_2$, such that $\Zaccini$ decreased monotonically with $A_2$. This is because both parameters have the same effect on the opacity in the solar interior. The correlations between $A_2$ and the other parameters (i.e., $\amlt$, $\fov$, and $\Yaccini$) are more complicated. This is because we used the simplex method to fit six observational constraints on which the input parameters affect differently \citep[see, e.g.,][]{Henyey+65,Kippenhahn+12}.

\subsection{Opacity increase and metal-poor accretion}
\label{sec:Zpoor-kap}

Finally, we performed simulations by including both opacity increase and metal-poor accretion (i.e., the $\Zacc$ evolution as shown in Fig.\,\ref{fig:Zacc}).
The blue and orange circles in Fig.\,\ref{fig:A2} show the results of models \kappla\ (with $A_3=0$) and \kapplb\ (with varying $A_3$), respectively, for $A_2=$ 0.12, 0.15, and 0.18.
We found that the minimized $\chi^2$ and rms$(\delcs)$ values were almost the same as those obtained using models \kapb\ and \kapa\ (i.e., with homogeneous $\Zacc$), as expected from the results presented in Sect.\,\ref{sec:Zpoor}.
This again points to the conclusion that planet formation processes do not affect the solar sound speed profile.

\subsection{The solar abundance problem}\label{sec:solarproblem}

\begin{figure}[!ht]
  \begin{center}
        \includegraphics[width=\hsize,keepaspectratio]{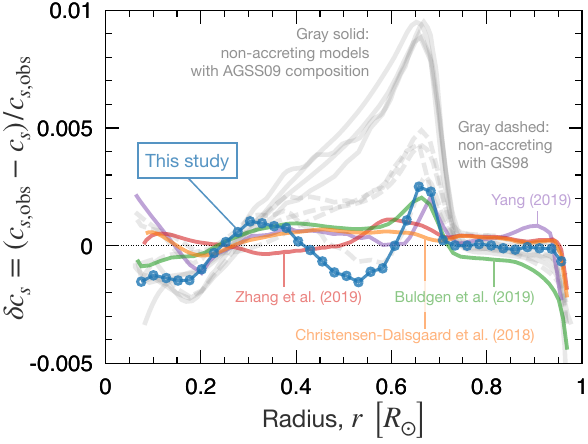}        
        \caption{\small{
        Radial profiles of observed minus calculated sound speed $\delcs$.
        The blue solid line with circles shows the result obtained in this study using model \kapb\ with $A_2=0.12$.
        The gray solid and dashed lines show the results of non-accreting models with \citetalias{GS98} and \citetalias{Asplund+09} compositions, respectively (see Fig.\,\ref{fig:non-accreting}).
        The other solid lines show the results in the literature: the orange, red, purple, and green lines show the results of: \citet[][see their Figure\,7]{Christensen-Dalsgaard+18},
        \citet[][see the magenta line in their Figure\,9]{Buldgen+19},
        Model AGSSr2a of \citet[][]{Yang19},
        and Model TWA of \citet[][]{Zhang+19}.
        }}
        \label{fig:r-cs}
    \end{center}
\end{figure}

It is interesting to note that the \kapb\ model with $A_2=0.12$ fits the observational constraints, and in particular, rms$(\delcs)$, even better than the models using \citetalias{GS98} abundances (see Fig.\,\ref{fig:A2}). This is better demonstrated in Fig.\,\ref{fig:r-cs}, where we plot $\delcs$ as a function of radius. For our fiducial model (i.e., model \kapb\ with $A_2=0.12$), the peak value at the base of the CZ is $\delcs=0.003$, which is much smaller than the peak value obtained for standard models with the \citetalias{Asplund+09} composition ($\delcs=0.009$), and even better than that obtained for models with the \citetalias{GS98} composition.

Figure\,\ref{fig:r-cs} also compares our model with other models in the literature that include various physical processes. These models are qualitatively equivalent fits to the sound speed constraint.
\citet{Zhang+19} claimed the importance of helium-poor accretion (however, see discussion in Sect.\,\ref{sec:Hepoor}) and the improved overshooting models (see Sect.\,\ref{sec:conv}).
In the models by \citet{Christensen-Dalsgaard+18} and \citet{Buldgen+19}, both an ad hoc opacity increase and extra mixing around the base of the CZ were considered.
\citet{Yang19} showed that rotational mixing is also promising. We note that in the present work, we did not investigate these possibilities, but Fig.\,\ref{fig:r-cs} shows that all the aforementioned models also fit the helioseismic constraints better than the models with the high-$Z$ \citetalias{GS98} abundances.

We note that the smoothness of our $\delcs$ profile differs from that of other models in the literature. We derived $\delcs$ simply by comparing our simulated $\cs$ profile with that given in \citet{Basu+09} (see Sect.\,\ref{sec:obs}). Conversely, in some of the studies, $\delcs$ was derived by comparing the oscillation modes from the calculated solar structure using an inversion method with the observed modes \citep[see, e.g.,][]{Buldgen+19}. We already confirmed that our $\delcs$ profiles for the non-accreting models are in good agreement with those of previous studies, and therefore, the difference in the derivation of $\delcs$ does not change the conclusions of this study.

\section{Consequences of planet formation}
\label{sec:discussion-planets}

Using our fiducial K2 model, shown in Sect.\,\ref{sec:results} to be our best fit to the observational constraints, we now examine the impact of an inhomogeneous $\Zacc$ on the following: present-day solar central metallicity $\Zc$ (see Sect.\,\ref{sec:discussion-Zc}), primordial Solar-System metallicity $\Zproto$ (see Sect.\,\ref{sec:discussion-Zini}), and primordial Solar-System helium abundance $\Yproto$ (see Sect.\,\ref{sec:discussion-Yini}).
Before discussing $\Zc$, $\Zproto$, and $\Yproto$, in Sect.\,\ref{sec:discussion-formation}, we first discuss the consequences of planet formation on the solar interior, as inferred from the results presented in Sect.\,\ref{sec:results}.

\subsection{Planet formation and the solar interior}
\label{sec:discussion-formation}

As discussed in Sect.\,\ref{sec:context}, planet formation is associated with both an increase in $\Zacc$ (i.e., the ``pebble wave'' phase) and a late metal-poor accretion phase during which planetesimals or planets are formed. The largely convective structure of the young Sun partly erases the signature of planet formation; however, a small trace remains and can be highlighted by comparing solar interior models calculated with and without planet formation mechanisms.  

\begin{figure*}[!ht]
  \begin{center}
        \includegraphics[width=0.48\hsize,keepaspectratio]{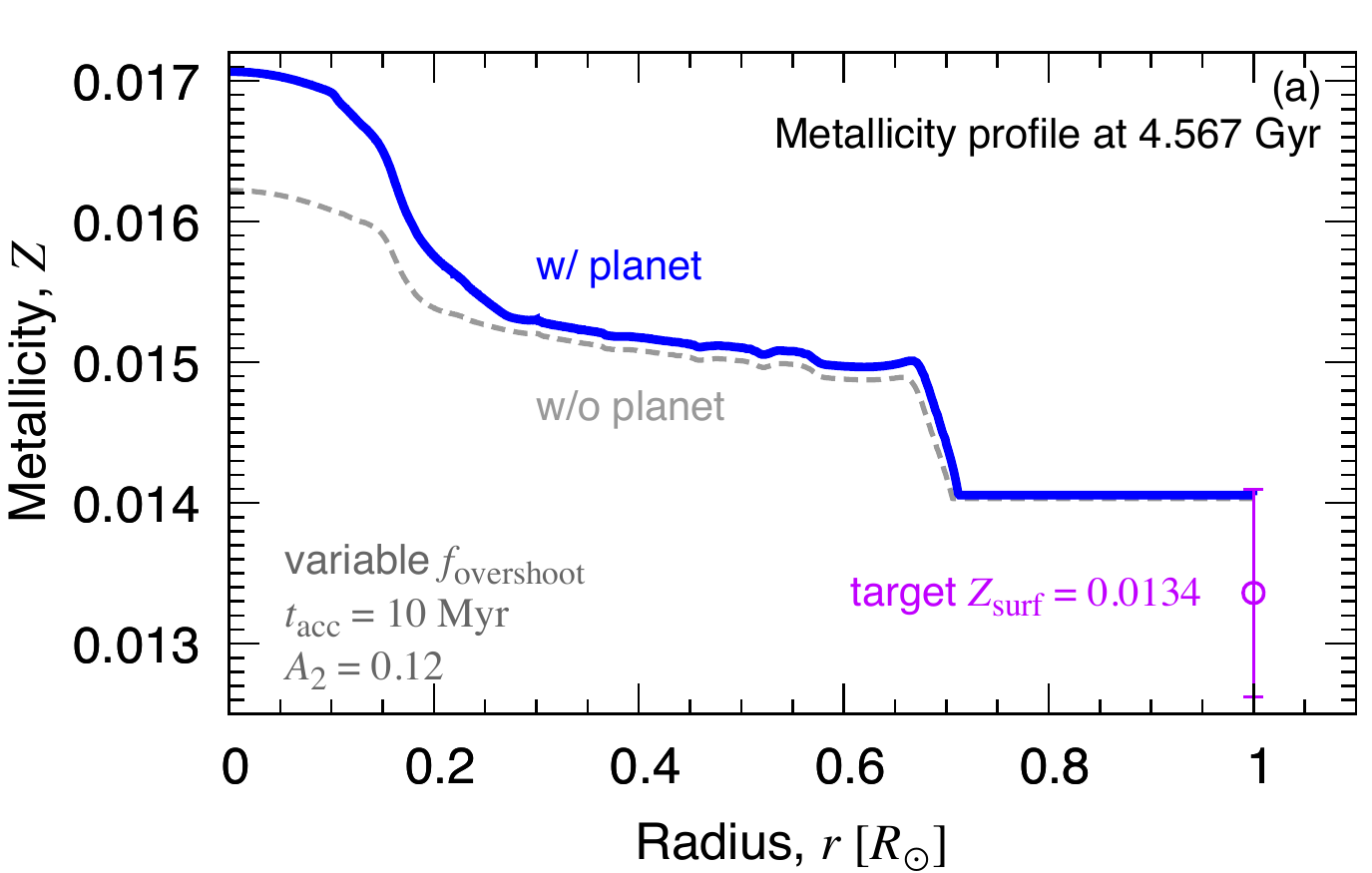}
        \includegraphics[width=0.48\hsize,keepaspectratio]{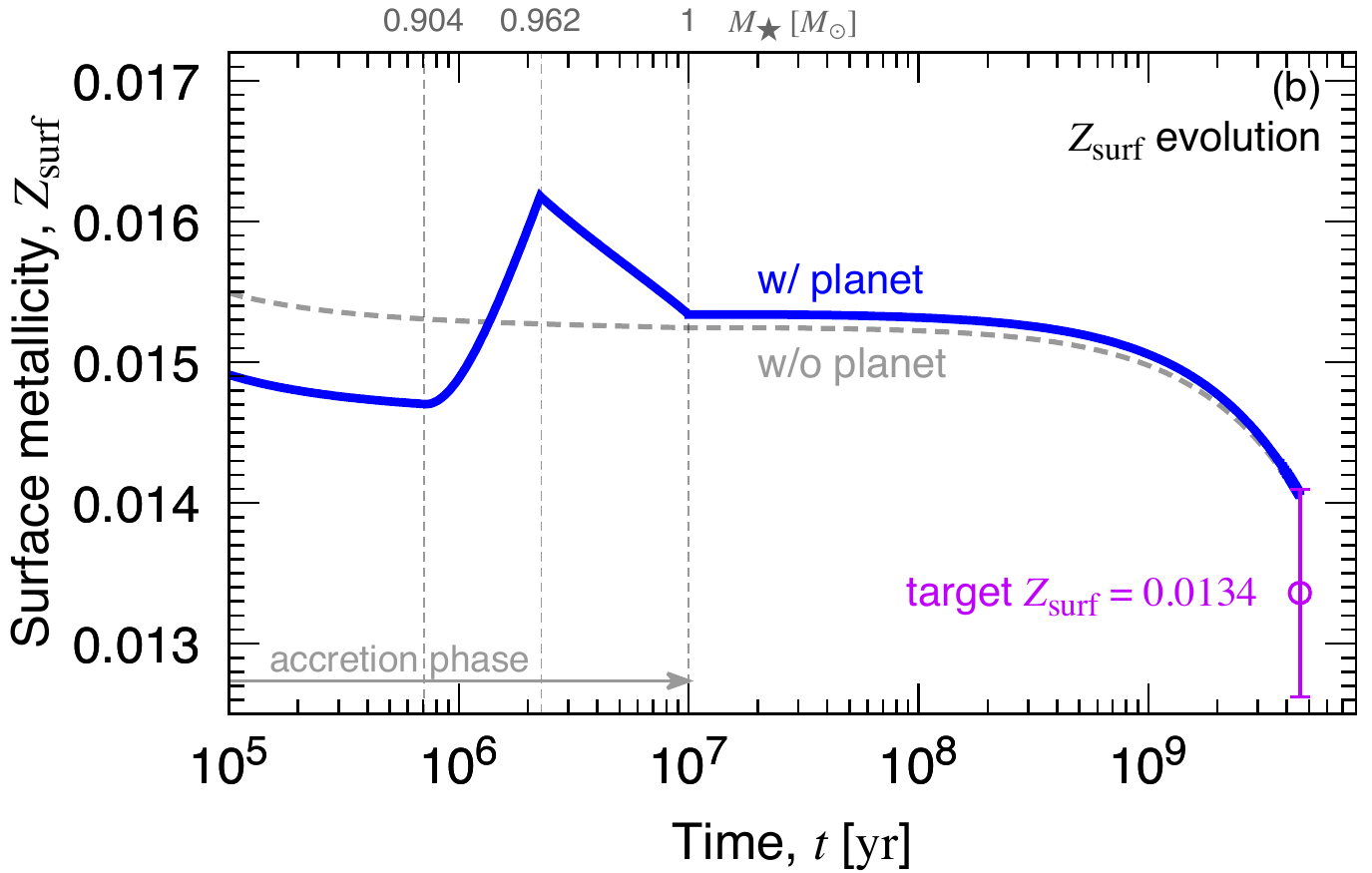} 
        \caption{\small{
        Comparison between models \kapb\ (gray dashed line) and \kappla\ (blue solid line) with $A_2 = 0.12$.
        \textit{Left panel}: Metallicity profile in the interior of present-day Sun.
        \textit{Right panel}: Time evolution of the surface metallicity $\Zs$.
        The gray vertical lines denote $M_1 (=0.904\,\Msun)$, $M_2 (=0.962\,\Msun)$, and the end of the accretion phase for model \kappla.
        The target value of $\Zs$ at 4.567\,Gyr is $0.0134\pm0.0007$ (see Table\,\ref{tab:targets}).
        }}
        \label{fig:r-Z-2}
    \end{center}
\end{figure*}

Figure\,\ref{fig:r-Z-2}(a) shows the metallicity profile in the interior of the present-day Sun for the models \kappla\ and \kapb\ (i.e., with and without a time-varying $\Zacc$, respectively) with $A_2=0.12$.
We observe that planet formation increases the central metallicity $\Zc$ by $9\times10^{-4}$ ($\simeq5\%$); however, in the outer part of the Sun ($r\ga0.2\,\Rsun$), the metallicity is almost the same.
This is because, as shown in Fig.\,\ref{fig:t-MCZ}, the proto-Sun at $\leq10\,$Myr has a small radiative core (i.e., a non-mixing region) and the signature of an initial high metallicity remains there until the present day.
Therefore, realistic protostellar and pre-MS evolution models are crucial for obtaining realistic $\Zc$ values.

Figure\,\ref{fig:r-Z-2}(b) shows the time evolution of $\Zs$.
For model \kappla\ with $A_2=0.12$, $\Zs$ varies considerably from 0.0147 to 0.0162 in the accretion phase because of planet formation; however, it decreases monotonically in the MS phase due to gravitational settling.
We note that $\Zs$ decreases slightly with time in the very early phase (e.g., $\Zs$ decreases from 0.0149 to 0.0147 during the time 0.1 to 0.7\,Myr). This is because the $0.1\,\Msun$ initial stellar seed has $Z=0.02$, which is larger than $\Zaccini=0.0140$.

\begin{table*}[!t]
	\begin{center}
	\caption{Parameter settings of the K2-MZ simulations for different planet formation scenarios.}
	\label{tab:MZ}
           \begin{tabular}{l|lll|lll}
            \hline
            \hline
            \noalign{\smallskip}
            Model name & \multicolumn{3}{c}{Parameter settings} & \multicolumn{3}{c}{Results} \\
            \noalign{\smallskip}
            \hline
            \noalign{\smallskip}
             & $M_1$ & $M_2$ & $\DZacc$ & $\Zaccini$ & $\Mlost$ & $\Mpl$\\
             & $[\Msun]$ & $[\Msun]$ & & & $[\Msun]$ & $[\Mearth]$   \\
            \noalign{\smallskip}
            \hline
            \noalign{\smallskip}
            MZ1 & 0.90 & 0.92 & 0.06 & 0.0155 & 0 & 213 \\
            MZ2 & 0.95 & 0.97 & 0.06 & 0.0144 & 0.04 & 150 \\
            MZ3 & 0.88 & 0.92 & 0.03 & 0.0155 & 0 & 213  \\
            MZ4 & 0.88 & 0.92 & 0.06 & 0.0148 & 0.03 & 150 \\
            MZ5 & 0.9703\tablefootmark{*} & 0.9703\tablefootmark{*} & -- & 0.0152 & 0 & 150 \\
            MZ6 & 0.86 & 0.92 & 0.02 & 0.0155 & 0 & 214  \\
            MZ7 & 0.86 & 0.92 & 0.06 & 0.0140 & 0.08 & 150 \\
            MZ8 & 0.91 & 0.97 & 0.06 & 0.0132 & 0.14 & 150 \\
            MZ9 & 0.92 & 0.94 & 0.06 & 0.0151 & 0.01 & 150 \\
            \noalign{\smallskip}
            \hline
            \noalign{\smallskip}
            \end{tabular} 
         \end{center}
         \tablefoot{
         $\fov=0.01$, $A_2=0.12$, and $\tacc=5$\,Myr (see Table\,\ref{tab:chi2}).
         The $\Mlost$ and $\Mpl$ values were calculated using the optimized $\Zaccini$ value (see text for details).
         $\DZacc=\Zaccmax-\Zaccini$. See Fig.\,\ref{fig:Zacc} for the definitions of $M_1$ and $M_2$.
         \tablefoottext{*}{In MZ5, we assumed that there is no pebble accretion phase (i.e., $M_1=M_2$), $\Mlost=0$, and $\Mpl=150\,\Mearth$; thus, $M_1$ was determined by $\Zaccini$ (see Eq.\,\eqref{eq:Mlost}).}
         }
\end{table*}

To evaluate the effects of planet formation on the properties of the solar interior, we examined models having different values of the three parameters, $M_1$, $M_2$, and $\DZacc\equiv \Zaccmax-\Zaccini$ (see Fig.\,\ref{fig:Zacc}), as listed in Table\,\ref{tab:MZ}. We refer to these new simulation models as K2-MZ. 
We set the fiducial $\tacc$ for these models to be 5\,Myr because the typical protoplanetary disk lifetime (i.e., half-life period) is several million years (see Sect.\,\ref{sec:interior}). We set $\tacc=10\,$Myr in Sect.\,\ref{sec:results} to investigate the maximum impact of planet formation on the $\cs$ profile. In contrast, in this section, our aim is to explore the realistic extent of the influence of planet formation.
In Sects.\,\ref{sec:discussion-Zc}--\ref{sec:discussion-Yini} we evaluate the effects of planet formation on the $\Zc$, $\Zproto$, and $\Yproto$ values.

It is important to link the parameters $M_1$, $M_2$, and $\DZacc$ to $\Mpl$, which is the mass retained by planets in the form of ``metals'' (i.e., excluding hydrogen and helium but including all the other elements) and to $ \Mlost$, which is the mass in metal-poor gas (i.e., containing hydrogen and helium only) that is lost due to selective photoevaporation and/or MHD disk winds (see Sect.\,\ref{sec:planets}). 
The conservation of the metal mass implies that
\begin{eqnarray}\label{eq:Mlost}
\Zaccini(1\,\Msun+\Mlost) \!&=&\! \Zaccini M_2+\frac12\DZacc (M_2-M_1) \nonumber \\
&& + \Mpl\,.
\end{eqnarray}
We note that in the above equation, $\Mlost$ and $\Mpl$ are degenerate, and hence, we assumed $\Mpl=150\,\Mearth$ following \citet{Kunitomo+18}, unless otherwise noted. Table\,\ref{tab:MZ} also lists the values obtained for $\Mlost$ using the optimized $\Zaccini$ and Eq.\,\eqref{eq:Mlost}. If we obtained $\Mlost<0$, we set $\Mlost=0$ and consequently increased $\Mpl$. We note that the value of $\Mpl$ in the Solar System is rather uncertain, but we estimate the upper limit to be $\simeq168\,\Mearth$  \citep{Kunitomo+18}.

\subsection{Impact on the central solar metallicity}
\label{sec:discussion-Zc}

\begin{figure*}[!ht]
  \begin{center}
        \includegraphics[width=\hsize,keepaspectratio]{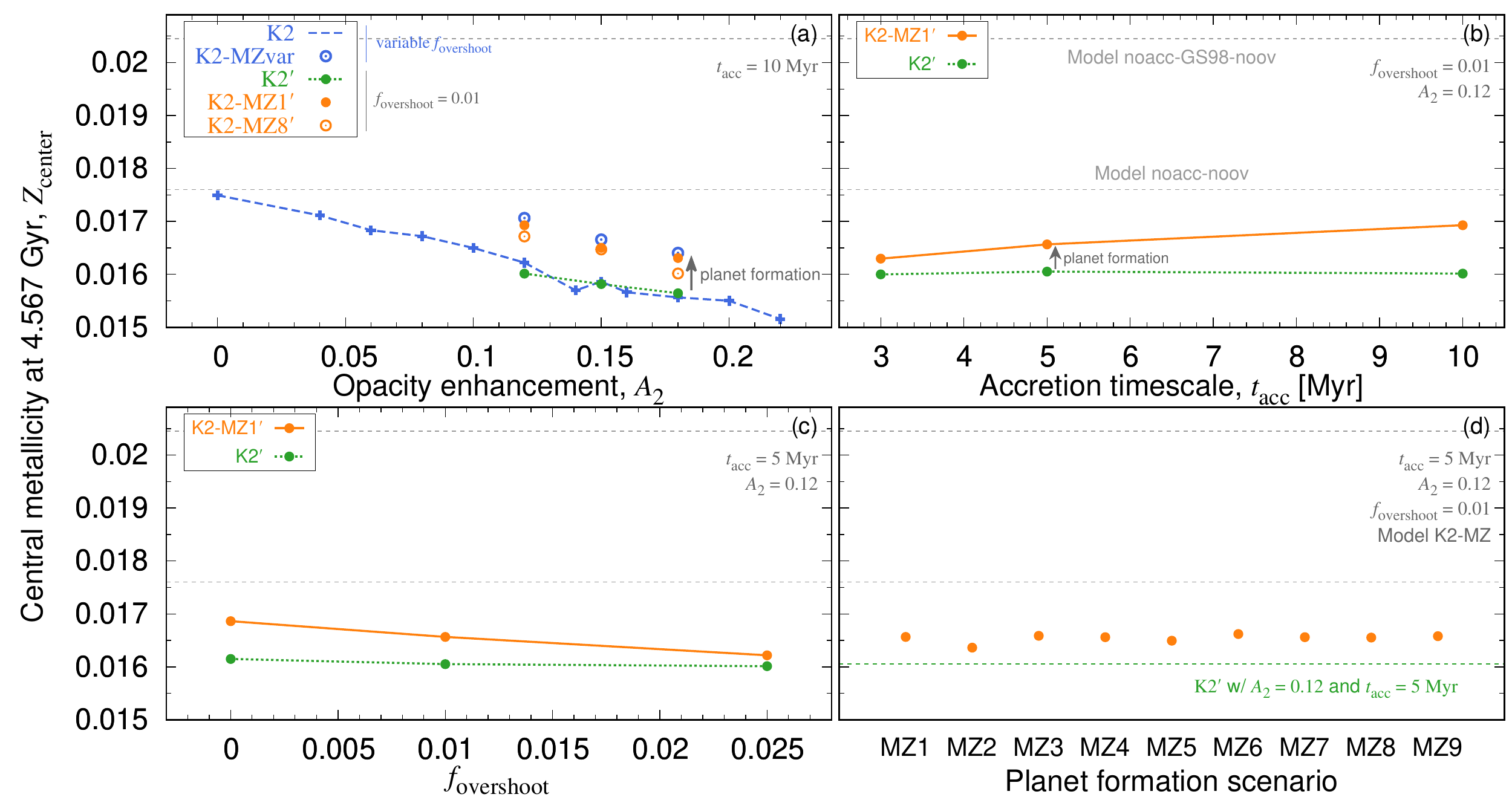}
        \caption{\small{
        Dependence of $\Zc$ on (a) $A_2$, (b) $\tacc$, (c) $\fov$, and (d) planet formation models (corresponding to different values of $M_1$, $M_2$, and $\Zaccmax$; see Table\,\ref{tab:MZ}). 
        In panel (a), the results of models K2 and K2-MZvar are shown, in addition to those of models K2$'$, K2-MZ1$'$, and K2-MZ8$'$, with $\fov=0.01$ and $\tacc=10\,$Myr (see Table\,\ref{tab:chi2}).
        In  panels (b) and (c), the green and orange curves indicate models K2$'$ and K2-MZ1$'$, respectively. The orange dots in  panel (d) are the K2-MZ simulation results.
        The two horizontal gray dashed lines in all the panels denote the results of the non-accreting models without overshooting for the \citetalias{GS98} (top) and \citetalias{Asplund+09} (bottom) compositions. The horizontal green dashed line in panel (d) denotes the result of model K2$'$ with $A_2=0.12$, $\tacc=5\,$Myr, and $\fov=0.01$.
        }}
        \label{fig:Zcore}
    \end{center}
\end{figure*}

To what extent can planet formation modify $\Zc$? To address this question, we focused on the results of the K2-MZ models (see Sect.\,\ref{sec:discussion-formation} and Table\,\ref{tab:MZ}) in addition to those of the K2 and K2-MZvar models presented in Sect.\,\ref{sec:results}. We explore in Fig.\,\ref{fig:Zcore} the dependence of $\Zc$ on the parameters $A_2$, $\tacc$, $\fov$, and planet formation scenarios.

Figure\,\ref{fig:Zcore}(a) shows that $\Zc$ is negatively correlated with $A_2$; moreover, $\Zc$ increases by $\simeq\!0.008$ because of planet formation processes, irrespective of the value of $A_2$. This negative correlation arises owing to the negative correlation between $\Zaccini$ and $A_2$ (see Sect.\,\ref{sec:kap-results}).
We note that, in the K2 and K2-MZvar models, $\fov$, $M_1$, $M_2$, and $\DZacc$ were allowed to vary in order to minimize $\chi^2$. We performed additional simulations (K2$'$, K2-MZ1$'$, and K2-MZ8$'$ with $\tacc=10\,$Myr; see Table\,\ref{tab:chi2}) by fixing these parameters. 
The $M_1$, $M_2$, and $\DZacc$ values of K2-MZ1$'$ and K2-MZ8$'$ are the same as those of K2-MZ1 and K2-MZ8, respectively (see Table\,\ref{tab:MZ}).
The results of the different models were quite similar, confirming that the final $\Zc$ value is mostly determined by the opacity ($A_2$) of the solar interior and planet formation processes (in particular the metal-poor accretion).

Next, we investigated the dependence of $\Zc$ on $\tacc$ (see Sect.\,\ref{sec:interior}).
Figure\,\ref{fig:Zcore}(b) shows the results of models K2-MZ1$'$ and K2$'$ for $\fov=0.01$ and $A_2=0.12$ with different $\tacc$ values. We find that for model K2-MZ1$'$, $\Zc$ increases with $\tacc$.
This is because the size of the radiative core in the accretion phase increases with time (Sect.\,\ref{sec:interior}); thus, in the longer $\tacc$ case, the metal-poor accretion (see Fig.\,\ref{fig:Zacc}) has a larger effect and $\Zaccini$ (and therefore $\Zc$) can be higher.
However, for model K2$'$ with a homogeneous $\Zacc$, $\Zc$ is independent of $\tacc$.
Although \citet[][see their Figures \,4 and 7]{Serenelli+11} and \citet[][see their Figures 15 and 18]{Zhang+19} have already shown that $\Zacc$ has the potential to modify $\Zc$, the accretion history in these studies \citep[and $\tacc$ of][]{Serenelli+11} is not based on the standard model of star formation (see Sect.\,\ref{sec:Mdot}). Considering the strong dependence of $\Zc$ on $\tacc$, a realistic accretion history is crucial for accurately evaluating $\Zc$.

Figure\,\ref{fig:Zcore}(c) shows that $\Zc$ decreases with $\fov$ for the model with planet formation. This is because if a star has a more vigorous overshooting, then the radiative core becomes smaller and is developed in a slightly later phase.
However, for the cases with homogeneous $\Zacc$, $\Zc$ is insensitive to $\fov$ because the internal metallicity profile is homogeneous until gravitational settling sets in (i.e., in $\sim1$\,Gyr).
Although recent studies have attempted to constrain the efficiency of overshooting by hydrodynamic simulations \citep{Freytag+96,Korre+19,Higl+21} and observations \citep[e.g., ][]{Deheuvels+16}, it remains uncertain. Further studies to constrain the mixing in stellar interiors are necessary.

Finally, Fig.\,\ref{fig:Zcore}(d) shows $\Zc$ for the K2-MZ models. We find that $\Zc$ is not sensitive to planet formation scenarios. This is related to the process used to satisfy the constraints in this work. We fixed $M_1$, $M_2$, and $\DZacc$, and chose $\Zaccini$ such that it satisfied the constraints (in particular, $\ZXs$; see Fig.\,\ref{fig:Z-MZ}(b)) imposed by the simplex method. 
Consequently, $\Zc$ does not have much variation with respect to different planet formation scenarios.

We note that in models with a higher $\Zc$, the central opacity is higher, leading to an increased radiative temperature gradient \citep[e.g.,][]{Kippenhahn+Weigert90}, and thus, to a higher central temperature. For models with $A_2 = [0.12, 0.18]$ and $A_3=0$, the central temperature ranges from $1.559\times10^7$ to $1.567\times10^7$\,K for values of $\Zc$ in the range $0.0155$--$0.0171$.

In summary, the value of $\Zc$ is larger for a lower $A_2$, longer $\tacc$, and lower $\fov$. For $A_2 =0.12$ and $\tacc=5\,$Myr, planet formation enhances $\Zc$ up to 0.01686.
The $\Zc$ values for models with $A_2 = [0.12, 0.18]$ (i.e., the models that reproduce the sound speed profile) and planet formation are $\simeq$5\% higher than those of homogeneous $\Zacc$ models (see Fig.\,\ref{fig:Zcore}(a)).
Interestingly, recent observations of solar neutrino fluxes reaching the Earth have also pointed to high values of the metallicity at the solar center \citep{Agostini+18,Borexino-Collaboration20}; however, their interpretation remains tentative at present. More accurate observations are required in the future.

\subsection{Impact on the constraints for the primordial metallicity}
\label{sec:discussion-Zini}

\begin{figure*}[htb]
  \begin{center}
        \includegraphics[width=\hsize,keepaspectratio]{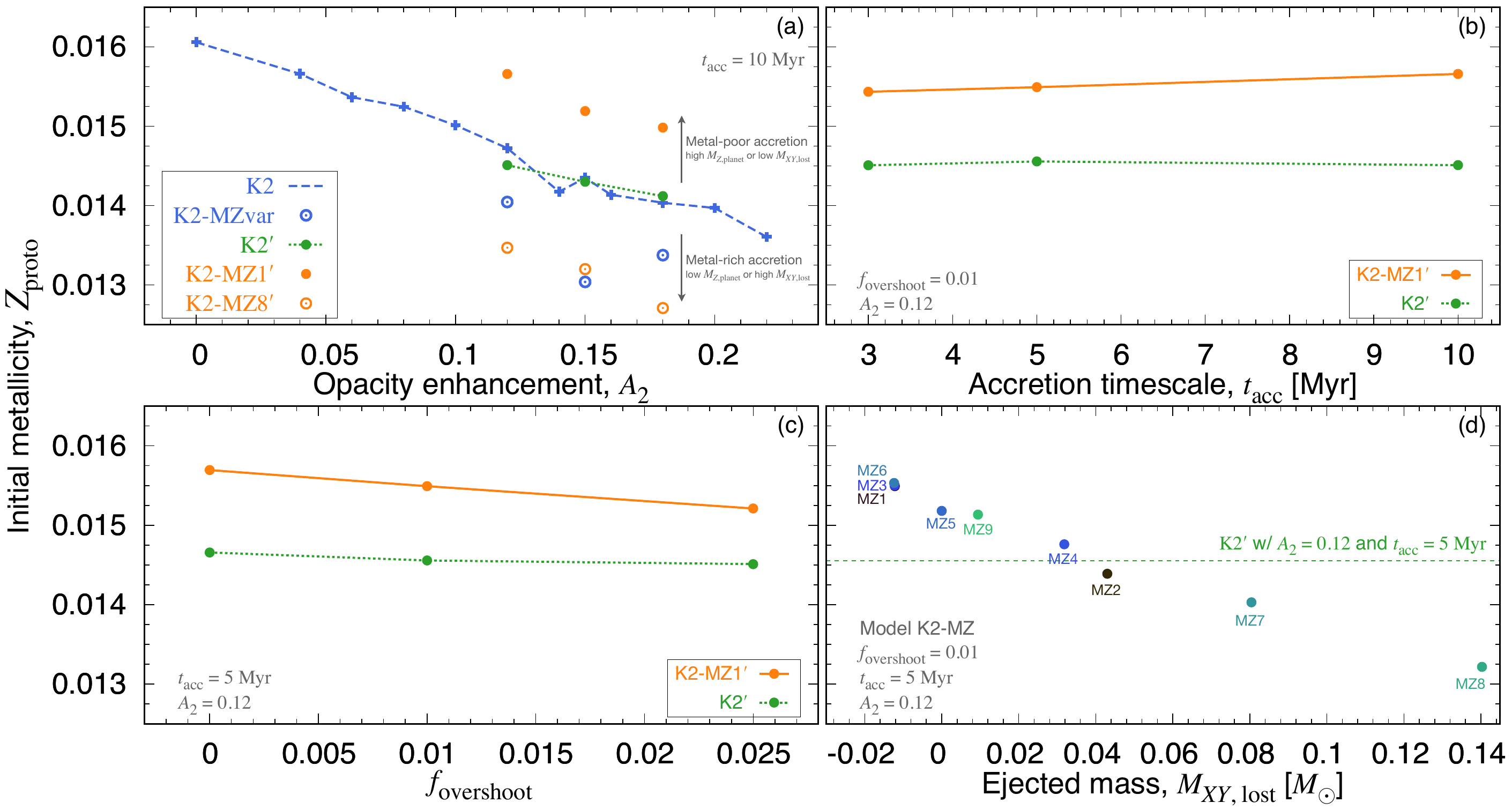}
        \caption{\small{
        Same as Fig.\,\ref{fig:Zcore} except the vertical axis is the initial metallicity $\Zaccini$, which corresponds to the primordial metallicity of the protosolar molecular cloud core; and  
        the horizontal axis of (d) is $\Mlost$, which was calculated using Eq.\,\eqref{eq:Mlost} and $\Mpl=150\,\Mearth$. We note that the unphysical $\Mlost<0$ values are due to our assumption that $\Mpl=150\,\Mearth$. These solutions are equivalent to solutions with $\Mlost=0$ and $\Mpl>150\,\Mearth$ (see Table\,\ref{tab:MZ}).
        }}
        \label{fig:Zini}
    \end{center}
\end{figure*}

\citet{Vinyoles+17} obtain constraints on the primordial metallicity $\Zproto$ that mostly depend on the assumed abundance table from their non-accreting models, that is, $\Zproto=0.0149$ for \citetalias{Asplund+09} and $\Zproto=0.0187$ for \citetalias{GS98}. For non-accreting models, we obtain $\Zproto=0.0163$ and $\Zproto=0.0187$, respectively. We find that the difference in the \citetalias{Asplund+09} case arises from the different number of target values, $N$. As described in Sect.\,\ref{sec:chi2}, $N=3$ in \citet[][and most previous studies]{Vinyoles+17}, whereas $N=6$ in this study, leading to a much better fit of $\Ys$. We stress that these solutions, however, do not represent the best fits to the observational constraints.

Figure\,\ref{fig:Zini} shows the variation in the primordial metallicity of the protosolar molecular cloud core $\Zaccini$ (see Sect.\,\ref{sec:planets}) for the same models and parameters as in Fig.\,\ref{fig:Zcore}, including our best models.
Figure\,\ref{fig:Zini}(a) shows that $\Zaccini$ is negatively correlated with $A_2$, as described in Sect.\,\ref{sec:kap-results}. Although this behavior is the same as that of $\Zc$, Fig.\,\ref{fig:Zini}(a) shows some differences compared to Fig.\,\ref{fig:Zcore}(a). Most importantly, while planet formation processes always increase $\Zc$, they can either increase or decrease $\Zaccini$. As shown by Figs.\,\ref{fig:Zini}(b) and (c), this cannot be explained by varying $\tacc$ or $\fov$, which have little impact on $\Zaccini$.

Figure\,\ref{fig:Zini}(d) shows that $\Zaccini$ decreases with $\Mlost$.
When the mass of the hydrogen and helium that is selectively removed from the disk is high, the primordial metallicity $\Zaccini$ must be low to compensate for the accretion of the disk gas that becomes relatively metal-rich and to account for the present-day observations. 
Conversely, in models with a low value of $\Mlost$, $\Zproto$ must be high to compensate for the metal-poor accretion. 
We note that the retention of heavy elements by planet formation has the opposite effect (see Eq.\,\eqref{eq:Mlost}) so that a low value of $\Mlost$ is equivalent to a high value of $\Mpl$ (we set $\Mpl = 150\,\Mearth$ in Fig.\,\ref{fig:Zini}(d); see Sect.\,\ref{sec:discussion-formation}).
If $\Mlost \simeq 0.03\,\Msun$, then $\Zaccini\Mlost$ compensates for $\Mpl = 150\,\Mearth$ and the value of $\Zproto$ is the same as that obtained for model K2$'$ (i.e., with homogeneous $\Zacc$).

The above behavior can also explain the non-monotonic relation between $\Zaccini$ and $A_2$ for the \kappla\ models in Fig.\,\ref{fig:Zini}(a). Indeed, the models with $A_2=0.12$, 0.15, and 0.18 correspond to $\Mlost=0.13$, $0.18$, and $0.13\,\Msun$, which explains the much lower $\Zaccini$ value for the \kappla\ model with $A_2=0.15$. 

For our preferred models with $A_2=[0.12, 0.18]$, which successfully reproduce the observational constraints (see Sect.\,\ref{sec:kap-results}), $\Zaccini$ ranges from 0.0127 to 0.0157. 
A slightly narrower range may be obtained by imposing tighter constraints on $\Mlost$ and $\Mpl$, namely, $\Mlost\la 0.05\,\Msun$ and $\Mpl=97$--168\,$\Mearth$ (see Sect.\,\ref{sec:planets}). In this case, $\Zaccini$ lies approximately in the range 0.0140--0.0153.

\subsection{Impact on the constraints for the primordial helium abundance}
\label{sec:discussion-Yini}

Estimating the primordial helium abundance (i.e., the helium abundance in the protosolar molecular cloud core $\Yproto$) accurately is of particular interest, {because} it provides reliable constraints on the internal structure and composition of Jupiter and Saturn \citep{Guillot+18}. \citet{Serenelli+Basu10} obtained a primordial helium abundance of $0.273\pm0.006$ by using solar evolution models with turbulent mixing below the CZ. {\citet{Vinyoles+17} obtained a lower value of $\Yproto=0.2613$ for the \citetalias{Asplund+09} composition, which resulted in a poor fit of $\Ys$ (see Sect.\,\ref{sec:chi2}). }

\begin{figure*}[!ht]
  \begin{center}
        \includegraphics[width=\hsize,keepaspectratio]{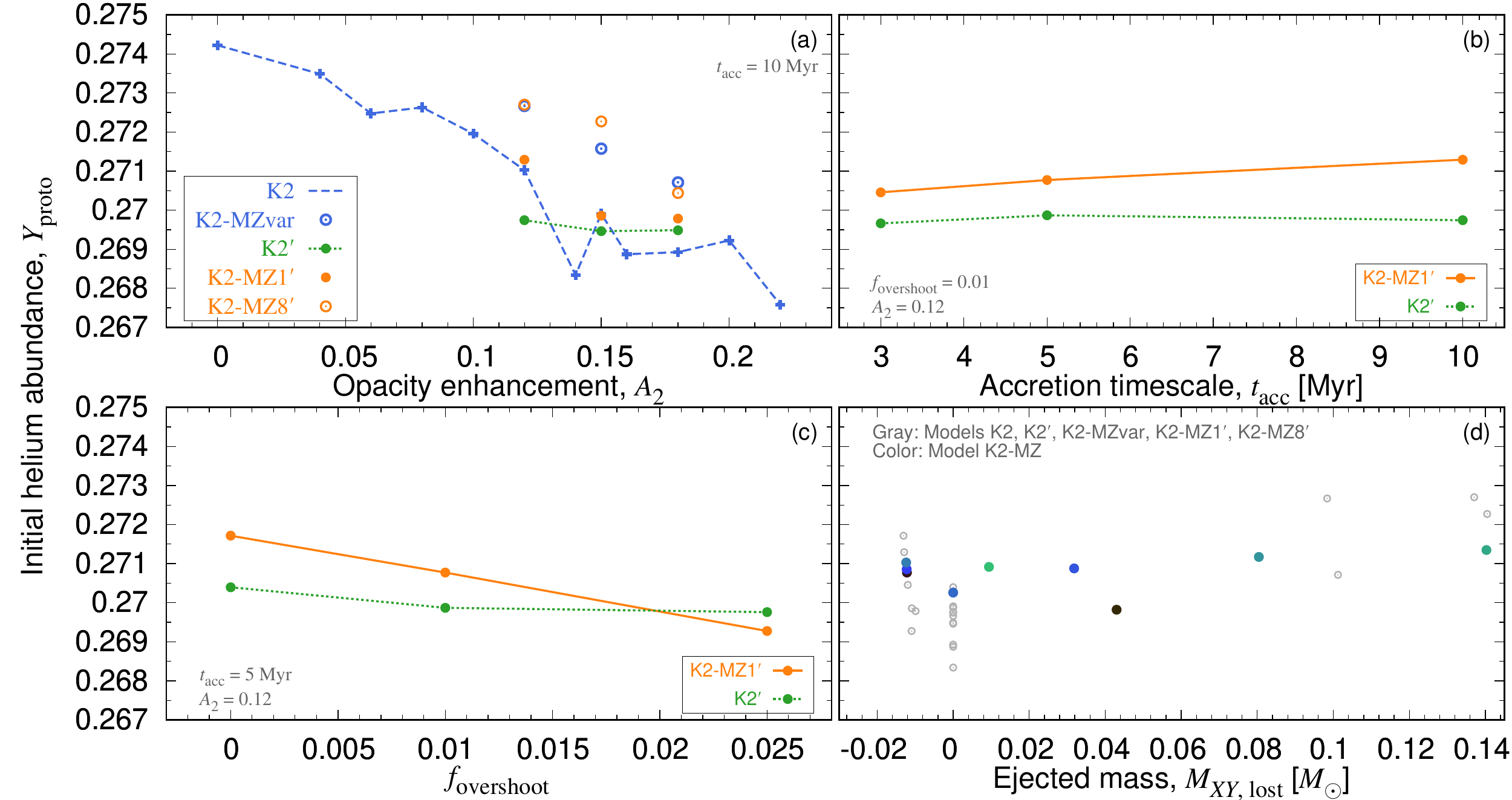}
        \caption{\small{
        Same as Fig.\,\ref{fig:Zini} except the vertical axis is the initial helium mass fraction $\Yaccini$, which corresponds to the primordial helium abundance of the protosolar molecular cloud core.
        The gray open circles in panel (d) show the results of models K2, K2$'$, K2-MZvar, K2-MZ1$'$, and K2-MZ8$'$ with $A_2=[0.12, 0.18]$.
        }}
        \label{fig:Yini}
    \end{center}
\end{figure*}

Figure\,\ref{fig:Yini} shows $\Yaccini$ for the same models as in Figs.\,\ref{fig:Zcore} and \ref{fig:Zini}. Again, the most significant correlation is obtained between $\Yaccini$ and the opacity-enhancement parameter $A_2$. We observe that $\Yaccini$ decreases from $0.274$ for the standard model ({K2 model} with $A_2=0$) to $0.267$ for the highest value of $A_2=0.22$. Figures\,\ref{fig:Yini}(b)--(d) show that $\Yaccini$ has a weak dependence on $\tacc$, $\fov$, and $\Mlost$ (for the models with planet formation).

However, Fig.\,\ref{fig:Yini}(a) shows that planet formation processes result in an increase in $\Yaccini$, independent of $\Mlost$. This is similar to what is observed for $\Zc$ but different from what is observed for $\Zaccini$. The reason for such behavior is threefold.
First, hydrogen and helium are believed to have a common evolutionary history that differs from that of metals (see Sect.\,\ref{sec:Zacc}). This implies that planet formation processes (i.e., pebble wave, planetesimal formation, and disk winds) affect $\Zacc$ while conserving the mass ratio of helium to hydrogen ($Y/X$) in the accreted material. 
Second, when planet formation results in a high $\Zc$, the central temperature $\Tc$ of the star increases because of the changes in the opacity, which in turn increases the radiative temperature gradient (see Sect.\,\ref{sec:discussion-Zc}).

Third, this increase in temperature affects the nuclear burning rate $r_{pp}$ because $r_{pp}\propto \Xc^2\Tc^4$ \citep{Kippenhahn+Weigert90}, where $\Xc\sim 1-\Yc$ is the hydrogen abundance in the Sun's nuclear burning core. Given the global constraints on the structure of the present-day Sun, to compensate for the higher $\Tc$, $\Xc$ needs to be lower on average on the MS, which in turn implies a lower $\Xaccini$ and therefore a higher $\Yaccini$. For these reasons, planet formation processes lead to higher values of $\Zc$, $\Tc$, and hence, $\Yaccini$.

In addition, $\Yaccini$ is positively correlated with $A_3$. Therefore, if the high-temperature opacities are changed, then $\Yaccini$ would be modified by an absolute factor, which is estimated to be $\delta\Yaccini\sim A_3/20$. 

Overall, our results for the models that fit the observational constraints (i.e., $A_2=[0.12, 0.18]$) imply that $\Yaccini$ ranges from 0.268 to 0.274.
This range compatible with, but somewhat more tightly constrained than the range obtained by \citet{Serenelli+Basu10}, which is 0.267--0.278. 
The $\Yaccini/(1-\Zaccini)$ value was found to range from 0.272 to 0.278.

\section{Conclusion}
\label{sec:conclusion}

In this work, we studied how the formation of the Solar System affected the composition and internal structure of the Sun. The Sun was formed owing to the collapse of a molecular cloud core, and it grew via accretion of the gas in the circumsolar disk over several million years while planet formation processes were at play. According to protoplanetary disk evolution models, when the proto-Sun was approximately between 90\% and 98\% of its final mass (see Fig.\,\ref{fig:Eacc}), a pebble wave led to a phase of accretion of high-metallicity gas. This was followed by a phase of metal-poor accretion due to the formation of planetesimals and planets, while the hydrogen and helium may have been selectively lost from the disk atmosphere via photoevaporation and/or MHD disk winds. Therefore, the Sun grew via the accretion of gas {with} an evolving composition. 

To study the evolution of the Sun and reproduce the present-day constraints on its structure and composition obtained from spectroscopy and helioseismology, we performed an extensive ensemble of simulations. Our simulations included accretion in the protostellar and pre-MS phases. The input parameters were adjusted using $\chi^2$ minimization to best reproduce the present-day constraints (i.e., after $4.567$\,Gyr of evolution) on the luminosity, effective temperature, surface composition ($\Ys$ and $\ZXs$), CZ radius ($\RCZ$), and sound speed profile of the Sun. The input parameters used were the mixing length parameter $\amlt$, overshooting parameter $\fov$, initial helium abundance $\Yaccini$, and initial metallicity $\Zaccini$. A second set of adjustable parameters were introduced to modify the opacity, namely, $A_1$, $A_2$, and $A_3$. A third set of parameters were introduced to modify the composition of the accreted material, $M_1$, $M_2$, $\Zaccmax$, and $\Yaccmin$.

Several scenarios were tested. Classical non-accreting models with old abundances (\citetalias{GS98}) are known to provide better fits to the helioseismic constraints than those using more recent abundances (\citetalias{Asplund+09}), leading to the so-called ``solar abundance problem.'' Models that consider the accretion of gas with an evolving composition (i.e., include planet formation processes) do not improve the $\chi^2$ fit significantly. This is because the proto-Sun has an almost fully convective interior in the accretion phase, implying that the accreted gas is heavily diluted and therefore the changes in the structure of the present-day Sun are quite limited. Models involving helium-poor accretion, as proposed by \citet{Zhang+19}, indeed improve the fit. However, such models are probably unlikely because the giant planets in our Solar System contain significant amounts of helium in their interiors. 
We note that other possibilities, such as extra mixing, were not tested in this work \citep[see][]{Christensen-Dalsgaard+18,Buldgen+19,Yang19}.
Our best models were found to be those with the \citetalias{Asplund+09} composition and a 12\%--18\% opacity increase centered at $T = 10^{6.4}$\,K. 
This is slightly higher but qualitatively in good agreement with the high iron opacities measured by \citet{Bailey+15} at this temperature range. 
The models with an opacity increase in the range 12\%--18\% represent better fits to the observations than those using old abundances, and are therefore a promising solution to the solar abundance problem.

The impact of planet formation on the solar structure can be examined by using the aforementioned models to globally fit the observational constraints and by modifying the parameters $M_1$, $M_2$, and $\Zaccmax$. 
We find that despite the negligibly small effect on the sound speed profile (and therefore on the helioseismic constraints), planet formation processes lead to a limited but real (up to $5\%$) increase in the metallicity $\Zc$ of the deep solar interior ($r\la0.2\,\Rsun$). 
The increase in $\Zc$ is smaller for a larger overshooting parameter $\fov$ but larger for a longer disk-accretion timescale $\tacc$. 
Qualitatively, this is in accordance with recent solar neutrino measurements that appear to favor high-$Z$ values in the deep solar interior \citep{Agostini+18, Borexino-Collaboration20}.
We will examine this issue in future work.

We also constrained the primordial composition of the molecular cloud core that gave birth to the Sun by using models that best reproduced all the present-day constraints. We found that the protosolar metallicity $\Zproto$ ranged from 0.0127 to 0.0157; moreover, the retention of heavy elements due to planet formation yielded higher values of $\Zproto$, while significant disk winds led to lower values. Similarly, the protosolar helium mass fraction $\Yproto$ 
was found to slightly increase because of planet formation processes and its value ranges from 0.268 to 0.274.

In conclusion, we expect that a combined investigation of the solar interior, solar evolution models (in particular, those that include improved opacities and a more sophisticated treatment for mixing), and observational constraints (from surface abundances, helioseismic observations, and neutrino fluxes) will help constrain the processes that led to the formation of the Solar System.

\begin{acknowledgements}
     This paper is dedicated to the memories of Rudolf Kippenhahn, whose diagrams were a source of inspiration for this work, and Jean-Paul Zahn, who taught one of us how to read them (upside-down!).
     We are grateful to Ga\"{e}l Buldgen, J{\o}rgen Christensen-Dalsgaard, Johan Appelgren, and Vardan Elbakyan for kindly providing their simulation results. We also thank 
     Lionel Bigot for fruitful discussions and comments.
      This work was supported by the JSPS KAKENHI Grant (number 20K14542), the Astrobiology Center Program of National Institutes of Natural Sciences (NINS; grant number AB301023), and a long-term visitor JSPS fellowship to TG.
      Numerical computations were carried out on the PC cluster at the Center for Computational Astrophysics, National Astronomical Observatory of Japan, and in part on the ``Mesocentre SIGAMM'' machine hosted by the Observatoire de la Cote d'Azur.
      \textit{Software}: \texttt{MESA} \citep[version 12115;  ][]{Paxton+11,Paxton+13,Paxton+15,Paxton+18, Paxton+19}; Numpy \citep{vanderWalt+11}; and Scipy \citep{Virtanen+20}. 
\end{acknowledgements}

%
%
\bibliographystyle{aa}

 \begin{appendix} 

\section{Details of simulation results} \label{app:non-accreting}

In this Appendix, we provide the details of our simulation results.
Tables\,\ref{tab:chi2-results-input} and \ref{tab:chi2-results-output} show the input parameters and results at 4.567\,Gyr, respectively, which were optimized using the simplex method.

\begin{figure*}[!t]
  \begin{center}
        \includegraphics[width=\hsize,keepaspectratio]{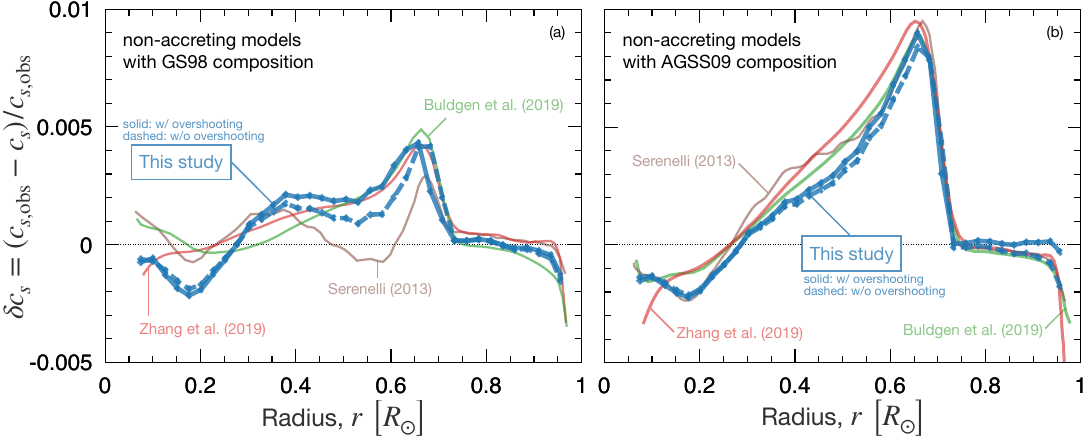}
        \caption{\small{
        Radial profiles of observed minus calculated sound speed $\delcs$ for the non-accreting models with the \citetalias{GS98} (left panel) and \citetalias{Asplund+09} (right) compositions.
        The blue solid and dashed lines with circles indicate the results with and without overshooting, respectively.
        The other solid lines show the results in the literature: the red, green, and brown lines indicate the results of \citet[][Model SSM98 for the left panel and SSM09 for the right panel]{Zhang+19}, \citet[][Models ``GS98-Free-OPAL'' and ``AGSS09-OPAL-OPAL''; see their Figure 1]{Buldgen+19}, and \citet[][see their Figure 1]{Serenelli13}.
        }}
        \label{fig:non-accreting}
    \end{center}
\end{figure*}

Figure\,\ref{fig:non-accreting} presents a comparison of the $\delcs$ profiles obtained by previous studies and the present study for the non-accreting models with the \citetalias{GS98} and \citetalias{Asplund+09} compositions.
As mentioned in Sect.\,\ref{sec:noacc}, our results agree well with those of previous studies.

\begin{figure}[!ht]
  \begin{center}
        \includegraphics[width=\hsize,keepaspectratio]{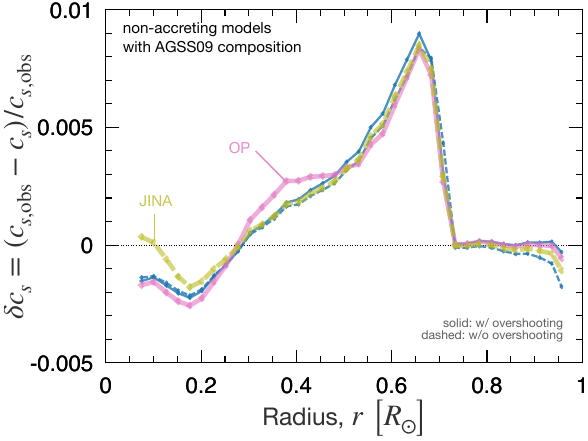}
        \caption{\small{
        Same as Fig.\,\ref{fig:non-accreting}(b) (i.e., non-accreting models with the \citetalias{Asplund+09} composition). The pink and gold lines indicate the cases with the OP opacity table (with overshooting) and JINA nuclear reaction table (without overshooting), respectively.
        The thin blue lines are the same as the thick blue lines in Fig.\,\ref{fig:non-accreting}(b).
        }}
        \label{fig:JINA}
    \end{center}
\end{figure}

Figure\,\ref{fig:JINA} shows the $\delcs$ profiles for non-accreting models with the \citetalias{Asplund+09} composition for different settings of the opacity and nuclear reaction tables.
Only marginal differences exist between the different models {\citep[see also][]{Serenelli+11,Buldgen+19,Salmon+21}}.

\begin{figure*}[!t]
  \begin{center}
        \includegraphics[width=0.49\hsize,keepaspectratio]{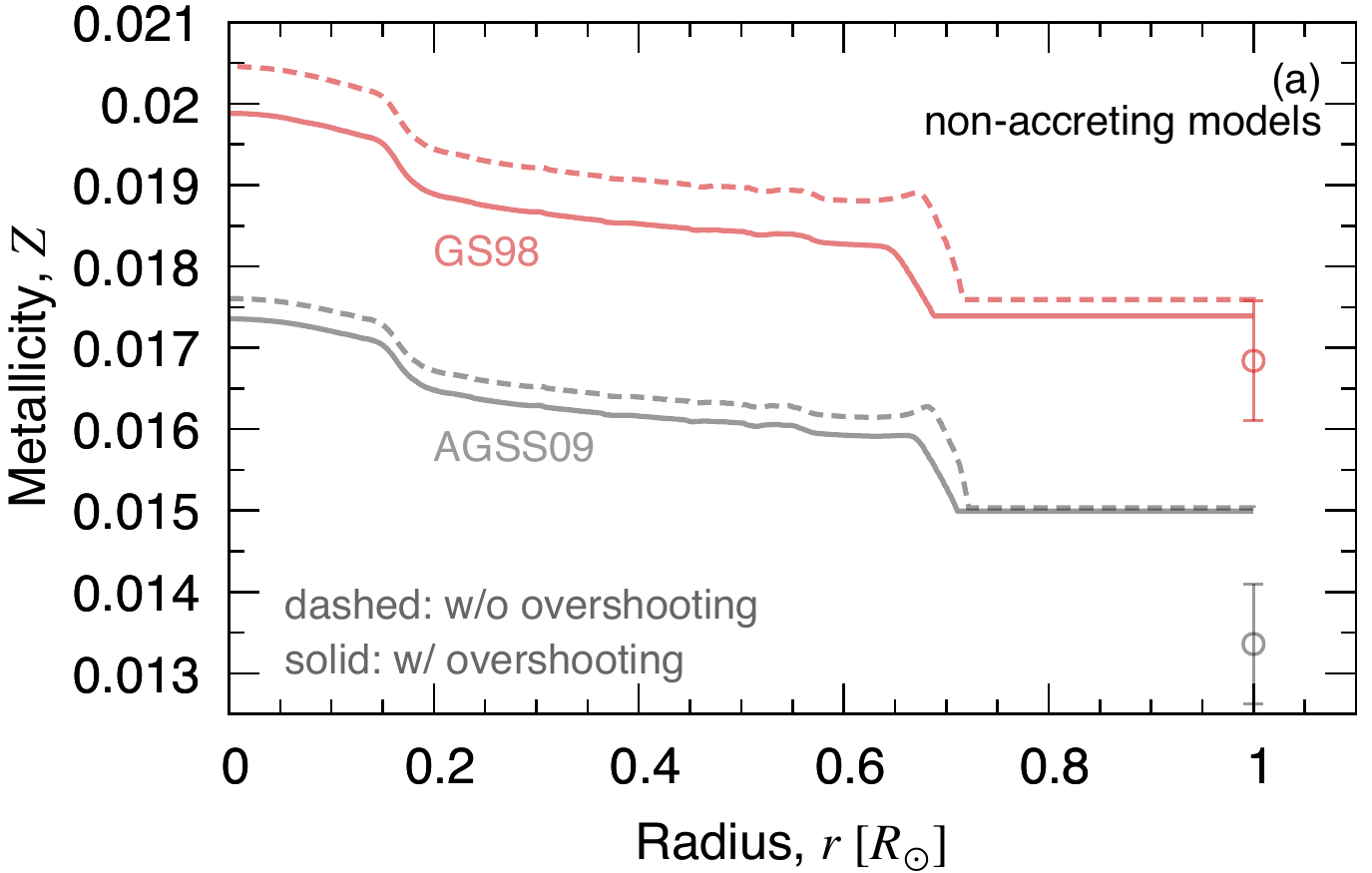}
        \includegraphics[width=0.49\hsize,keepaspectratio]{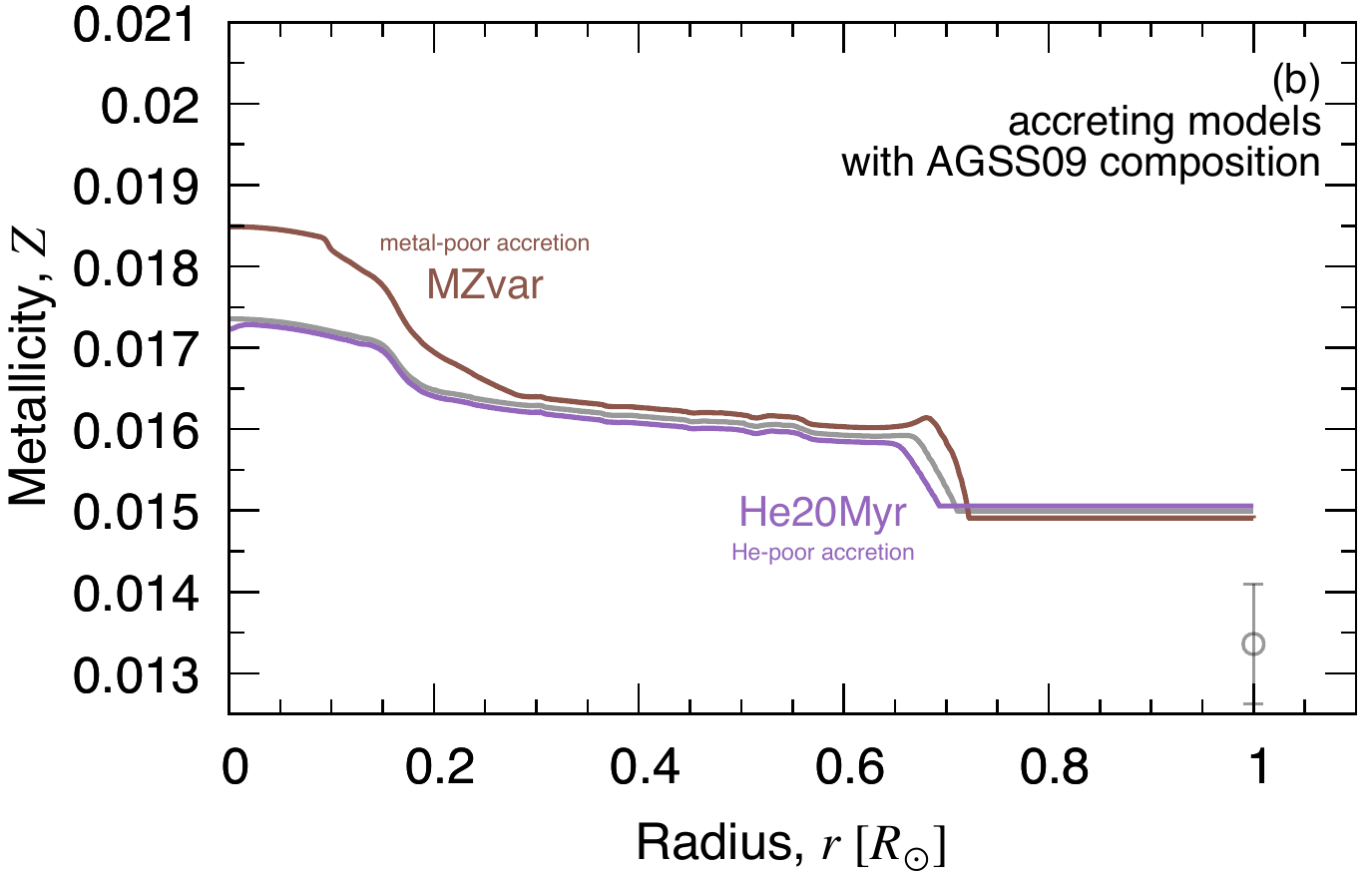}
        \caption{\small{
        Radial metallicity profiles in the solar interior at 4.567\,Gyr for non-accreting models (left panel) and accreting models (right panel).
        The red and gray lines denote the cases with the \citetalias{GS98} and \citetalias{Asplund+09} compositions, respectively (models noacc-GS98-noov, noacc-GS98, noacc-noov, and noacc, from top to bottom).
        The target values of $\Zs$ at 4.567\,Gyr are $0.0168\pm0.0007$ and $0.0134\pm0.0007$, respectively (see Table\,\ref{tab:targets}).
        The brown and purple lines in the right panel show the results of models MZvar and He20Myr, respectively (see Table\,\ref{tab:chi2} and Fig.\,\ref{fig:other}).
        }}
        \label{fig:r-Z-other}
    \end{center}
\end{figure*}

Figure\,\ref{fig:r-Z-other} shows the metallicity profiles at 4.567\,Gyr.
Since the observed $\ZXs$ value for the \citetalias{GS98} composition is higher than that for the \citetalias{Asplund+09} composition, the corresponding $\Zc$ is also higher.
As shown in Sect.\,\ref{sec:discussion-planets}, a metal-rich structure is obtained for the case with metal-poor accretion (see model \FULL). 

\begin{figure*}[!t]
  \begin{center}
        \includegraphics[width=0.49\hsize,keepaspectratio]{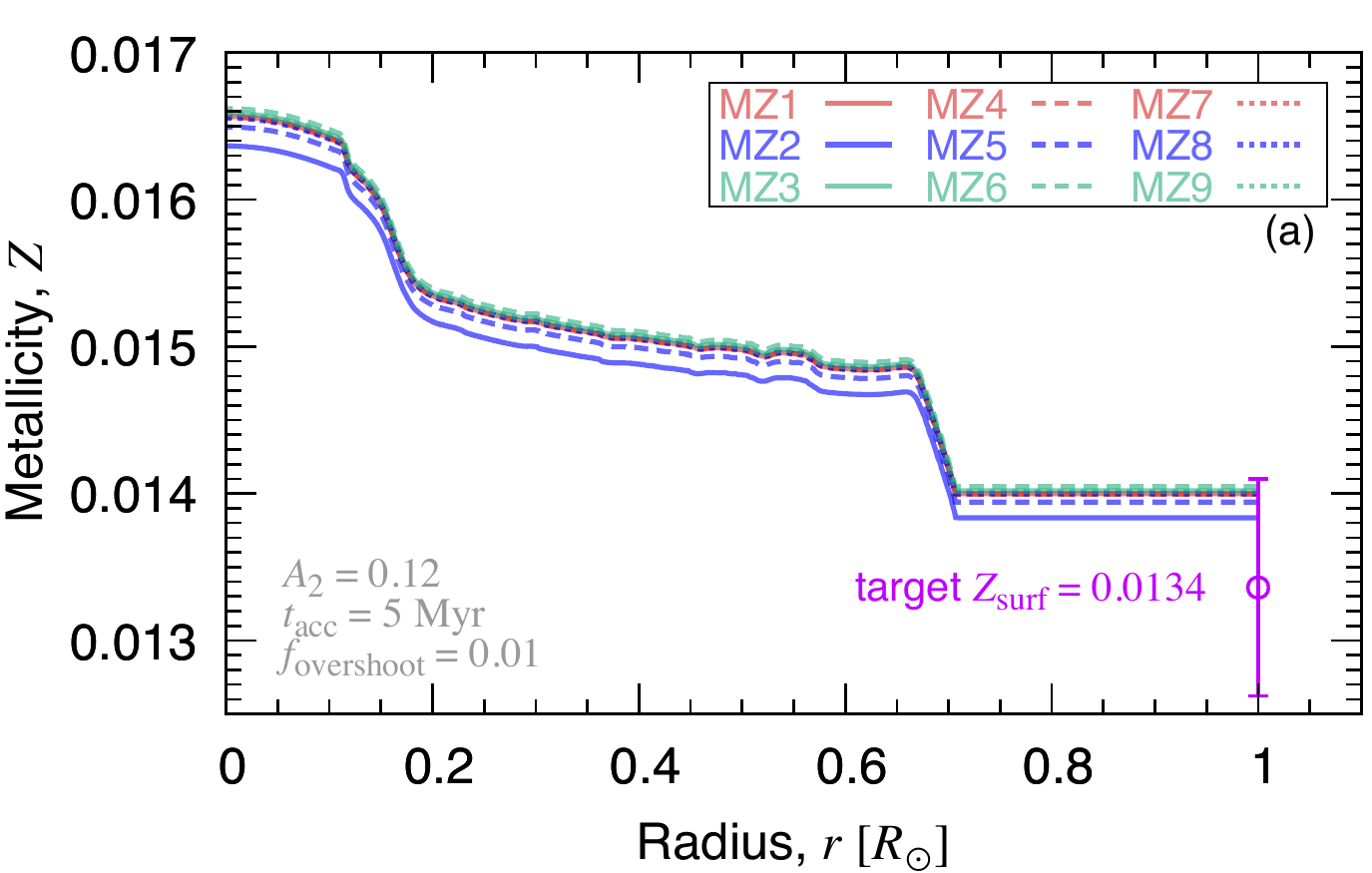}
        \includegraphics[width=0.49\hsize,keepaspectratio]{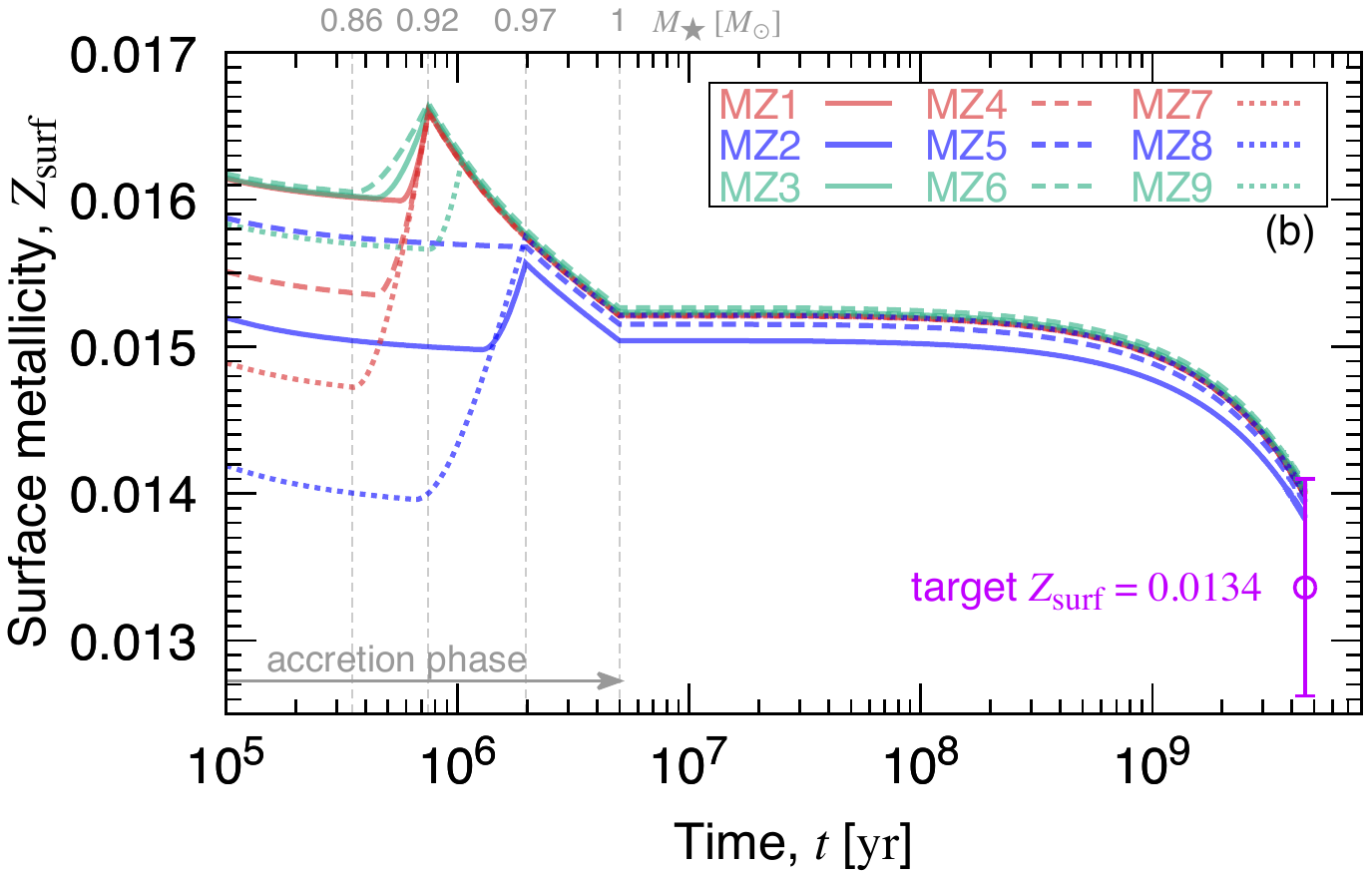}
        \caption{\small{
        Results of the K2-MZ model with different $\Zacc$ (models MZ1--MZ9; see Tables\,\ref{tab:chi2} and \ref{tab:MZ}). The left and right panels show the radial metallicity profile in the solar interior at 4.567\,Gyr and the time evolution of the surface metallicity $\Zs$, respectively.
        }}
        \label{fig:Z-MZ}
    \end{center}
\end{figure*}

\begin{figure}[!t]
  \begin{center}
        \includegraphics[width=\hsize,keepaspectratio]{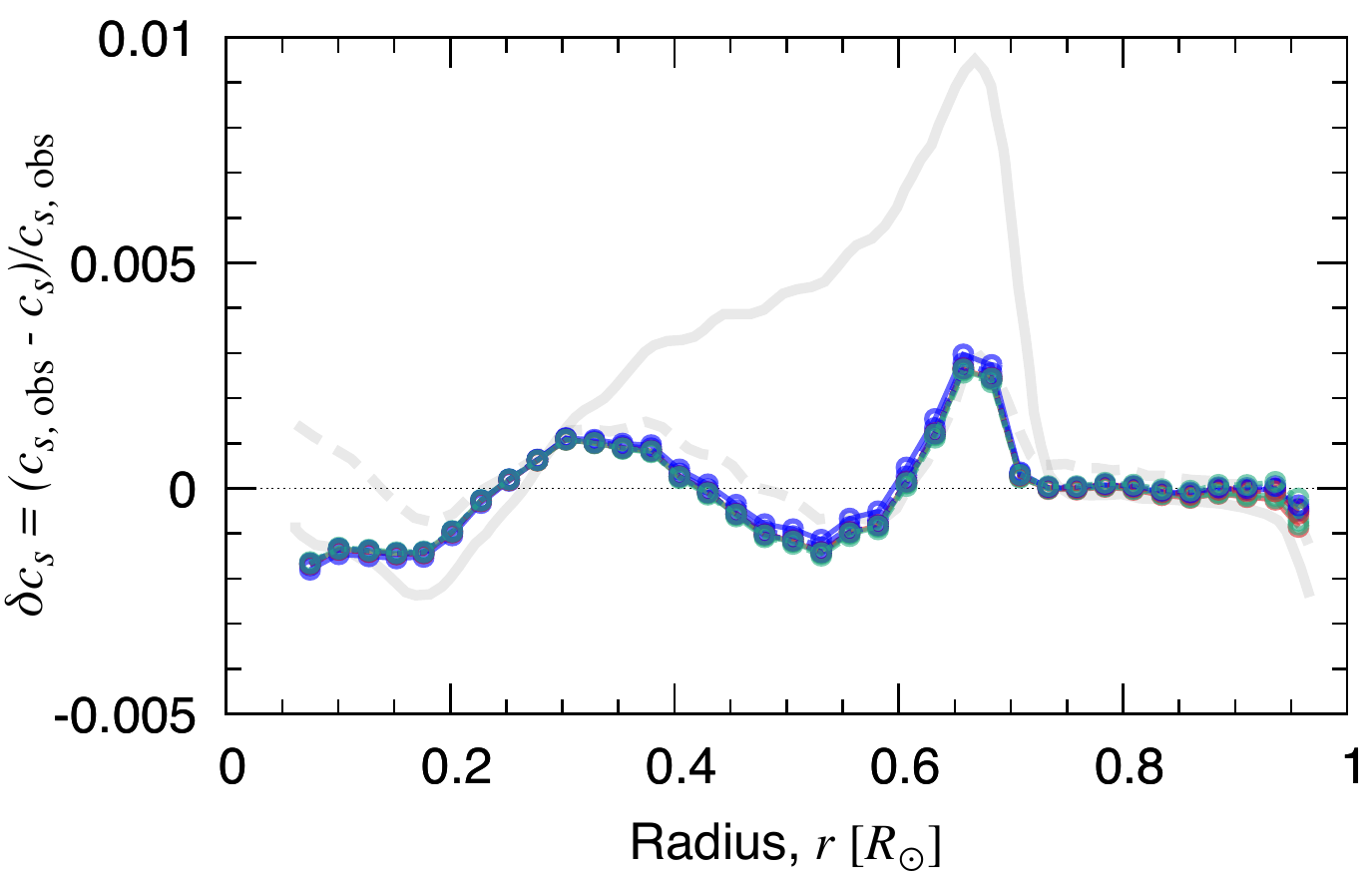}
        \caption{\small{
        Radial sound speed profiles of the K2-MZ model with different $\Zacc$; the lines follow the same color-coding as in Fig.\,\ref{fig:Z-MZ}.
        The gray solid and dashed lines show the results of the non-accreting models with overshooting and with \citetalias{Asplund+09} and \citetalias{GS98} abundances (i.e., noacc and noacc-GS98 in Table\,\ref{tab:chi2}), respectively.
        }}
        \label{fig:r-cs-MZ}
    \end{center}
\end{figure}

Figure\,\ref{fig:Z-MZ} shows the metallicity profiles at 4.567\,Gyr and the time evolution of $\Zs$ for the K2-MZ simulations with different $\Zacc$ models (see Table\,\ref{tab:MZ}).
Although there exists a substantial difference in the $\Zs$ evolution in the accretion phase, it converges in the MS phase to meet the observed constraints.
The $\delcs$ profiles are also shown in Fig.\,\ref{fig:r-cs-MZ}. As expected, the different $\Zacc$ models do not affect the $\delcs$ profile (see Sect.\,\ref{sec:Zpoor}).

\begin{figure*}[!t]
  \begin{center}
        \includegraphics[width=0.497\hsize,keepaspectratio]{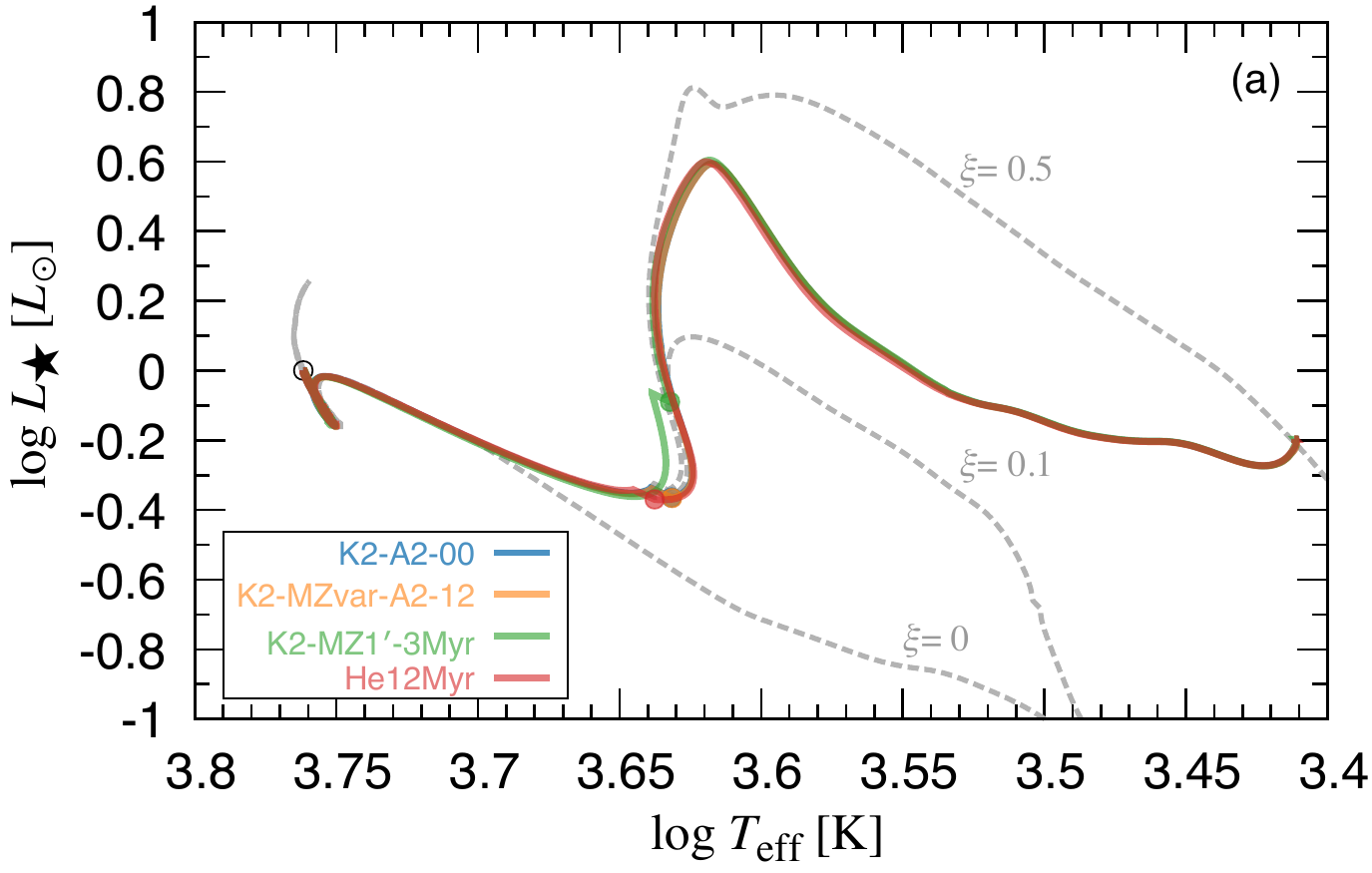}
        \includegraphics[width=0.497\hsize,keepaspectratio]{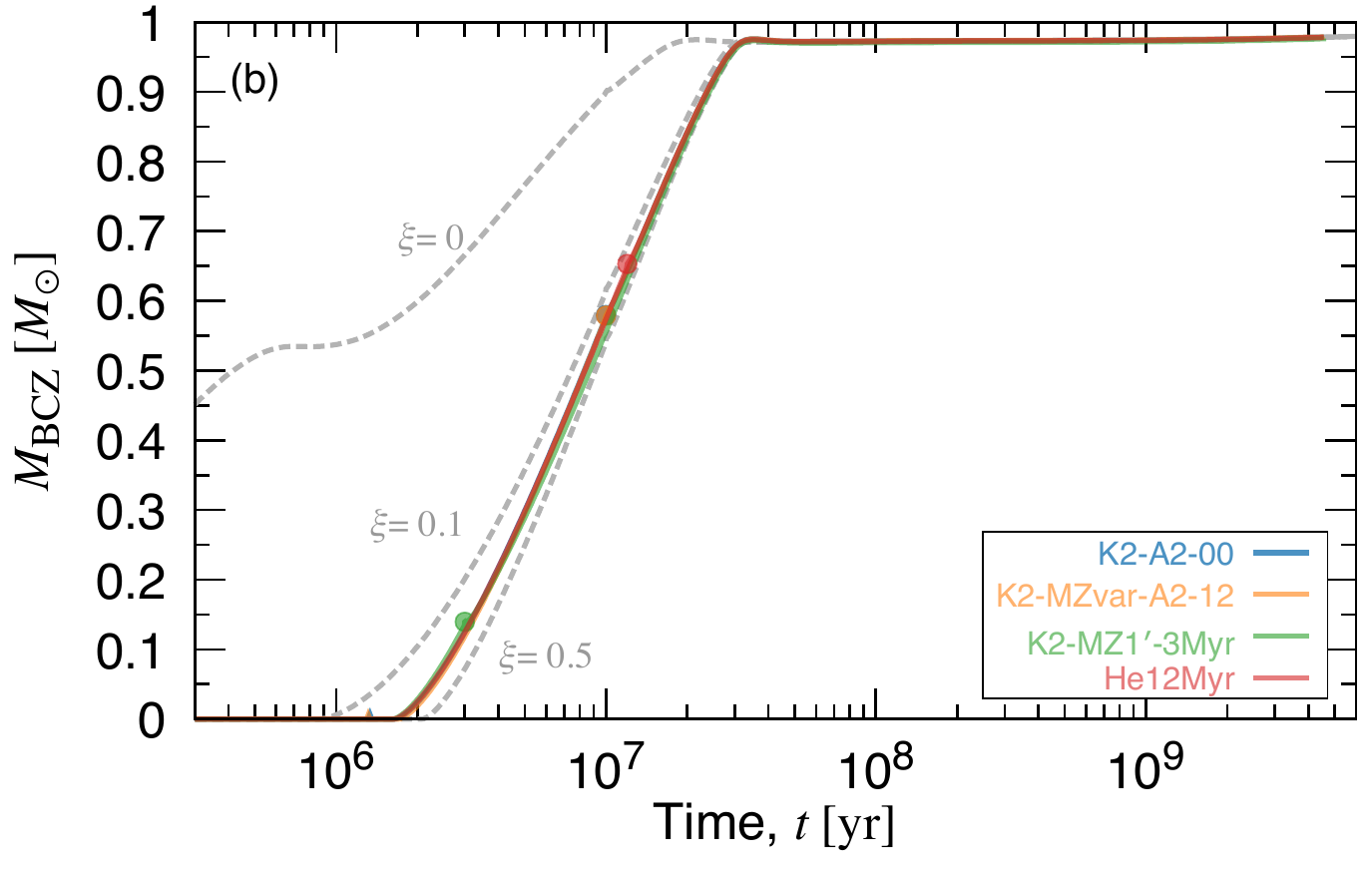}
        \caption{\small{
        Evolutionary tracks on the Hertzsprung-Russell diagram (left panel) and time evolution of the mass coordinate of the base of the CZ $\Mrad$ (right), for models \kapb\ with $A_2=0$ (blue line), \kappla\ with $A_2=0.12$ (orange line), K2-MZ1$'$ with $\fov=0.01$ and $\tacc=3\,$Myr (green line), and He12Myr (red line) (see Table\,\ref{tab:chi2}).
        The filled circles indicate the times at which  accretion ends. The open circle in the left panel indicates the position of the Sun. The gray dashed lines show the evolutionary tracks obtained by \citet[][for $\xi=0.5$, 0.1, and 0]{Kunitomo+18}.
        }}
        \label{fig:HR}
        \label{fig:t-Mrad}
    \end{center}
\end{figure*}

Figures\,\ref{fig:HR}(a) and (b) show various evolutionary tracks on the Hertzsprung-Russell diagram and time evolution of the mass coordinate of the base of the CZ $\Mrad$, respectively.
We found that the tracks are insensitive to $A_2$, $\tacc$, $\Zacc$, and $\Yacc$.
In Fig.\,\ref{fig:HR}, we also show the tracks obtained by \citet[][calculated with a $0.01\,\Msun$ seed using the \texttt{MESA} code version 6596; see their Table\,C.1]{Kunitomo+18}\footnote{See \url{http://cdsarc.u-strasbg.fr/viz-bin/qcat?J/A+A/618/A132}\,.}.
As described in Sect.\,\ref{sec:init}, our initial condition corresponds to a high-entropy $0.1\,\Msun$ stellar seed. Therefore, the initial location of the evolutionary track (i.e., $\log\Teff=3.41$ and $\log\Lstar=-0.2$) coincides with the track for $\xi=0.5$ obtained by \citet{Kunitomo+18}.
Because we adopted $\xi=0.1$ in this work, our tracks deviate from the one with $\xi=0.5$, and the difference in luminosity can be up to $\simeq1$\,dex. However, after the stellar mass reaches $\simeq1\,\Msun$ and they reach their Hayashi track, the tracks agree well.
We note that one finds a small offset in $\Teff$, which arises due to the switching of the outer boundary conditions \citep[see Section 2.7 of ][]{Kunitomo+18}.
Figure\,\ref{fig:t-Mrad}(b) shows that the evolution of $\Mrad$ is also insensitive to $A_2$, $\tacc$, $\Zacc$, and $\Yacc$, and is almost same as that for the case with $\xi=0.5$ obtained by \citet{Kunitomo+18}.

Additional supplemental materials are available online (see the links provided in the footnote on the first page). These include a csv file, summarizing the optimized input parameters and results of all the simulation models, and structure and evolution data of the optimized cases of models K2, K2-MZ5, and K23-MZvar with $A_2=0.12$. In addition, plots showing how the input parameters and results are changed with runs by the simplex method, animations of the evolutions of the optimized case of each model, and the input files of the \texttt{MESA} code are also available at the \href{https://doi.org/10.5281/zenodo.5506424}{Zenodo}.
In the csv file, we list the values of $\Mlost\,[\Msun]$, assuming $\Mpl=150\,\Mearth$, and $\Mpl\,[\Mearth]$, assuming $\Mlost=0$, for the models with planet formation.
Neutrino fluxes, calculated using the subroutine provided in \citet{Farag+20}, are also included in the csv file.

\onecolumn

\begin{longtable}{l|llllllllllll}
	\caption{
	\label{tab:chi2-results-input}
	Input parameters minimized using chi-squared simulations.
	}\\
 	\hline \hline
            Model name & $\amlt$ & $\fov$ & $\Xaccini$ & $\Yaccini$ & $\Zaccini$ & $\Zaccmax$  & $M_1$  & $M_2$   & $A_1$   & $A_2$   & $A_3$   & $\Yaccmin$ \\
            & & & & & & & [$\Msun$] & [$\Msun$] \\
            \hline
            \endfirsthead
            \caption{continued.}\\
            \hline\hline
            Model name & $\amlt$ & $\fov$ & $\Xaccini$ & $\Yaccini$ & $\Zaccini$ & $\Zaccmax$  & $M_1$  & $M_2$   & $A_1$   & $A_2$   & $A_3$   & $\Yaccmin$ \\
            & & & & & & & [$\Msun$] & [$\Msun$] \\
            \hline
            \endhead
            \hline
            \endfoot
            noacc &  1.808 &  0.0132 &  0.710 &  0.274 &  0.0163 &  \unused &  \unused &  \unused &  \unused &  \unused &  \unused &  \unused \\
            noacc-GS98 &  1.866 &  0.0320 &  0.709 &  0.273 &  0.0187 &  \unused &  \unused &  \unused &  \unused &  \unused &  \unused & \unused \\
            noacc-noov &  1.815 &  \unused &  0.708 &  0.276 &  0.0166 &  \unused &  \unused &  \unused &  \unused &  \unused &  \unused & \unused \\
            nooac-GS98-noov &  1.881 &  \unused &  0.706 &  0.275 &  0.0193 &  \unused &  \unused &  \unused &  \unused &  \unused &  \unused & \unused \\
            He12Myr &  1.819 &  0.0238 &  0.697 &  0.288 &  0.0157 &  \unused &  0.87 &  \unused &  \unused &  \unused &  \unused &  0.18 \\
            He20Myr &  1.856 &  0.0266 &  0.695 &  0.289 &  0.0158 &  \unused &  0.91 &  \unused &  \unused &  \unused &  \unused &  0.11 \\
            MZvar &  1.801 &  0.0012 &  0.709 &  0.275 &  0.0165 &  0.0533 &  0.90 &  0.94 &  \unused &  \unused &  \unused &  \unused \\
            MZvar-noov &  1.808 &  \unused &  0.709 &  0.277 &  0.0146 &  0.0701 &  0.89 &  0.97 &  \unused &  \unused &  \unused &  \unused \\
            K2-A2-00 &  1.808 &  0.0005 &  0.710 &  0.274 &  0.0161 &  \unused & \unused & \unused &  \fixed{0.000} &  \fixed{0.000} &  \fixed{0.000} &  \unused \\
            K2-A2-04 &  1.816 &  0.0052 &  0.711 &  0.273 &  0.0157 &  \unused & \unused & \unused &  \fixed{0.000} &  \fixed{0.040} &  \fixed{0.000} &  \unused \\
            K2-A2-06 &  1.814 &  0.0049 &  0.712 &  0.272 &  0.0154 &  \unused & \unused & \unused &  \fixed{0.000} &  \fixed{0.060} &  \fixed{0.000} &  \unused \\
            K2-A2-08 &  1.820 &  0.0033 &  0.712 &  0.273 &  0.0152 &  \unused & \unused & \unused &  \fixed{0.000} &  \fixed{0.080} &  \fixed{0.000} &  \unused \\
            K2-A2-10 &  1.821 &  0.0039 &  0.713 &  0.272 &  0.0150 &  \unused & \unused & \unused &  \fixed{0.000} &  \fixed{0.100} &  \fixed{0.000} &  \unused \\
            K2-A2-12 &  1.821 &  0.0103 &  0.714 &  0.271 &  0.0147 &  \unused & \unused & \unused &  \fixed{0.000} &  \fixed{0.120} &  \fixed{0.000} &  \unused \\
            K2-A2-14 &  1.811 &  0.0057 &  0.717 &  0.268 &  0.0142 &  \unused & \unused & \unused &  \fixed{0.000} &  \fixed{0.140} &  \fixed{0.000} &  \unused \\
            K2-A2-15 &  1.821 &  0.0097 &  0.716 &  0.270 &  0.0143 &  \unused & \unused & \unused &  \fixed{0.000} &  \fixed{0.150} &  \fixed{0.000} &  \unused \\
            K2-A2-16 &  1.817 &  0.0087 &  0.717 &  0.269 &  0.0141 &  \unused & \unused & \unused &  \fixed{0.000} &  \fixed{0.160} &  \fixed{0.000} &  \unused \\
            K2-A2-18 &  1.820 &  0.0093 &  0.717 &  0.269 &  0.0140 &  \unused & \unused & \unused &  \fixed{0.000} &  \fixed{0.180} &  \fixed{0.000} &  \unused \\
            K2-A2-20 &  1.825 &  0.0064 &  0.717 &  0.269 &  0.0140 &  \unused & \unused & \unused &  \fixed{0.000} &  \fixed{0.200} &  \fixed{0.000} &  \unused \\
            K2-A2-22 &  1.820 &  0.0075 &  0.719 &  0.268 &  0.0136 &  \unused & \unused & \unused &  \fixed{0.000} &  \fixed{0.220} &  \fixed{0.000} &  \unused \\
            K23-A2-00 &  1.862 &  0.0094 &  0.719 &  0.265 &  0.0160 &  \unused & \unused & \unused &  \fixed{0.000} &  \fixed{0.000} & -0.084 &  \unused \\
            K23-A2-04 &  1.843 &  0.0336 &  0.718 &  0.266 &  0.0154 &  \unused & \unused & \unused &  \fixed{0.000} &  \fixed{0.040} & -0.058 &  \unused \\
            K23-A2-06 &  1.841 &  0.0092 &  0.717 &  0.268 &  0.0153 &  \unused & \unused & \unused &  \fixed{0.000} &  \fixed{0.060} & -0.043 &  \unused \\
            K23-A2-08 &  1.832 &  0.0040 &  0.717 &  0.268 &  0.0150 &  \unused & \unused & \unused &  \fixed{0.000} &  \fixed{0.080} & -0.033 &  \unused \\
            K23-A2-10 &  1.822 &  0.0040 &  0.712 &  0.273 &  0.0151 &  \unused & \unused & \unused &  \fixed{0.000} &  \fixed{0.100} &  0.001 &  \unused \\
            K23-A2-12 &  1.819 &  0.0029 &  0.714 &  0.271 &  0.0147 &  \unused & \unused & \unused &  \fixed{0.000} &  \fixed{0.120} &  0.004 &  \unused \\
            K23-A2-14 &  1.820 &  0.0014 &  0.714 &  0.271 &  0.0146 &  \unused & \unused & \unused &  \fixed{0.000} &  \fixed{0.140} &  0.004 &  \unused \\
            K23-A2-15 &  1.803 &  0.0062 &  0.717 &  0.269 &  0.0140 &  \unused & \unused & \unused &  \fixed{0.000} &  \fixed{0.150} &  0.013 &  \unused \\
            K23-A2-16 &  1.807 &  0.0057 &  0.717 &  0.269 &  0.0140 &  \unused & \unused & \unused &  \fixed{0.000} &  \fixed{0.160} &  0.011 &  \unused \\
            K23-A2-18 &  1.794 &  0.0032 &  0.712 &  0.274 &  0.0141 &  \unused & \unused & \unused &  \fixed{0.000} &  \fixed{0.180} &  0.049 &  \unused \\
            K23-A2-20 &  1.795 &  0.0002 &  0.712 &  0.274 &  0.0139 &  \unused & \unused & \unused &  \fixed{0.000} &  \fixed{0.200} &  0.052 &  \unused \\
            K23-A2-22 &  1.783 &  0.0014 &  0.712 &  0.274 &  0.0137 &  \unused & \unused & \unused &  \fixed{0.000} &  \fixed{0.220} &  0.067 &  \unused \\
            K23$'$-A3-05 &  1.790 &  0.0042 &  0.709 &  0.276 &  0.0150 &  \unused & \unused & \unused &  \fixed{0.000} &  \fixed{0.100} &  \fixed{0.050} &  \unused \\
            K23$'$-A3--05 &  1.854 &  0.0119 &  0.717 &  0.268 &  0.0151 &  \unused & \unused & \unused &  \fixed{0.000} &  \fixed{0.100} & \fixed{-0.050} &  \unused \\
            K2$'$ &  1.814 &  \fixed{0.0100} &  0.716 &  0.270 &  0.0146 &  \unused & \unused & \unused &  \fixed{0.000} &  \fixed{0.120} &  \fixed{0.000} &  \unused \\
            K2$'$-3Myr &  1.814 &  \fixed{0.0100} &  0.716 &  0.270 &  0.0145 &  \unused & \unused & \unused &  \fixed{0.000} &  \fixed{0.120} &  \fixed{0.000} &  \unused \\
            K2$'$-10Myr &  1.814 &  \fixed{0.0100} &  0.716 &  0.270 &  0.0145 &  \unused & \unused & \unused &  \fixed{0.000} &  \fixed{0.120} &  \fixed{0.000} &  \unused \\
            K2$'$-A2-15-10Myr &  1.818 &  \fixed{0.0100} &  0.716 &  0.269 &  0.0143 &  \unused & \unused & \unused &  \fixed{0.000} &  \fixed{0.150} &  \fixed{0.000} &  \unused \\
            K2$'$-A2-18-10Myr &  1.823 &  \fixed{0.0100} &  0.716 &  0.269 &  0.0141 &  \unused & \unused & \unused &  \fixed{0.000} &  \fixed{0.180} &  \fixed{0.000} &  \unused \\
            K2$'$-fov-0 &  1.816 &  \fixed{0.0000} &  0.715 &  0.270 &  0.0147 &  \unused & \unused & \unused &  \fixed{0.000} &  \fixed{0.120} &  \fixed{0.000} &  \unused \\
            K2$'$-fov-025 &  1.813 &  \fixed{0.0250} &  0.716 &  0.270 &  0.0145 &  \unused & \unused & \unused &  \fixed{0.000} &  \fixed{0.120} &  \fixed{0.000} &  \unused \\
            K2-MZvar-A2-12 &  1.817 &  0.0042 &  0.713 &  0.273 &  0.0140 &  0.0652 &  0.90 &  0.96 &  \fixed{0.000} &  \fixed{0.120} &  \fixed{0.000} &  \unused \\
            K2-MZvar-A2-15 &  1.817 &  0.0067 &  0.715 &  0.272 &  0.0130 &  0.0653 &  0.89 &  0.97 &  \fixed{0.000} &  \fixed{0.150} &  \fixed{0.000} &  \unused \\
            K2-MZvar-A2-18 &  1.819 &  0.0064 &  0.716 &  0.271 &  0.0134 &  0.0660 &  0.91 &  0.96 &  \fixed{0.000} &  \fixed{0.180} &  \fixed{0.000} &  \unused \\
            K23-MZvar-A2-12 &  1.818 &  0.0054 &  0.714 &  0.273 &  0.0136 &  0.0674 &  0.90 &  0.97 &  \fixed{0.000} &  \fixed{0.120} & -0.001 &  \unused \\
            K23-MZvar-A2-15 &  1.810 &  0.0057 &  0.714 &  0.273 &  0.0136 &  0.0729 &  0.90 &  0.96 &  \fixed{0.000} &  \fixed{0.150} &  0.012 &  \unused \\
            K23-MZvar-A2-18 &  1.796 &  0.0056 &  0.713 &  0.274 &  0.0133 &  0.0637 &  0.91 &  0.96 &  \fixed{0.000} &  \fixed{0.180} &  0.038 &  \unused \\
            K2-MZ1 &  1.819 &  \fixed{0.0100} &  0.714 &  0.271 &  0.0155 &  0.0755 &  \fixed{0.90} & \fixed{0.92} & \fixed{0.000} &  \fixed{0.120} &  \fixed{0.000} &  \unused \\
            K2-MZ2 &  1.813 &  \fixed{0.0100} &  0.716 &  0.270 &  0.0144 &  0.0744 &  \fixed{0.95} &  \fixed{0.97} &  \fixed{0.000} &  \fixed{0.120} &  \fixed{0.000} &  \unused \\
            K2-MZ3 &  1.819 &  \fixed{0.0100} &  0.714 &  0.271 &  0.0155 &  0.0455 &  \fixed{0.88} &  \fixed{0.92} &  \fixed{0.000} &  \fixed{0.120} &  \fixed{0.000} &  \unused \\
            K2-MZ4 &  1.818 &  \fixed{0.0100} &  0.714 &  0.271 &  0.0148 &  0.0748 &  \fixed{0.88} &  \fixed{0.92} &  \fixed{0.000} &  \fixed{0.120} &  \fixed{0.000} &  \unused \\
            K2-MZ5 &  1.815 &  \fixed{0.0100} &  0.715 &  0.270 &  0.0152 &  0.0152 &  0.97 &  0.97 &  \fixed{0.000} &  \fixed{0.120} &  \fixed{0.000} &  \unused \\
            K2-MZ6 &  1.819 &  \fixed{0.0100} &  0.713 &  0.271 &  0.0155 &  0.0355 &  \fixed{0.86} &  \fixed{0.92} &  \fixed{0.000} &  \fixed{0.120} &  \fixed{0.000} &  \unused \\
            K2-MZ7 &  1.819 &  \fixed{0.0100} &  0.715 &  0.271 &  0.0140 &  0.0740 &  \fixed{0.86} &  \fixed{0.92} &  \fixed{0.000} &  \fixed{0.120} &  \fixed{0.000} &  \unused \\
            K2-MZ8 &  1.818 &  \fixed{0.0100} &  0.715 &  0.271 &  0.0132 &  0.0732 &  \fixed{0.91} & \fixed{0.97} & \fixed{0.000} &  \fixed{0.120} &  \fixed{0.000} &  \unused \\
            K2-MZ9 &  1.819 &  \fixed{0.0100} &  0.714 &  0.271 &  0.0151 &  0.0751 &  \fixed{0.92} &  \fixed{0.94} &  \fixed{0.000} &  \fixed{0.120} &  \fixed{0.000} &  \unused \\
            K2-MZ1$'$-3Myr &  1.818 &  \fixed{0.0100} &  0.714 &  0.270 &  0.0154 &  0.0754 &  \fixed{0.90} & \fixed{0.92} & \fixed{0.000} &  \fixed{0.120} &  \fixed{0.000} &  \unused \\
            K2-MZ1$'$-10Myr &  1.814 &  \fixed{0.0100} &  0.713 &  0.271 &  0.0157 &  0.0757 &  \fixed{0.90} & \fixed{0.92} & \fixed{0.000} &  \fixed{0.120} &  \fixed{0.000} &  \unused \\
            K2-MZ1$'$-A2-15-10Myr &  1.813 &  \fixed{0.0100} &  0.715 &  0.270 &  0.0152 &  0.0752 &  \fixed{0.90} & \fixed{0.92} & \fixed{0.000} &  \fixed{0.150} &  \fixed{0.000} &  \unused \\
            K2-MZ1$'$-A2-18-10Myr &  1.817 &  \fixed{0.0100} &  0.715 &  0.270 &  0.0150 &  0.0750 &  \fixed{0.90} & \fixed{0.92} & \fixed{0.000} &  \fixed{0.180} &  \fixed{0.000} &  \unused \\
            K2-MZ1$'$-fov-0 &  1.822 &  \fixed{0.0000} &  0.713 &  0.272 &  0.0157 &  0.0757 &  \fixed{0.90} & \fixed{0.92} & \fixed{0.000} &  \fixed{0.120} &  \fixed{0.000} &  \unused \\
            K2-MZ1$'$-fov-025 &  1.810 &  \fixed{0.0250} &  0.716 &  0.269 &  0.0152 &  0.0752 &  \fixed{0.90} & \fixed{0.92} & \fixed{0.000} &  \fixed{0.120}&  \fixed{0.000} &  \unused \\
            K2-MZ8$'$-10Myr &  1.818 &  \fixed{0.0100} &  0.714 &  0.273 &  0.0135 &  0.0735 &  \fixed{0.91} & \fixed{0.97} & \fixed{0.000} &  \fixed{0.120} &  \fixed{0.000} &  \unused \\
            K2-MZ8$'$-A2-15-10Myr &  1.822 &  \fixed{0.0100} &  0.715 &  0.272 &  0.0132 &  0.0732 &  \fixed{0.91} & \fixed{0.97} & \fixed{0.000} &  \fixed{0.150} &  \fixed{0.000} &  \unused \\
            K2-MZ8$'$-A2-18-10Myr &  1.818 &  \fixed{0.0100} &  0.717 &  0.270 &  0.0127 &  0.0727 &  \fixed{0.91} & \fixed{0.97} & \fixed{0.000} &  \fixed{0.180} &  \fixed{0.000} &  \unused \\
         \end{longtable}
         \tablefoot{
         The numbers in roman font represent the values obtained by the minimization process. 
         The numbers in italics represent the values set a priori (see Table\,\ref{tab:chi2} for the parameter settings of each model).
         This table is available in electric form at the CDS and \href{https://doi.org/10.5281/zenodo.5506424}{Zenodo}.
        }
         

\begin{longtable}{l|llllllllll}
	\caption{
	\label{tab:chi2-results-output}
	Results minimized by the chi-squared simulations.}\\
    \hline\hline
            Model name & $\chi^2$ & rms($\delcs$) & $\ZXs$ & $\Ys$ & $\RCZ$ & $\log\,\Lstar$ & $\Teff$ & $\Zc$ & $\Zs$ & $\Zc/\Zs$\\
             & & & & & [$\Rsun$] & [$\Lsun$] & [K]\\
            \hline
            \endfirsthead
            \caption{continued.}\\
            \hline\hline
            Model name & $\chi^2$ & rms($\delcs$) & $\ZXs$ & $\Ys$ & $\RCZ$ & $\log\,\Lstar$ & $\Teff$ & $\Zc$ & $\Zs$ & $\Zc/\Zs$\\
             & & & & & [$\Rsun$] & [$\Lsun$] & [K]\\
            \hline
            \endhead
            \hline
            \endfoot
            noacc &  2.93 & 3.4e-03 &  0.0203 & {\bf  0.248} & {\bf  0.722} & {\bf 4e-04} & {\bf 5778} &  0.0174 &  0.0150 &  1.1575 \\
            noacc-GS98 & {\bf  0.69} & 1.8e-03 & {\bf  0.0237} & {\bf  0.249} & {\bf  0.718} & {\bf 4e-04} & {\bf 5777} &  0.0199 &  0.0174 &  1.1431 \\
            noacc-noov &  2.81 & 3.2e-03 &  0.0204 & {\bf  0.247} &  0.723 & {\bf 1e-03} & {\bf 5779} &  0.0176 &  0.0150 &  1.1711 \\
            nooac-GS98-noov & {\bf  0.62} & 1.6e-03 & {\bf  0.0239} & {\bf  0.248} & {\bf  0.717} & {\bf 4e-04} & {\bf 5777} &  0.0205 &  0.0176 &  1.1626 \\
            He12Myr &  2.20 & 2.8e-03 &  0.0202 & {\bf  0.247} & {\bf  0.721} & {\bf 7e-04} & {\bf 5779} &  0.0171 &  0.0149 &  1.1489 \\
            He20Myr &  1.48 & 1.4e-03 &  0.0203 &  0.244 & {\bf  0.718} & {\bf 1e-03} & {\bf 5780} &  0.0172 &  0.0151 &  1.1443 \\
            MZvar &  3.18 & 3.7e-03 &  0.0202 & {\bf  0.247} &  0.723 & {\bf -4e-06} & {\bf 5776} &  0.0185 &  0.0149 &  1.2402 \\
            MZvar-noov &  3.17 & 3.5e-03 &  0.0205 & {\bf  0.248} &  0.723 & {\bf 6e-04} & {\bf 5779} &  0.0181 &  0.0151 &  1.2002 \\
            K2-A2-00 &  2.86 & 3.4e-03 &  0.0202 & {\bf  0.246} &  0.723 & {\bf 6e-04} & {\bf 5778} &  0.0175 &  0.0149 &  1.1732 \\
            K2-A2-04 &  1.52 & 2.3e-03 &  0.0199 & {\bf  0.247} & {\bf  0.720} & {\bf 6e-04} & {\bf 5778} &  0.0171 &  0.0147 &  1.1636 \\
            K2-A2-06 &  1.06 & 1.9e-03 &  0.0196 & {\bf  0.246} & {\bf  0.719} & {\bf 2e-04} & {\bf 5777} &  0.0168 &  0.0145 &  1.1631 \\
            K2-A2-08 & {\bf  0.72} & 1.4e-03 &  0.0194 & {\bf  0.247} & {\bf  0.717} & {\bf 6e-04} & {\bf 5778} &  0.0167 &  0.0144 &  1.1636 \\
            K2-A2-10 & {\bf  0.50} & 1.1e-03 &  0.0192 & {\bf  0.246} & {\bf  0.716} & {\bf 4e-04} & {\bf 5778} &  0.0165 &  0.0142 &  1.1620 \\
            K2-A2-12 & {\bf  0.36} & {\bf 9.9e-04} & {\bf  0.0190} & {\bf  0.246} & {\bf  0.715} & {\bf 5e-04} & {\bf 5778} &  0.0162 &  0.0140 &  1.1561 \\
            K2-A2-14 & {\bf  0.51} & 1.1e-03 & {\bf  0.0182} &  0.244 & {\bf  0.715} & {\bf -8e-05} & {\bf 5776} &  0.0157 &  0.0135 &  1.1596 \\
            K2-A2-15 & {\bf  0.37} & 1.2e-03 & {\bf  0.0185} & {\bf  0.246} & {\bf  0.714} & {\bf 6e-04} & {\bf 5779} &  0.0159 &  0.0137 &  1.1556 \\
            K2-A2-16 & {\bf  0.45} & 1.2e-03 & {\bf  0.0183} &  0.245 & {\bf  0.714} & {\bf 2e-04} & {\bf 5777} &  0.0157 &  0.0135 &  1.1562 \\
            K2-A2-18 & {\bf  0.58} & 1.6e-03 & {\bf  0.0182} &  0.245 & {\bf  0.713} & {\bf 1e-04} & {\bf 5777} &  0.0156 &  0.0135 &  1.1550 \\
            K2-A2-20 & {\bf  0.84} & 2.0e-03 & {\bf  0.0181} &  0.245 & {\bf  0.713} & {\bf 3e-04} & {\bf 5778} &  0.0155 &  0.0134 &  1.1562 \\
            K2-A2-22 &  1.15 & 2.3e-03 & {\bf  0.0177} &  0.244 & {\bf  0.712} & {\bf -9e-05} & {\bf 5776} &  0.0152 &  0.0131 &  1.1552 \\
            K23-A2-00 &  1.81 & 1.4e-03 &  0.0202 &  0.241 & {\bf  0.717} & {\bf 9e-04} & {\bf 5779} &  0.0174 &  0.0150 &  1.1569 \\
            K23-A2-04 & {\bf  0.91} & {\bf 9.9e-04} &  0.0199 &  0.245 & {\bf  0.717} & {\bf 9e-04} & {\bf 5779} &  0.0168 &  0.0147 &  1.1442 \\
            K23-A2-06 & {\bf  0.87} & 1.1e-03 &  0.0195 &  0.244 & {\bf  0.716} & {\bf 7e-04} & {\bf 5779} &  0.0168 &  0.0145 &  1.1569 \\
            K23-A2-08 & {\bf  0.79} & 1.0e-03 &  0.0191 &  0.243 & {\bf  0.716} & {\bf 5e-05} & {\bf 5777} &  0.0164 &  0.0142 &  1.1614 \\
            K23-A2-10 & {\bf  0.51} & 1.1e-03 &  0.0194 & {\bf  0.247} & {\bf  0.716} & {\bf -2e-05} & {\bf 5776} &  0.0166 &  0.0143 &  1.1616 \\
            K23-A2-12 & {\bf  0.41} & 1.1e-03 & {\bf  0.0188} & {\bf  0.246} & {\bf  0.716} & {\bf 9e-04} & {\bf 5779} &  0.0162 &  0.0139 &  1.1628 \\
            K23-A2-14 & {\bf  0.39} & 1.1e-03 & {\bf  0.0187} & {\bf  0.246} & {\bf  0.715} & {\bf 2e-04} & {\bf 5777} &  0.0161 &  0.0138 &  1.1638 \\
            K23-A2-15 & {\bf  0.46} & 1.1e-03 & {\bf  0.0181} &  0.244 & {\bf  0.716} & {\bf 4e-05} & {\bf 5777} &  0.0155 &  0.0134 &  1.1595 \\
            K23-A2-16 & {\bf  0.44} & 1.1e-03 & {\bf  0.0181} &  0.244 & {\bf  0.715} & {\bf -2e-04} & {\bf 5776} &  0.0156 &  0.0134 &  1.1594 \\
            K23-A2-18 & {\bf  0.25} & 1.2e-03 & {\bf  0.0183} & {\bf  0.248} & {\bf  0.715} & {\bf -1e-04} & {\bf 5776} &  0.0157 &  0.0135 &  1.1622 \\
            K23-A2-20 & {\bf  0.31} & 1.3e-03 & {\bf  0.0179} & {\bf  0.248} & {\bf  0.715} & {\bf 2e-04} & {\bf 5778} &  0.0154 &  0.0133 &  1.1650 \\
            K23-A2-22 & {\bf  0.35} & 1.4e-03 & {\bf  0.0177} & {\bf  0.248} & {\bf  0.715} & {\bf -3e-04} & {\bf 5775} &  0.0152 &  0.0131 &  1.1645 \\
            K23'-A3-05 &  1.04 & 2.2e-03 &  0.0192 & {\bf  0.249} & {\bf  0.719} & {\bf 7e-05} & {\bf 5777} &  0.0164 &  0.0141 &  1.1649 \\
            K23'-A3--05 & {\bf  0.80} & 1.3e-03 &  0.0194 &  0.244 & {\bf  0.713} & {\bf 7e-04} & {\bf 5779} &  0.0166 &  0.0144 &  1.1528 \\
            K2$'$ & {\bf  0.39} & 1.0e-03 & {\bf  0.0188} &  0.245 & {\bf  0.716} & {\bf -2e-04} & {\bf 5776} &  0.0161 &  0.0139 &  1.1555 \\
            K2$'$-3Myr & {\bf  0.39} & 1.0e-03 & {\bf  0.0187} &  0.245 & {\bf  0.716} & {\bf 1e-05} & {\bf 5777} &  0.0160 &  0.0139 &  1.1549 \\
            K2$'$-10Myr & {\bf  0.40} & 1.1e-03 & {\bf  0.0187} &  0.245 & {\bf  0.716} & {\bf 4e-04} & {\bf 5777} &  0.0160 &  0.0138 &  1.1566 \\
            K2$'$-A2-15-10Myr & {\bf  0.37} & 1.1e-03 & {\bf  0.0185} &  0.245 & {\bf  0.714} & {\bf 6e-05} & {\bf 5777} &  0.0158 &  0.0137 &  1.1554 \\
            K2$'$-A2-18-10Myr & {\bf  0.56} & 1.6e-03 & {\bf  0.0183} & {\bf  0.246} & {\bf  0.713} & {\bf 4e-04} & {\bf 5778} &  0.0156 &  0.0136 &  1.1543 \\
            K2$'$-fov-0 & {\bf  0.48} & 1.1e-03 & {\bf  0.0187} &  0.245 & {\bf  0.716} & {\bf -4e-04} & {\bf 5775} &  0.0162 &  0.0139 &  1.1647 \\
            K2$'$-fov-025 & {\bf  0.29} & {\bf 9.2e-04} & {\bf  0.0189} & {\bf  0.247} & {\bf  0.716} & {\bf 2e-04} & {\bf 5777} &  0.0160 &  0.0140 &  1.1474 \\
            K2-MZvar-A2-12 & {\bf  0.39} & 1.1e-03 & {\bf  0.0190} & {\bf  0.247} & {\bf  0.716} & {\bf 3e-04} & {\bf 5778} &  0.0171 &  0.0141 &  1.2140 \\
            K2-MZvar-A2-15 & {\bf  0.33} & 1.1e-03 & {\bf  0.0186} & {\bf  0.246} & {\bf  0.714} & {\bf 3e-04} & {\bf 5777} &  0.0167 &  0.0138 &  1.2091 \\
            K2-MZvar-A2-18 & {\bf  0.48} & 1.5e-03 & {\bf  0.0183} & {\bf  0.246} & {\bf  0.713} & {\bf 4e-04} & {\bf 5778} &  0.0164 &  0.0135 &  1.2113 \\
            K23-MZvar-A2-12 & {\bf  0.38} & 1.1e-03 & {\bf  0.0190} & {\bf  0.247} & {\bf  0.715} & {\bf 3e-04} & {\bf 5778} &  0.0169 &  0.0141 &  1.2007 \\
            K23-MZvar-A2-15 & {\bf  0.28} & 1.1e-03 & {\bf  0.0187} & {\bf  0.247} & {\bf  0.715} & {\bf 2e-04} & {\bf 5777} &  0.0168 &  0.0138 &  1.2149 \\
            K23-MZvar-A2-18 & {\bf  0.25} & 1.2e-03 & {\bf  0.0183} & {\bf  0.248} & {\bf  0.715} & {\bf 2e-04} & {\bf 5777} &  0.0164 &  0.0135 &  1.2146 \\
            K2-MZ1 & {\bf  0.36} & 1.0e-03 & {\bf  0.0189} & {\bf  0.246} & {\bf  0.715} & {\bf 5e-04} & {\bf 5778} &  0.0166 &  0.0140 &  1.1834 \\
            K2-MZ2 & {\bf  0.39} & 1.1e-03 & {\bf  0.0187} &  0.245 & {\bf  0.716} & {\bf 2e-04} & {\bf 5777} &  0.0164 &  0.0138 &  1.1829 \\
            K2-MZ3 & {\bf  0.35} & 1.0e-03 & {\bf  0.0190} & {\bf  0.246} & {\bf  0.715} & {\bf 4e-04} & {\bf 5778} &  0.0166 &  0.0140 &  1.1832 \\
            K2-MZ4 & {\bf  0.35} & 1.0e-03 & {\bf  0.0189} & {\bf  0.246} & {\bf  0.715} & {\bf 3e-04} & {\bf 5777} &  0.0166 &  0.0140 &  1.1832 \\
            K2-MZ5 & {\bf  0.37} & 1.0e-03 & {\bf  0.0188} & {\bf  0.246} & {\bf  0.716} & {\bf -1e-04} & {\bf 5776} &  0.0165 &  0.0139 &  1.1831 \\
            K2-MZ6 & {\bf  0.35} & {\bf 9.9e-04} & {\bf  0.0190} & {\bf  0.247} & {\bf  0.715} & {\bf 3e-04} & {\bf 5778} &  0.0166 &  0.0140 &  1.1832 \\
            K2-MZ7 & {\bf  0.36} & 1.0e-03 & {\bf  0.0189} & {\bf  0.246} & {\bf  0.715} & {\bf 5e-04} & {\bf 5778} &  0.0166 &  0.0140 &  1.1833 \\
            K2-MZ8 & {\bf  0.36} & 1.0e-03 & {\bf  0.0189} & {\bf  0.246} & {\bf  0.715} & {\bf 2e-04} & {\bf 5777} &  0.0166 &  0.0140 &  1.1825 \\
            K2-MZ9 & {\bf  0.35} & 1.0e-03 & {\bf  0.0189} & {\bf  0.246} & {\bf  0.715} & {\bf 4e-04} & {\bf 5778} &  0.0166 &  0.0140 &  1.1832 \\
            K2-MZ1$'$-3Myr & {\bf  0.36} & 1.0e-03 & {\bf  0.0189} & {\bf  0.246} & {\bf  0.715} & {\bf 6e-05} & {\bf 5777} &  0.0163 &  0.0140 &  1.1644 \\
            K2-MZ1$'$-10Myr & {\bf  0.37} & 1.1e-03 & {\bf  0.0189} & {\bf  0.247} & {\bf  0.716} & {\bf 3e-04} & {\bf 5778} &  0.0169 &  0.0140 &  1.2098 \\
            K2-MZ1$'$-A2-15-10Myr & {\bf  0.32} & 1.1e-03 & {\bf  0.0184} & {\bf  0.246} & {\bf  0.715} & {\bf 3e-04} & {\bf 5777} &  0.0165 &  0.0137 &  1.2080 \\
            K2-MZ1$'$-A2-18-10Myr & {\bf  0.45} & 1.5e-03 & {\bf  0.0183} & {\bf  0.246} & {\bf  0.713} & {\bf 3e-04} & {\bf 5778} &  0.0163 &  0.0135 &  1.2061 \\
            K2-MZ1$'$-fov-0 & {\bf  0.41} & 1.0e-03 & {\bf  0.0190} & {\bf  0.246} & {\bf  0.716} & {\bf 4e-04} & {\bf 5778} &  0.0169 &  0.0140 &  1.2018 \\
            K2-MZ1$'$-fov-025 & {\bf  0.29} & {\bf 9.5e-04} & {\bf  0.0188} & {\bf  0.247} & {\bf  0.716} & {\bf 4e-05} & {\bf 5777} &  0.0162 &  0.0139 &  1.1667 \\
            K2-MZ8$'$-10Myr & {\bf  0.38} & 1.0e-03 &  0.0191 & {\bf  0.247} & {\bf  0.716} & {\bf 6e-04} & {\bf 5779} &  0.0167 &  0.0141 &  1.1853 \\
            K2-MZ8$'$-A2-15-10Myr & {\bf  0.34} & 1.2e-03 & {\bf  0.0188} & {\bf  0.247} & {\bf  0.714} & {\bf 7e-04} & {\bf 5779} &  0.0165 &  0.0139 &  1.1843 \\
            K2-MZ8$'$-A2-18-10Myr & {\bf  0.46} & 1.5e-03 & {\bf  0.0183} & {\bf  0.246} & {\bf  0.713} & {\bf 3e-04} & {\bf 5778} &  0.0160 &  0.0135 &  1.1836 \\
\end{longtable}
\tablefoot{
         The numbers highlighted in bold indicate the values that satisfy the constraints listed in Table\,\ref{tab:targets}
         or have $\chi^2<1$.
         This table is available in electric form at the CDS and \href{https://doi.org/10.5281/zenodo.5506424}{Zenodo}.
         }

\end{appendix}

\end{document}